\documentclass[letterpaper,12pt]{article}
\pdfoutput=1

\usepackage{enumerate}
\usepackage{jheppub}
\usepackage{amsmath}
\usepackage{graphicx}
\usepackage{subfig}
\usepackage{cancel}
\usepackage[dvipsnames]{xcolor}
\usepackage{color}
\usepackage{mathrsfs}
\usepackage{amssymb}
\usepackage{epstopdf}
\usepackage{dsfont}
\usepackage{float}

\graphicspath{{./Figures/}}

\newcommand*{\cyl}{\vcenter{\hbox{\includegraphics[width=1.8em]{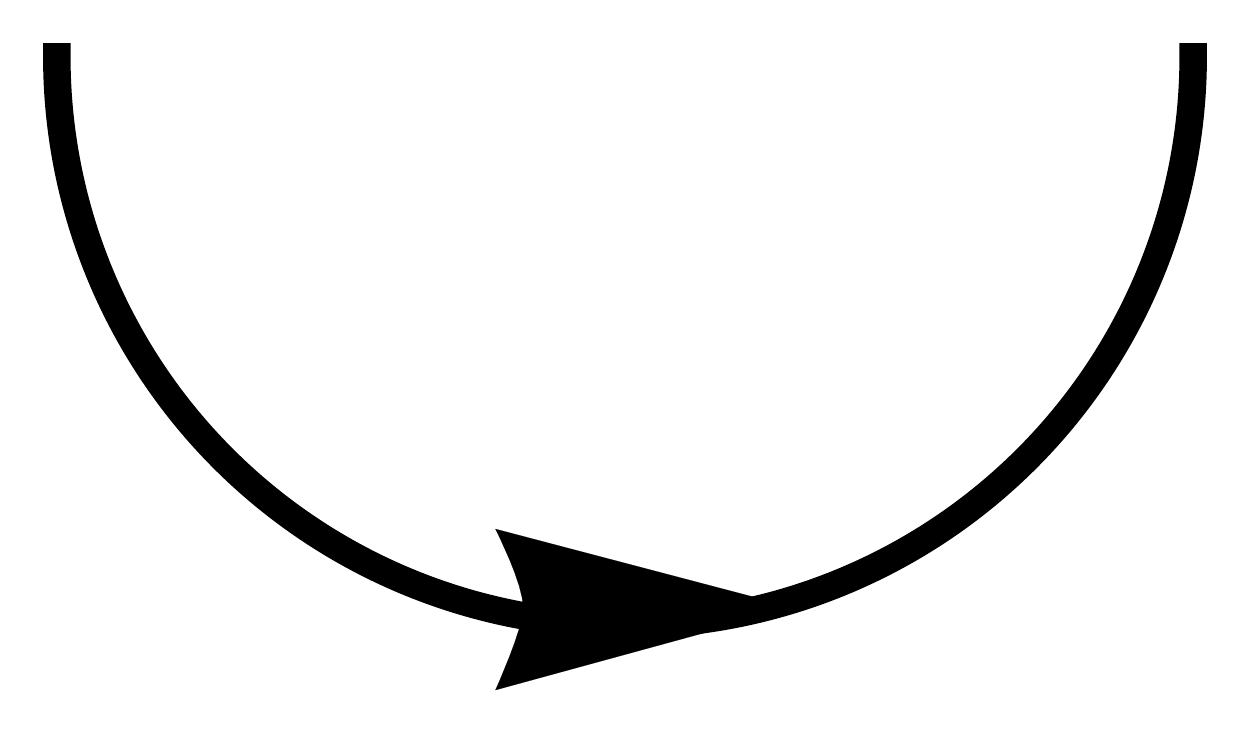}}}}
\newcommand*{\hilb}{\mathcal{H}}		% Hilbert space
\newcommand*{\hbu}{\mathcal{H}_\text{BU}}	% Baby universe Hilbert space
\newcommand*{\NN}{\mathbb{N}}		% Natural numbers
\newcommand*{\RR}{\mathbb{R}}		% Real numbers
\newcommand*{\CC}{\mathbb{C}}		% Complex numbers

\renewcommand*{\Re}{\operatorname{Re}}
\renewcommand*{\Im}{\operatorname{Im}}
\renewcommand*{\emptyset}{\varnothing}
\DeclareMathOperator{\Tr}{Tr}
\DeclareMathOperator{\rank}{rank}

\newcommand*{\HH}{\mathrm{HH}}	% Hartle-Hawking
\newcommand*{\dmanifold}{\mathcal{M}}		% asymptotic boundary manifold

\title{
Transcending the ensemble: baby universes, spacetime wormholes, and the order and disorder of black hole information}

\author{Donald Marolf}
\author{and Henry Maxfield}

\affiliation{Department of Physics, University of California, Santa Barbara, CA 93106, USA}

\emailAdd{marolf@physics.ucsb.edu}

\emailAdd{hmaxfield@physics.ucsb.edu}

\abstract{
In the 1980's, work by Coleman and by Giddings and Strominger linked the physics of spacetime wormholes to `baby universes' and an ensemble of theories.  We revisit such ideas, using features associated with a negative cosmological constant and asymptotically AdS boundaries to strengthen the results, introduce a change in perspective, and connect with recent  replica wormhole discussions of the Page curve. A key new feature is an emphasis on the role of null states.   We explore this structure in detail in simple topological models of the bulk that allow us to compute the full spectrum of associated boundary theories. The dimension of the asymptotically AdS Hilbert space turns out to become a random variable $Z$, whose value can be less than the naive number $k$ of independent states in the theory.  For $k>Z$, consistency arises from an exact degeneracy in the inner product defined by the gravitational path integral, so that many a priori independent states differ only by a null state.  We argue that a similar property must hold in any consistent gravitational path integral.  We also comment on other aspects of extrapolations to more complicated models, and on possible implications for the black hole information problem in the individual members of the above ensemble.}

\begin{document}
\maketitle

\section{Introduction}
\label{Introduction}

The past year has seen several interesting developments in the study of black hole information.  In particular, it has been well-known for some time that the von Neumann entropy $S_\text{rad}$ of emitted Hawking radiation as a function of time gives an important diagnostic of whether and to what degree information is preserved or lost in evaporating black holes \cite{Page:1993df}. Familiar effective field theory would give an entropy that increases monotonically throughout the evaporation, even though the black hole's Bekenstein-Hawking entropy $S_\text{BH} = \frac{A}{4G}$ monotonically decreases to a value near zero.  In contrast, a model in which the black hole is a standard quantum system with density of states $S_\text{BH}$ coupled unitarily to the radiation field would --- when the initial state is pure --- require  $S_\text{rad} \le S_\text{BH}$ at all times.  As a result, in such models $S_\text{rad}$ generally increases to a maximum, at which time it nearly equals $S_\text{BH}$, and then decreases monotontically thereafter.  The final phase with decreasing $S_\text{rad}$ describes the return of information to the external universe from the black hole.

Despite many arguments suggesting that the latter so-called `Page curve' should accurately approximate the result of black hole evaporation, for many years it was unclear how such a result could be obtained from a controlled gravitational calculation; see e.g.\ reviews in \cite{Jacobson:2003vx,Mathur:2009hf,Harlow:2014yka,Unruh:2017uaw,Marolf:2017jkr}.  The plethora of proposals for new physics that might be associated with obtaining this Page curve (including \cite{Susskind:1993if,Susskind:1993ki,Susskind:1993aa,Chapline:2000en,Maldacena:2001kr,Mazur:2001fv,FW,Horowitz:2003he,Hawking:2005kf,Horowitz:2006mr,Mathur:2005zp,Giddings:2006sj,Mathur:2008nj,Mathur:2009hf,Giddings:2011ks,Davidson:2011eu,Giddings:2012gc,Almheiri:2012rt,Mathur:2012jk,Papadodimas:2012aq,Verlinde:2012cy,Nomura:2012sw,Verlinde:2013uja,Mathur:2013gua,Giddings:2013kcj,Maldacena:2013xja,Silverstein:2014yza,Rovelli:2014cta,Haggard:2014rza,Giveon:2015cma,Hawking:2016msc,Christodoulou:2016vny,Itzhaki:2019cgg,Amadei:2019ssp}) were thus all properly viewed as speculative and contained at least some optimistic extrapolation or ad hoc ingredient.\footnote{As an example, the firewall proposal of \cite{Almheiri:2012rt,Almheiri:2013hfa,Marolf:2013dba} did nothing to explain the dynamics from which the supposed firewall might arise.}

Recently, however, it was noted  that the `unitary' Page curve, including the turnover of $S_\text{rad}$, could be obtained by combining ideas from holography with effective field theory \cite{Penington:2019npb,Almheiri:2019psf} --- or equivalently with quantum field theory in curved space.   In particular, under very general conditions \cite{Penington:2019npb,Almheiri:2019psf}  argued that one could obtain this result by computing the generalized entropy $S_\text{gen} = \frac{A}{4G} + S_\text{bulk}$ of an appropriate comdimension-2 quantum extremal surface (QES) , where the surface is chosen so that holography suggests this might represent $S_\text{rad}$.  Here $S_\text{bulk}$ is the von Neumann entropy of bulk fields outside the codimension-2 QES.  See also further explorations of this idea in \cite{Almheiri:2019hni,Almheiri:2019yqk,Almheiri:2019psy,Chen:2019uhq}.

Critically, \cite{Almheiri:2019qdq,Penington:2019kki} then pointed out that --- at least in some contexts --- this seemingly-hybrid recipe in fact follows from replica trick calculations of $S_\text{rad}$ using the gravitational path integral (and in particular that this was implicit in earlier derivations of the quantum corrected Ryu-Takayanagi \cite{Ryu:2006ef,Ryu:2006bv} and Hubeny-Rangamani-Takayanagi \cite{Hubeny:2007xt} entropy formulae \cite{Faulkner:2013ana,Dong:2017xht}).  While at this level the physical mechanisms behind such results remain somewhat mysterious, the derivation from the gravitational path integral nevertheless implies that
the explicit addition of novel physics is not required.  Indeed, it instead  suggests that fundamental lessons might be revealed  by carefully dissecting the relevant calculations and studying the path integral in more detail.

A starting point for such further investigation is the observation of \cite{Penington:2019kki} that the above replica trick results appear to be inconsistent with one might normally call a single well-defined theory.  In particular, rather than taking single well-defined values, partition-function-like quantities seem to have both a mean value and a non-zero variance.  This feature is associated with the fact that dominant saddles in the replica computations involve connected bulk spacetimes with {\it disconnected} asymptotically AdS boundaries.  Such geometries have been termed spacetime wormholes, or Euclidean wormholes when the geometry is Euclidean.

This relation will be reviewed below, but is familiar from older discussions \cite{Coleman:1988cy,Giddings:1988cx,Giddings:1988wv,Maldacena:2004rf}.  In particular, refs. \cite{Coleman:1988cy,Giddings:1988cx,Giddings:1988wv} argued that spacetime wormholes require the gravitational Hilbert space to include spacetimes with compact Cauchy surfaces, and thus for which space at a moment of time has no asymptotically AdS boundary.  This part of the gravitational Hilbert space was called the baby universe sector.  Furthermore, it was argued that entanglement with this sector typically led the rest of the theory (here the asymptotically AdS sector) to act as if it were part of an ensemble of theories.  However, a particular member of the ensemble could be chosen by selecting an appropriate baby universe state.

Our goal here is to combine the above ideas to better understand the ensembles associated with replica trick computations and to extract implications for particular members of such ensembles.  We begin in section \ref{sec:wormholesandBU} by reviewing the connection between spacetime wormholes and ensemble-like properties, and by revisiting the baby universe ideas of
\cite{Coleman:1988cy,Giddings:1988cx,Giddings:1988wv}.  In doing so, we incorporate features associated with a negative cosmological constant and asymptotically AdS boundaries.  This both strengthens the results and allows a useful change in perspective. In particular, we avoid the use of `third quantized perturbation theory' and emphasize that certain results follow exactly from any well-defined path integral.  We also focus on the key role played by null states.

 The output is a description of how (say, partition-function-like) quantities at asymptotically AdS boundaries have a spectrum of possible values determined by the gravitational path integral.  Below, we focus on quantities $Z[\tilde J^*, J]$ that might be interpreted as computing the inner product of a state created by a source $J$ on the past half of the Euclidean AdS boundary with another state created by a source $\tilde J  = (\tilde J^*)^*$ on the future half of a Euclidean AdS boundary, where $*$ denotes CPT conjugation.  However, the most general partition-function-like quantities allowed by our formalism include quantities that in a dual CFT would describe matrix elements of operators as well as e.g. $\Tr\rho^n$ for a wide variety of density matrices.  The R\'enyi entropies of \cite{Almheiri:2019hni,Penington:2019kki} are then functions of these quantities.  In accordwith the original works \cite{Coleman:1988cy,Giddings:1988cx,Giddings:1988wv}, our analysis will show that one may generally describe such quantities as bring drawn from an ensemble of their possible values with the particular ensemble specified by the choice of baby universe state.

After describing this framework in section \ref{sec:wormholesandBU}, section \ref{sec:models} introduces some simple toy models in which the gravitational path integral can be performed exactly including the full sum over possible topologies.  The toy models are topological and involve finite-dimensional Hilbert spaces.  An interesting feature of the models is that the dimension of the asymptotically AdS Hilbert space becomes a random variable $Z$, whose value can be \emph{less} than the naive number $k$ of independent states in the theory.  For $k > Z$, consistency turns out to arise from an exact degeneracy in the inner product defined by the gravitational path integral.  This degeneracy means that many a priori independent states differ by a null state, and so should be regarded as linearly dependent in the gravitational Hilbert space.  Section \ref{sec:relation} relates this degeneracy to diffeomorphism invariance, black holes, and the Page curve, arguing in particular that the replica computations of \cite{Penington:2019kki,Almheiri:2019qdq} will imply a corresponding degeneracy in more general contexts. In section \ref{sec:3q}, we describe the approximation in which wormhole effects are small, analogous to the third quantised formalism of \cite{Giddings:1988wv}, and emphasise that the appearance of null states is associated with the failure of this approximation. We close with some summary and final discussion in section \ref{disc}.

\section{The gravitational path integral with spacetime wormholes}
\label{sec:wormholesandBU}

\subsection{Path integrals and ensembles}

We begin by describing a natural set of observables in any theory of gravity. For definiteness and convenience, we will assume locally AdS$_{d+1}$ asymptotics.  This is the context in which we have the most control and the clearest interpretation in terms of possible CFT duals.

Our theory will be defined by the path integral over a set of fields (including a metric) denoted collectively by $\Phi$, with action $S[\Phi]$. Each boundary is associated with a set of admissible boundary conditions labelled by $J$, describing the behaviour of the fields $\Phi\sim J$ near the given boundary. In particular, $J$ includes a $d$-dimensional boundary metric on a boundary manifold $\dmanifold$. We will focus on the case where the boundary metric has Euclidean signature, but Lorentzian or complex metrics are also allowed.  We will generally take each $\dmanifold$ to be connected, and introduce disconnected boundary manifolds by specifying multiple such boundaries, each with its own $J$.  However, there is no harm in letting $\dmanifold$ be disconnected, and the notation below remains consistent.  For each field other than the metric, $J$ typically includes a function on the $d$-dimensional boundary $\dmanifold$ specifying an appropriate boundary condition for that field; e.g., it will typically specify what in the AdS/CFT context is known as the ``non-normalisable part'' of the field. In all cases, by $S[\Phi]$, we then mean the holographically renormalised action with boundary condition $J$.

Now, the gravitational path integral with asymptotically AdS boundary conditions specified by $J$ is usually interpreted as computing a partition function $Z[J]$.  This is particularly familiar in the AdS/CFT context \cite{Gubser:1998bc,Witten:1998qj} where it gives the partition function of the dual CFT\footnote{We emphasize, however, that we allow very general notions of `sources' and thus very general notions of `partition functions.'  In particular, one may use sources to prepare initial and final states and to insert operators, so that one should be able to represent any matrix element of any operator in the dual CFT should as some $Z[J]$. In the same way, any R\'enyi entropy of any state that can be prepared by sources (and perhaps restricted to any region) should again be some $Z[J]$.}, but the identification of this quantity as a partition function in fact dates back to the first discussions of Euclidean approaches to black hole thermodynamics (see e.g. \cite{Gibbons:1976ue}).  Motivated by this interpretation, with an eye toward the ideas of \cite{Coleman:1988cy,Giddings:1988cx,Giddings:1988wv}, and following \cite{Saad:2019lba}, we introduce the following notation  for the path integral defined by an asymptotic boundary with $n$ connected components, each with an associated $J_i$:
\begin{equation}\label{eq:defPI}
	\Big\langle Z[J_1]\cdots Z[J_n] \Big\rangle := \int_{\Phi\sim J} \mathcal{D}\Phi \,e^{-S[\Phi]}
\end{equation}
This equation \emph{defines} the left hand side as the path integral over all configurations with $n$ asymptotic boundaries with boundary conditions specified by $J_1,\ldots,J_n$. The notation is chosen to be suggestive of a particular interpretation to be described below.

\begin{figure}[h]
\centering
\includegraphics[width =0.8\textwidth]{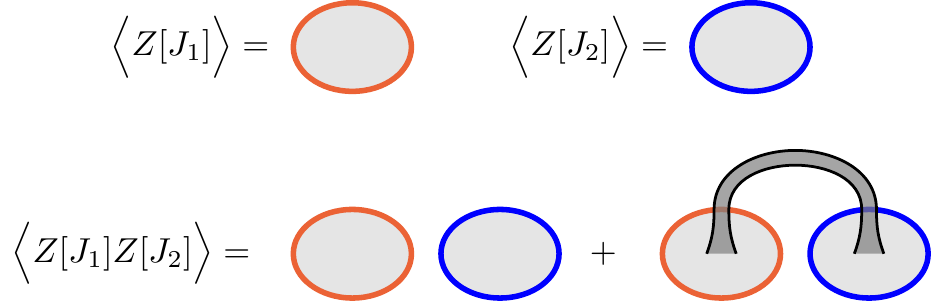}\\
\caption{The gravitational path integral with spacetime wormholes does not factorize.  The top line gives a diagramatic representation of the path integrals $\big\langle Z[J]_1 \big\rangle$ and $\big\langle Z[J_2]\big\rangle$ that would naively define partition functions $Z[J_1]$ and $Z[J_2]$.  The natural path integral $\big\langle Z[J_1]Z[J_2]\big\rangle$ associated with a pair of boundaries yields all terms generated by multiplying $\big\langle Z[J_1] \big\rangle \big\langle Z[J_2]\big\rangle$, but also contains additional connected contributions schematically shown as the second term in the bottom line.  }
\label{fig:nofac}
\end{figure}

The presence of spacetime wormholes in the path integral now leads to a phenomenon which is very puzzling from the standard AdS/CFT point of view \cite{Maldacena:2004rf,ArkaniHamed:2007js} (see \cite{Coleman:1988cy,Giddings:1988cx} for earlier discussions of the asymptotically flat analogue in which $S$-matrix elements play the role of our partition functions). The path integral \eqref{eq:defPI} does generally not factorize over disconnected boundaries:
\begin{equation}
\label{eq:nofac}
	\Big\langle Z[J_1] Z[J_2] \Big\rangle \neq \Big\langle Z[J_1]\Big\rangle \Big\langle Z[J_2] \Big\rangle.
\end{equation}
The difference between right and left sides arises because the sum over topologies in the Euclidean path integral for $\Big\langle Z[J_1]Z[J_2] \Big\rangle$ not only yields terms of the form $T_1T_2$ for any pair $T_1,T_2$ of terms associated separately with $\Big\langle Z[J_1]\Big\rangle$ and $\Big\langle Z[J_2]\Big\rangle$, but also contains additional contributions from terms in which the two boundaries lie in the same connected component of the bulk manifold; see figure \ref{fig:nofac}.  We use the term spacetime wormhole, or sometimes Euclidean wormhole, to refer to any such connection.  Note that spacetime wormholes are generally localized in both space and time, and thus differ qualitatively from spatial wormholes like the familiar Einstein-Rosen bridge that exist on every smooth Cauchy slice of the maximally extended Lorentz signature Schwarzschild spacetime.

The two sides of \eqref{eq:nofac} must thus differ unless the contributions with extra connections exactly cancel among themselves, or unless such contributions are excluded.  The first option appears to require fine tuning, and the second the imposition of non-local constraints that undermine the presumed local nature of the theory.  It is also difficult to see how one might introduce useful such constraints without destroying other apparent successes of the Euclidean path integral, such as the description of the Hawking-Page transition for AdS black holes, which is associated with a change in the topology of the dominant Euclidean saddle.
We therefore allow terms with extra connections, and at least for the moment  assume that they lead to a non-zero difference between the two sides of \eqref{eq:nofac}. It follows that we cannot simply interpret $\big\langle Z[J_1]\big\rangle$, $\big\langle Z[J_2] \big\rangle$ as partition functions with product $\big\langle Z[J_1] Z[J_2] \big\rangle$.

From the bulk point of view, the extra connections appear to describe dynamical interactions between a priori independent asymptotic regions.  This point of view is not naturally compatible with standard AdS/CFT, but it may instead be consistent to interpret $\big\langle Z[J_1] Z[J_2]\cdots \big\rangle$ as the expectation value of a product of partition functions in an ensemble of boundary dual theories.  In this interpretation, the connected contributions would describe probabilistic correlations from the ensemble average rather than dynamical interactions.

While these two interpretations may at first seem to be in tension, in analogous settings it was argued by \cite{Coleman:1988cy,Giddings:1988cx,Giddings:1988wv} that they are in fact consistent.  The rest of section \ref{sec:wormholesandBU} will be dedicated to providing a version of this discussion that incorporates features associated with asymptotically AdS boundaries. We find that using these new features allow strengthened conclusions, and perhaps as a result we will take a slightly different perspective than that of \cite{Coleman:1988cy,Giddings:1988cx,Giddings:1988wv}.

\begin{figure}[h]
\centering
\includegraphics[width =0.6\textwidth]{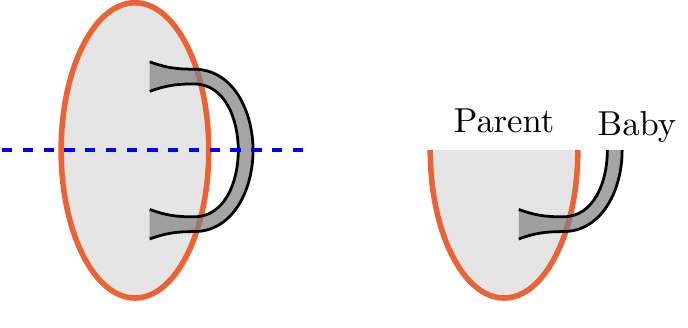}\\
\caption{Slicing open a spacetime with a boundary and a handle (left) can give a disconnected geometry on the slice, including a closed `baby universe' that has become detached from the parent asymptotically AdS universe.  The baby universe does not intersect the asymptotically AdS boundary (red line) at the moment of time described by the indicated slice.}
\label{fig:BUbranch}
\end{figure}

Before turning to the detailed discussion in section \ref{sec:BUH}, it is useful to provide a brief overview.
As in \cite{Coleman:1988cy,Giddings:1988cx,Giddings:1988wv}, the connection between the above two interpretations is motivated by realizing that summing over arbitrary topologies in our path integrals, and in particular over manifolds with arbitrary numbers of connected components, means that generic terms in $\big\langle Z[J_1] Z[J_2]\cdots \big\rangle$ contain factors associated with compact spacetimes having no boundaries whatsoever. The idea that the Hilbert space of a theory can be identified by cutting open the path integral then suggests that we should also slice open such compact spacetimes.  Doing so identifies a new sector not associated on this slice with any of the asymptotically AdS boundaries, but which is instead associated with spatially compact universes; see figure \ref{fig:BUbranch}.  We call this the baby universe sector following \cite{Coleman:1988cy,Giddings:1988cx,Giddings:1988wv}, where the name comes from the idea that one can in many cases  \cite{Hawking:1987mz,Hawking:1988ae,Giddings:1987cg,Lavrelashvili:1987jg} think of the closed universe having been emitted by a (here asymptotically AdS) parent universe.

The discussion of baby universes is simplest in the context of Euclidean path integrals with boundary conditions $J_i$ given by Euclidean metrics, but our discussion does not exclude more general contexts.  In particular, one can choose boundary conditions with Lorentzian pieces of the metric, using a Schwinger-Keldysh type formalism in which Euclidean sections of the metric are used to prepare states and Lorentzian sections give real time evolution. In such a case, it is useful to think of the gravitational path integral as involving complex metrics.

Such constructions allow us to describe quite general observables that might be associated with a putative dual CFT.
Indeed, the set of observables we are using is also sufficient to describe coupling to an auxiliary quantum system, as is important in \cite{Almheiri:2019psf,Almheiri:2019hni,Almheiri:2019yqk,Almheiri:2019psy,Chen:2019uhq,Almheiri:2019qdq}. To do this, we can simply allow sources $J$ to be operators in the auxiliary system, and then include a corresponding auxlliary path integral to compute the effects of such operators. We discuss this construction in more detail in section \ref{sec:relation}.

Note that the ability of Euclidean or complex universes to split and join as shown in figure \ref{fig:BUbranch} indicates that baby universes can affect the physics of universes with asymptotically AdS boundaries. In this context, it becomes clear that the definition of our path integral \eqref{eq:defPI} includes an implicit choice of the initial and final state of closed baby universes. Most naturally, the path integral computes expectation values in the Hartle-Hawking no-boundary state \cite{Hartle:1983ai}, defined by the absence of additional boundaries besides those required by the $Z[J]$ insertions. But this is not the only choice of baby universe state that we can describe with our gravitational path integral, and other choices will be associated with different ensembles. In particular, we will construct special `$\alpha$-states' of baby universes in which the factorisation property is restored, and no ensemble is required.

One further comment is in order before turning to the details.  In the above discussion we have written our amplitudes as if the path integral gives some definite, finite value. However, in all but the very simplest contexts, gravitational path integrals have been defined only as asymptotic expansions (perhaps with nonperturbative contributions) in some small coupling. Both loop expansions and sums over nonperturbative sectors will typically fail to converge, and there may be no obvious, natural or unique way to define a finite result. The distinction between exact quantities with finite values of parameters and asymptotic expansions may well be important, and we will return to this issue in section \ref{disc}. Nonetheless, for the remainder of this section we will treat the path integral in \eqref{eq:defPI} as if it gives well-defined exact results.

\subsection{The baby universe Hilbert space}
\label{sec:BUH}

As described above, one can obtain a natural Hilbert space interpretation by cutting open the path integral \eqref{eq:defPI}.   In particular, we split each history over which we sum  into a `past' and `future' that meet on some slice where we imagine summing  over a complete set of intermediate states. There is a choice of how we cut, constrained by the way in which the asymptotic boundaries are labelled past or future. For now, we will choose to place each connected component of the boundary either entirely to the past or entirely entirely to future of our cut, so that our intermediate slice intersects no asymptotically AdS boundaries (generalizing in section \ref{sec:moreHilbert}).  We thus identify the relevant Hilbert space as the space of closed universes in the theory. We call this the `baby universe' Hilbert space $\hbu$ for the reasons described above.

One might hope to describe elements of the baby universe Hilbert space as wavefunctions of all possible spatial metrics (and field configurations on those metrics). A complication is that, as usual in a gravitational theory, diffeomorphism invariance forbids a notion of  universal time that might be used specify precisely where the past/future cut is to be made. Proceeding in this manner would thus require imposing the gravitational constraints (the Wheeler-DeWitt equation) on the resulting wavefunctions. This is made particularly challenging in the current context where spacetime wormholes are important, so that the associated splitting and joining of universes should modify these constraints \cite{Giddings:1988wv}.

However, we can bypass these difficulties entirely by using our asymptotic boundaries to define states in the baby universe Hilbert space.
Given a set $\{J_1,\ldots,J_m\}$ of boundary conditions, there is a state
\begin{equation}\label{eq:BUstate}
			\Big|Z[J_1]\cdots Z[J_m]\Big\rangle	\in \hbu,
\end{equation}
defined by the specified boundary conditions for the path integral.
This is particularly natural for sources defining Euclidean signature boundary metrics and in the presence of a negative cosmological constant.  While a negative cosmological constant tends to cause universes to collapse in Lorentzian time evolution (perhaps with a sinusoidal form), after Wick rotation to Euclidean signature it tends to cause accelerated expansion with respect to Euclidean time.  As a result, such closed cosmologies naturally have Euclidean signature asymptotically AdS boundaries at infinite Euclidean times.

We will think of the boundary conditions associated with the state \eqref{eq:BUstate} as living `in the past.' They can then be paired with bra-vectors living `in the future' --- though one should understand that these are simply names without intrinsic meaning.
Note that the ordering of the $Z[J_i]$ in \eqref{eq:BUstate} is not important.  Reordering the sources gives equivalent boundary conditions for the path integral, and so must define the same state.  An important special case is $m=0$, giving the Hartle-Hawking state with no boundary in the past:
\begin{equation}
	\text{No boundaries }(m=0) \longrightarrow \Big| \text{HH} \Big\rangle	\in \hbu.
\end{equation}
Here we emphasize that this is not just a state on a single universe, but that it instead represents a state of the full collection of an indefinite number of baby universes.

States of the form \eqref{eq:BUstate} defined by different sources, or even with different numbers of sources $m$, are generally not mutually orthogonal in any useful sense.  Note that the physical notion of inner product cannot simply be assumed to have any particular form, but is something we must compute from the theory.  It must thus follow from an appropriate path integral.  Now, some readers may be confused by the fact that in quantum field theory one typically uses first-quantized path integrals to compute Green's functions and not to compute inner products.  However, as explained in e.g. \cite{Halliwell:1990qr}, in defining the gravitational path integral one must make a choice --- in some languages, associated with specifying the contour of integration --- as to whether it fully imposes the gravitational constraints or instead defines a Green's function.   We simply choose the former, and we take the correlators \eqref{eq:defPI} to be computed with the same specifications.  With this understanding, the path integral indeed computes the inner product\footnote{In the language of Dirac constraint quantization \cite{Dirac}, \eqref{eq:BUIP} corresponds to taking two arbitrary `kinematic' states (which may not satisfy the constraints), projecting them onto the space of states satisfying the constraints, and computing the physical inner product of the resulting projections.  See \cite{Landsman:1993xe,Marolf:1994wh,Ashtekar:1995zh,Marolf:2000iq,Shvedov:2001ai} for further comments, and \cite{Marolf:1996gb,Reisenberger:1996pu,Hartle:1997dc} for connections to path integrals. As in \cite{Marolf:1996gb,Reisenberger:1996pu}, using \eqref{eq:BUIP} corresponds to simply skipping to the final answer without going through the intermediate steps inherent in \cite{Dirac}.} which is then given by
\begin{equation}\label{eq:BUIP}
			\Big\langle Z[\tilde{J}_1]\cdots Z[\tilde{J}_n]\Big|Z[J_1]\cdots Z[J_m]\Big\rangle = \Big\langle Z[\tilde{J}_1^*]\cdots Z[\tilde{J}_n^*] Z[J_1]\cdots Z[J_m]\Big\rangle.
\end{equation}
Here the right hand side is just the amplitude defined in \eqref{eq:defPI} with boundary conditions $Z[J]$ and $Z[\tilde{J}^*]$, and where $*$ is the CPT conjugate operation on boundary conditions $J$. This operation should have the property that if we act with $*$ on every boundary, the amplitude is complex conjugated:
\begin{equation}
	\Big\langle Z[J_1^*]\cdots Z[J_n^*] \Big\rangle = \Big\langle Z[J_1]\cdots Z[J_n] \Big\rangle^*.
\end{equation}
This guarantees that the inner product \eqref{eq:BUIP} is Hermitian. If we can interpret $Z[J]$ as random variables with correlation functions $\Big\langle Z[J_1]\cdots Z[J_n] \Big\rangle$, then \eqref{eq:BUIP} reduces to a standard construction in probability theory, in which the covariance matrix of pairs of random variables defines an inner product.  In particular, showing that the amplitudes follow from expectation values of a distribution with nonnegative probabilities would imply that our inner product is positive semi-definite.

Note that the states \eqref{eq:BUstate} need not be normalised. In particular, the norm of the Hartle-Hawking state is given by what one might call the cosmological partition function $\mathfrak{Z}$, defined by the path integral over all spacetimes without boundary:
\begin{equation}\label{eq:HHnorm}
	\mathfrak{Z} = \big\langle 1 \big\rangle = \big\langle\,\HH\,\big| \,\HH\,\big\rangle =  \mkern-16mu\int\displaylimits_{\text{no boundary}}  \mkern-16mu \mathcal{D}\Phi \,e^{-S[\Phi]}.
\end{equation}
For most purposes, it would be sufficient to consider normalised amplitudes, where we divide by $\mathfrak{Z}$. This is equivalent to performing the path integral excluding closed components of spacetime which do not connect to any asymptotic boundary.

We now have a space of states defined by (finite) linear combinations of the states \eqref{eq:BUstate} in correspondence with formal polynomials of `partition functions' $Z[J]$, and an inner product defined by extending \eqref{eq:BUIP} sesquilinearly. This is almost enough to construct a baby universe Hilbert space. The missing ingredient is a single property that we demand of our path integral \eqref{eq:defPI}, namely reflection positivity. This can be stated as the requirement that \eqref{eq:BUIP} defines a positive semidefinite inner product on finite linear combinations of states \eqref{eq:BUstate}:
\begin{equation}\label{eq:RP}
	\big\Vert\Psi\big\Vert^2 := \big\langle\Psi|\Psi\big\rangle\geq 0
	\text{ for all }\big|\Psi\big\rangle = \sum_{i=1}^N c_i \Big|Z[J_{i,1}]\cdots Z[J_{i,m_i}]\Big\rangle.
\end{equation}
Thus is clearly required if our gravitational path integral is to define a standard quantum theory, though it is cumbersome to verify directly for all states.  While this can be done for the simple toy models studied in section \ref{sec:models}, for more complicated systems it would be very useful to find properties that imply \eqref{eq:RP} but are easier to check.

Assuming \eqref{eq:RP}, we now \emph{define} the baby universe Hilbert space $\hbu$ though a standard construction, as the completion of the space of linear combinations of states \eqref{eq:BUstate} with the inner product \eqref{eq:BUIP}. Roughly speaking, states of $\hbu$ are infinite sums over states \eqref{eq:BUstate} with finite norm defined by \eqref{eq:BUIP}.\footnote{$\hbu$ is the set of equivalence classes of Cauchy sequences $\left\{|\Psi_i\rangle\right\}_{i\in\NN}$, where two sequences $\left\{|\Psi_i\rangle\right\}$, $\left\{|\Phi_j\rangle\right\}$ are equivalent if $\Vert\,|\Psi_i\rangle-|\Phi_j\rangle\,\Vert^2\to 0$ as $i,j\to\infty$. Recall that a sequence is Cauchy when $\Vert\,|\Psi_i\rangle-|\Psi_j\rangle\,\Vert^2\to 0$ as $i,j\to\infty$.  The inner product between two such sequences is defined by the limit of the inner products of the terms, which exists and is the same for all members of the equivalence class. $\hbu$ is then a Hilbert space, so in particular is complete and the inner product is positive definite. It is separable as long as the set of possible sources $J$ has a countable dense subset (assuming that amplitudes are continuous in $J$).} Importantly, however, infinite sums with different terms and coefficients may not give rise to distinct states in $\hbu$.  Equivalently, some infinite sums may be identified with the zero state in $\hbu$; i.e., for appropriate coefficients $c_i$ one may find
\begin{equation}\label{eq:nullstates}
	\sum_{i=1}^\infty c_i \Big|Z[J_{i,1}]\cdots Z[J_{i,m_i}]\Big\rangle=0.
\end{equation}

\begin{figure}
	\centering
	\includegraphics[width=.5\textwidth]{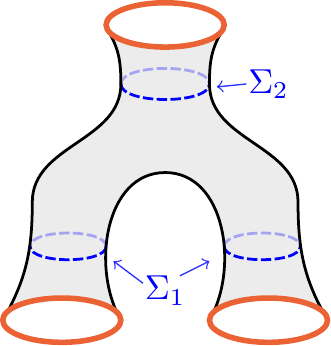}
	\caption{In the presence of spacetime wormholes, different spatial slices of a spacetime may have different number of connected components. Here, on the slice $\Sigma_1$ we have two circular universes, but on $\Sigma_2$ we have only one. These may be thought of as different gauge choices for the same state.
	\label{fig:slices}}
\end{figure}

Naively, the Hilbert space $\hbu$ may appear to consist of formal power series in the objects $Z[J]$ with some convergence property.  But it is in fact smaller since the construction divides out by the set of `null states' \eqref{eq:nullstates}. This may seem like a minor technical point.  Of course, from one perspective the inner product defined by any gravitational path integrals naturally leads to a large set of such null states due to the gravitational gauge symmetry.  But we usually expect that symmetry to act trivially at the asymptotically AdS boundaries where our sources $J$ are defined; i.e., natural sources $J$ are invariant under familiar gravitational gauge symmetries.  As a result, one might expect the null states to simply encode possible senses in which one may have accidentally introduced an overcomplete set of sources.  However, one should expect the sum over topologies to modify the gravitational gauge invariance so that it no longer corresponds precisely to familiar diffeomorphisms.  As illustrated in figure  \ref{fig:slices}, one expects different slices of the same spacetime to describe gauge equivalent states.  But including a sum over topologies means that two such slices may no longer be related by a diffeomorphism, and in fact that they need not even contain the same number of connected components for space at the given time.  It will thus be important to compute the effects of this modified gauge symmetry rather than to assume that they take a familiar form.  In particular, while one might naively expect the effect of such modifications to be small, we will find  sections \ref{sec:models} and \ref{sec:relation} that in certain circumstances they lead to dramatic physical consequences.

The above construction of $\hbu$ is very similar to the construction of the Hilbert space of a quantum field theory from its correlation functions in the Wightman \cite{Wightman:1956zz} or Osterwalder-Schrader (see \textbf{Theorem  3-7} of \cite{streater2016pct}, \cite{Osterwalder:1973dx}) reconstuction theorems. In this analogy, our objects $Z[J]$ correspond to (smeared) local operators inserted in the Euclidean past, and the inner products between states with finitely many operator insertions are given by the (Euclidean) Wightman functions.  The Hilbert space is again defined by the above completion construction.

\subsection{Operators and $\alpha$-eigenstates}
\label{sec:ops}

Having  constructed the baby universe Hilbert space $\hbu$, we now introduce a set of operators acting on it. Here we once again find asymptotic boundaries useful. In particular, we take any boundary $Z[J]$ to define an operator $\widehat{Z[J]}$ on $\hbu$. The matrix elements of this operator are defined by a path integral over all configurations with boundaries specified by some initial and final states with an additional boundary $Z[J]$.

Since the labelling of boundaries as past, future, and in between does not affect the value of the path integral, the defining relation of the operator $\widehat{Z[J]}$ is
\begin{equation}
\label{eq:moveop}\Big\langle Z[\tilde J_1]\cdots Z[\tilde J_{\tilde m}]	\Big|\widehat{Z[J]}\Big|Z[J_1]\cdots Z[J_m]\Big\rangle = \Big\langle Z[\tilde J_1]\cdots Z[\tilde J_{\tilde m}]	\Big|Z[J]Z[J_1]\cdots Z[J_m]\Big\rangle.
\end{equation}
Since the span of the bra-vectors in \eqref{eq:moveop} is dense in $\hbu$, we may write the action of such operators as
\begin{equation}
\label{eq:opdef}
	\widehat{Z[J]}\Big|Z[J_1]\cdots Z[J_m]\Big\rangle = \Big|Z[J]Z[J_1]\cdots Z[J_m]\Big\rangle,
\end{equation}
extending the action of such operators to the full Hilbert space $\hbu$ by continuity\footnote{Strictly speaking, this is the case for bounded functions of the $Z[J_i]$.  As usual, unbounded operators can be defined only on somewhat smaller domains.}. For later use, we note that \eqref{eq:opdef} implies that our defining states may be created by acting with the $\widehat{Z[J]}$ operators on the Hartle-Hawking no-boundary state,
\begin{equation}
\label{eq:opHH}
	\Big|Z[J_1]\cdots Z[J_m]\Big\rangle = \widehat{Z[J_1]}\cdots \widehat{Z[J_m]}\Big|HH\Big\rangle,
\end{equation}
and thus by combining \eqref{eq:BUIP} and \eqref{eq:opdef} that we may identify our original path integral as computing correlators in $\Big|HH\Big\rangle$ as advertised earlier:
\begin{equation}
\label{eq:HHensemble}
	\Big\langle Z[J_1]\cdots Z[J_m]\Big\rangle = \Big\langle \HH \Big|\widehat{Z[J_1]}\cdots \widehat{Z[J_m]}\Big|\HH\Big\rangle.
\end{equation}
We also see that the Hermitian conjugate of $\widehat{Z[J]}$ is given by taking the CPT conjugate of the source:
\begin{equation}
\label{eq:Zconj}
	\widehat{Z[J]}^\dag = \widehat{Z[J^*]}
\end{equation}

Thus far, we have really defined the $\widehat{Z[J]}$ as operators on the baby universe pre-Hilbert space (before taking the quotient by null vectors \eqref{eq:nullstates}).  To show that $\widehat{Z[J]}$ is well-defined on $\hbu$, we must show that it maps null states to null states.  But this follows immediately from either \eqref{eq:Zconj} or \eqref{eq:opHH}.  In particular, for any null state $\big| \mathcal{N} \big \rangle$ and an arbitrary state $\big | \Psi\big \rangle$, we may define $\big | \Psi' \big \rangle = \widehat{Z[J^*]} \big | \Psi\big \rangle$ to write
\begin{equation}
\label{eq:Zpreservesnull}
\big\langle \Psi \big | \widehat{Z[J]} \big| \mathcal{N} \big \rangle = \big\langle \Psi' \big| \mathcal{N} \big \rangle =0.
\end{equation}
The last equality follows from the fact that $\big| \mathcal{N} \big\rangle$ is null, and since   $\big | \Psi\big \rangle$ is arbitrary we see that
$\widehat{Z[J]} \big| \mathcal{N} \big \rangle $ is also null as desired.

The set of operators $\widehat{Z[J]}$ for all possible $J$ turns out to have a powerful set of properties. Firstly, since the states $\big|Z[J_1]\cdots Z[J_m]\big\rangle$ are unchanged by permutations of the sources $J_i$, it follows immediately from \eqref{eq:opdef} that all $\widehat{Z[J]}$ mutually commute\footnote{A similar result was derived in \cite{Coleman:1988cy,Giddings:1988cx,Giddings:1988wv} using an additional assumption about locality of induced couplings.  Crucially, this assumption played no role in our argument above.}:
\begin{equation}\label{eq:Zcommute}
	\left[\widehat{Z[J]},\widehat{Z[J']}\right]=0.
\end{equation}
In particular, this implies that each $\widehat{Z[J]}$ is normal (that is, it commutes with its Hermitian conjugate), so that we may apply the spectral theorem.  It then follows from \eqref{eq:Zcommute} that the Hilbert space $\hbu$ has a basis of orthonormal states $|\alpha\rangle$ which are simultaneous eigenvectors for all $\widehat{Z[J]}$ operators:
\begin{equation}
	\widehat{Z[J]} |\alpha\rangle =  Z_\alpha[J] |\alpha\rangle \quad \forall J.
\end{equation}

Following \cite{Coleman:1988cy}, we call these $\alpha$-eigenstates, or $\alpha$-states for short. The spectrum $\left\{Z_\alpha[J]\right\}_\alpha$ of $\widehat{Z[J]}$ may be either discrete or continuous.  In the latter case the $|\alpha\rangle$ are not normalisable states, but are instead delta function normalized.  However, for simplicity we use notation in either case as if $|\alpha\rangle$ are normalisable eigenvectors, writing
\begin{equation}
	\left\langle \alpha' \middle| \alpha \right\rangle = \delta_{\alpha'\alpha}\, ,
\end{equation}
leaving the appropriate modifications for continuous spectrum implicit.

It turns out that the set $\{\widehat{Z[J]}\}$ for all possible $J$ in fact defines a {\it complete} commuting set of operators on $\hbu$, as the state $|\alpha\rangle$ is determined up to a phase by its eigenvalues $Z_\alpha[J]$. To see this, note that we can determine all matrix elements of $|\alpha\rangle$ via
\begin{equation}
\begin{aligned}
	\Big\langle Z[J_1]\cdots Z[J_n]\Big|\,\alpha\Big\rangle &= \Big\langle\HH\,\Big| \widehat{Z[J_1]}^\dag\cdots \widehat{Z[J_n]}^\dag\Big|\,\alpha\,\Big\rangle \\
	&= Z_\alpha[J_1^*]\cdots Z_\alpha[J_n^*] \big\langle\, \HH\,\big|\,\alpha\,\big\rangle.
\end{aligned}	
\end{equation}
This means that the $\alpha$-states define a preferred orthonormal basis for $\hbu$; we can even fix phases by choosing $\big\langle\HH\big|\alpha\big\rangle>0$.

The above calculation of the matrix elements also shows that the Hartle-Hawking state has non-zero overlap with every $\alpha$-state, $\langle\text{HH}|\alpha\rangle\neq 0$. Otherwise $|\alpha\rangle$ has vanishing overlap with a dense set of states, and hence must be the zero state. If we define $p_\alpha$ by these overlaps according to
\begin{equation}
	p_\alpha = \frac{\left|\left\langle\,\mathrm{HH}\,\middle|\,\alpha\,\right\rangle\right|^2}{\left\langle\,\mathrm{HH}\,\middle|\,\mathrm{HH}\,\right\rangle}, \qquad .
\end{equation}
we find
\begin{equation}\label{eq:palpha}
	p_\alpha>0,\quad \sum_\alpha p_\alpha =1,
\end{equation}
where the second follows from completeness and orthonormality of the $\alpha$ basis. Now, by inserting complete sets of $\alpha$-states, we can compute the general amplitude \eqref{eq:defPI}:
\begin{align}\label{eq:probability}
	\Big\langle Z[J_1]\cdots Z[J_n] \Big\rangle &= \mkern-12mu \sum_{\alpha_0,\alpha_1,\ldots,\alpha_n}	 \mkern-12mu \big\langle\,\HH\,\big|\alpha_0\big\rangle\big\langle \alpha_0|Z[J_1]\big|\alpha_1\big\rangle\cdots \big\langle \alpha_{n-1}|Z[J_n]\big|\alpha_n\big\rangle\big\langle \alpha_n| \,\HH\,\big\rangle \nonumber \\
	&= \mathfrak{Z} \enspace \sum_\alpha p_\alpha Z_\alpha[J_1]\cdots Z_\alpha[J_n].
\end{align}
The normalising factor $\mathfrak{Z}$ is the norm of the Hartle-Hawking state \eqref{eq:HHnorm}.

Equation \eqref{eq:probability}, along with \eqref{eq:palpha}, tells us that a gravitational path integral \eqref{eq:defPI} is quite generally compatible with an ensemble interpretation, exemplified by the matrix ensemble dual to JT gravity in \cite{Saad:2019lba}, and analogous to the random couplings of \cite{Coleman:1988cy,Giddings:1988cx}.  Specifically, the parameters $\alpha$ label the various theories in the ensemble, the eigenvalues $Z_\alpha[J]$ give definite values for observables in the theory associated with the particular label $\alpha$, and $p_\alpha$ gives the probability of selecting $\alpha$ from the ensemble. The states $|\alpha\rangle$ making up our preferred eigenbasis of $\hbu$ are in one-to-one correspondence with members of the ensemble. A less extreme example of $\alpha$-states is provided by the `eigenbranes' described in \cite{Blommaert:2019wfy} in the context of JT gravity, which act to constrain the eigenvalues of $\widehat{Z[J]}$, thus partially diagonalizing these operators. Note that we arrived at a classical probability distribution because the relevant operators are mutually commuting \eqref{eq:Zcommute}. The only property required of the gravitational path integral (besides its existence) was reflection positivity, to guarantee nonnegative probabilities.

With our new Hilbert space point of view, it is now clear that the ensemble described above is not unique.  Instead, through \eqref{eq:HHensemble} it was associated with the implicit choice of the Hartle-Hawking state in $\hbu$.  While the Hartle-Hawking state is a particularly simple and natural choice, we are nevertheless free to select any state we like. In particular, if the initial state of the baby universes is an $\alpha$-state, this selects a single member of the ensemble so that amplitudes factorize:
\begin{equation}\label{eq:alphafactorise}
	\big\langle\alpha\big|\widehat{Z[J_1]}\widehat{Z[J_2]}\big|\alpha\rangle = \big\langle\alpha\big|\widehat{Z[J_1]}\big|\alpha\rangle \big\langle\alpha\big|\widehat{Z[J_2]}\big|\alpha\rangle = Z_\alpha[J_1]Z_\alpha[J_2].
\end{equation}
Any other state $|\Psi\rangle$ is a superposition of $\alpha$-states, and describes an ensemble with probabilities $p_\alpha=|\langle\alpha|\Psi\rangle|^2$. Classical probabilities are sufficient to describe the ensemble, since relative phases between different $\alpha$-states in the superposition are irrelevant for correlation functions of the commuting operators $\widehat{Z[J]}$.  In other words, with respect to the algebra of the $\widehat{Z[J]}$, the $\alpha$-states define superselection sectors.

If the path integral \eqref{eq:defPI} already defines factorising amplitudes, so that our theory of gravity has a single boundary dual, we have a trivial special case of the formalism described here. In that case, the operators $\widehat{Z[J]}$ are constants $Z[J]$, and the Hilbert space of closed universes $\hbu$ is one-dimensional, spanned by the Hartle-Hawking state, which is also the unique $\alpha$-state. We discuss this possibility further in section \ref{disc}.

\subsection{More Hilbert spaces}\label{sec:moreHilbert}

The above discussion concerned the Hilbert space $\hbu$ of closed `baby' universes.  We constructed $\hbu$ by cutting amplitudes in such a way that any given asymptotic boundary lies completely on one side of the cut. We now generalize this construction to allow cuts that intersect one or more components of the asymptotic boundary, thus splitting such boundary components into two parts.   This gives us many different Hilbert spaces depending on the boundary conditions at the intersection, and in particular on the choice of a $(d-1)$-dimensional (perhaps oriented) spatial boundary geometry $\Sigma$.  We thus call the resulting Hilbert space $\hilb_\Sigma$, leaving implicit the other sources $J$ on $\Sigma$.  Note that $\Sigma$ can have any number of connected components, and if $\Sigma$ is empty we find again the Hilbert space $ \hilb_{\Sigma=\emptyset} =\hbu$ of closed baby universes described above.

The construction of $\hilb_\Sigma$ proceeds much as for $\hbu$, except that in addition to closed asymptotic boundary conditions denoted by $Z[J]$ we also have objects $\psi[J]$ defining boundary conditions on a piece $\dmanifold$ of an asymptotic boundary with $\partial\dmanifold = \Sigma$. As before, the manifold $\dmanifold$, and in particular its boundary $\Sigma$, is implicitly included in the sources $J$. For example, in the right panel of figure \ref{fig:BUbranch}, $\dmanifold$ is the solid black semicircle forming the past asymptotically AdS boundary and $\Sigma$ consists of the right and left endpoints. In a dual interpretation, $\psi[J]$ would define a state on the CFT Hilbert space with spatial geometry $\Sigma$, as the wavefunction for a given CFT field configuration on $\Sigma$ would be computed by a path integral on $\dmanifold$ with sources $J$.

As before, we may choose $\dmanifold$ to be connected. Note that this does not imply  $\Sigma=\partial\dmanifold$ to be connected.  When $\Sigma$ is not, it can be useful to write
$\Sigma$ as the disjoint union $\Sigma = \Sigma_1 \sqcup\cdots\sqcup \Sigma_m$ of components $\Sigma_i$ (where the ordering of the components is meaningful, in case they have the same geometry). Generalizing \eqref{eq:BUstate}, we then have states
\begin{equation}
\label{eq:multicomponentstate}
	\Big| \psi[J_1]\cdots \psi[J_{m}] Z[J'_1]\cdots Z[J'_n] \Big\rangle \in \hilb_\Sigma,
\end{equation}
where $\psi[J_i]$ is associated with component $\Sigma_i$ for any source $J$.  While this notation is useful, it is also somewhat awkward if we take a given $\psi[J_i]$ to be associated with a connected $\dmanifold_i$, whose boundary $\partial \dmanifold_i = \Sigma_i$ may again be disconnected.   As a result, one will sometimes need to use a number of distinct decompositions $\Sigma = \Sigma_1 \sqcup\cdots\sqcup \Sigma_m$  (perhaps with different values of $m$) for a given $\hilb_\Sigma$.

The inner product on $\hilb_\Sigma$ generalizes \eqref{eq:BUIP} in a natural way if we note that a boundary condition $\psi[\tilde{J_i}]$  in the `bra' (on some $\tilde \dmanifold_i$ with $\partial \tilde \dmanifold_i = \Sigma_i$)  can be paired with a boundary condition  $\psi[J_i]$ in the `ket' (again on some $\dmanifold_i$ with $\partial  \dmanifold_i = \Sigma_i$) to define a boundary condition $Z[\tilde J^*, J]$ associated with the closed boundary manifold $\tilde \dmanifold_i^* \dmanifold_i$ constructed by taking the manifold $\tilde \dmanifold_i^*$ (formed from $\tilde \dmanifold_i$ by reversing the orientation) and sewing $\tilde \dmanifold_i^*$ to $\dmanifold_i$ along $\Sigma_i$. In $Z[\tilde J^*, J]$,  $*$ again denotes CPT conjugation of sources, and the sources on $\tilde \dmanifold^* \dmanifold$ are given locally by $\tilde J^*, J$.  One may also wish to restrict the allowed sources to vanish sufficiently quickly at $\Sigma_i$ so that the sources defined on $\tilde \dmanifold_i \dmanifold_i$ by such sewings are sufficiently smooth.

It is important that the above sewing is uniquely defined even when $\Sigma_i$ admits isometries.  In particular, recall that the above discussion fixed a manifold $\Sigma \supseteq \Sigma_i$ from the beginning, and at no point was there a quotient by diffeomorphisms of $\Sigma$.  The individual points of $\Sigma$ should thus be thought of as carrying definite labels, defining the unique sewing of $\tilde \dmanifold$ to $\dmanifold$.  In particular, the notation in \eqref{eq:multicomponentstate} is not invariant under reordering of the $\Sigma_i$.

We shall write the pairing as
$Z[\tilde J^*,  J] = \left(\psi[\tilde J],\psi[J]\right)$.
This notation is chosen be suggestive of an inner product $(\cdot,\cdot)$  of states in the dual CFT Hilbert space. The distinguishability of points in $\Sigma$ is motivated either by a dual CFT perspective, or from familiar gravitational boundary conditions at asymptotically AdS boundaries.    The extended inner product is then defined by using the above pairing and and evaluating the resulting path integral as before:
  \begin{equation}\label{eq:psipsiZ}
 	\Big\langle\psi[\tilde J]\Big| \psi[J] \Big\rangle = \Big\langle \left(\psi[\tilde J],\psi[J]\right) \Big\rangle =  \Big\langle Z[\tilde J^*,  J] \Big\rangle
 \end{equation}

We emphasize again that if $\Sigma$ contains identical connected components $\Sigma_1,\Sigma_2$, the components are treated as distinguished and canonically ordered.  Thus in the notation of \eqref{eq:multicomponentstate}, $\big|\psi[J_1]\psi[J_2]\big\rangle \neq \big|\psi[J_2]\psi[J_1]\big\rangle$. While the norms of these states will agree, the inner product of these states with generic other kets will not (for example, $\big\langle\psi[J_2]\psi[J_1]\big|\psi[J_1]\psi[J_2]\big\rangle=\big\langle Z[J_2^*,J_1] Z[J_1^*,J_2]\big\rangle \neq \big\langle Z[J_2^*,J_2] Z[J_1^*,J_1]\big\rangle $, even if $\Sigma_1=\Sigma_2$ so this pairing makes sense).  This is a special case of the statement that states need not be invariant under symmetries of $\Sigma$.

As in the discussion of $\hbu$, the structure above is properly described as being pre-Hilbert space.  The actual Hilbert space $\hilb_{\Sigma}$  is then constructed as a completion, which includes a quotient with respect to the space of null vectors.  This procedure succeeds when the path integral is appropriately reflection positive, by which we mean that the inner product it defines on the pre-Hilbert space is positive semi-definite.  The inner product on the final $\hilb_{\Sigma}$ is then positive definite as desired. Note that reflection positivity on $\hilb_\Sigma$ is an additional requirement we impose on the path integral, not necessarily implied by reflection positivity on $\hbu$; this will prove to be relevant for the toy model discussed in section \ref{sec:models}.

As before, we have operators $\widehat{Z[J]}$ acting on the Hilbert spaces $\hilb_\Sigma$, and in particular which preserve the space of null states in the pre-Hilbert space for the same reason as before.  Again, these operators mutually commute. But now we also have a plethora of new operators which can map between Hilbert spaces with different boundaries.  In particular, if $\psi[J]$ is associated with $\dmanifold$ having $\partial \dmanifold = \Sigma$, then for any $\tilde \Sigma$ there is an operator
\begin{gather}
\label{eq:psiop}
	\widehat{\psi[J]} : \hilb_{\tilde \Sigma} \to \hilb_{\Sigma \sqcup \tilde \Sigma}, \\
	\text{with }\quad \widehat{\psi[J]} \Big| \tilde \psi[\tilde J] Z[J'_1]\cdots Z[J'_n] \Big\rangle = \Big| \psi[J] \tilde \psi[\tilde J] Z[J'_1]\cdots Z[J'_n] \Big\rangle,
\end{gather}
where in ${ \Sigma \sqcup \tilde \Sigma}$ we define the components of $\Sigma$ to be ordered before components of $\tilde \Sigma$.  We may use \eqref{eq:psiop}  even when $\dmanifold, \tilde \dmanifold$ are disconnected. Note, however, that (when $\Sigma\neq\Sigma'$) it does not make sense to ask whether
$\widehat{\psi[J]}, \widehat{\psi[J']}$ commute, as $\widehat{\psi[J]}\widehat{\psi[J']} $ maps $\hilb_{\tilde \Sigma} \to \hilb_{\Sigma \sqcup \Sigma' \sqcup \tilde \Sigma }$, while $\widehat{\psi[J']}\widehat{\psi[J]} $ maps $\hilb_{\tilde \Sigma} \to \hilb_{ \Sigma' \sqcup \Sigma \sqcup \tilde \Sigma}$.

Nevertheless, one can build a dense set of states in $\hilb_{\Sigma}$ by acting with such operators on $\hilb_{\emptyset} = \hbu$.  As a result, the fact that $\widehat{\psi[J]}$ preserves the null space, and thus is truly well-defined on $\hilb_{\Sigma}$, follows from \eqref{eq:psipsiZ} and the corresponding property for $\widehat{Z[\tilde J^*,J]}$.

The adjoint operator $\widehat{\psi[J]}{}^\dag$ maps from $\hilb_{\tilde \Sigma \sqcup \Sigma}$ to $\hilb_{\tilde \Sigma}$ by taking the boundary conditions defined by the state on which it acts, and gluing to boundary conditions of the CPT conjugate source $J^*$ along the manifold $\Sigma$.

Since the $\widehat{Z[J]}$  commute, it is again useful to diagonalize them using $\alpha$-states.  Thus the Hilbert space splits as
\begin{equation}
	\mathcal{H}_\Sigma = \bigoplus_\alpha \hilb_{\Sigma}^{\alpha}.
\end{equation}
One can explicitly build the spaces $\hilb_{\Sigma}^{\alpha}$ from the $\alpha$-states of $\hbu$, as we may define
\begin{equation}
\label{eq:buildSigma}
	\Big|\psi[J_1]\cdots \psi[J_m];\alpha\Big\rangle :=
	\widehat{\psi[J_1]}\cdots \widehat{\psi[J_m]}\Big|\alpha\Big\rangle
	\in \hilb_{\Sigma}^{\alpha},
\end{equation}
and, the states \eqref{eq:buildSigma} are dense in $\hilb_{\Sigma}^{\alpha}$. In the special case $\Sigma=\emptyset$ corresponding to $\hbu$, each $\hilb_\emptyset^\alpha$ is one dimensional, consisting of multiples of $|\alpha\rangle$.
It follows that all of our boundary operators leave $\alpha$ unchanged. For example, evaluating the analogue of \eqref{eq:psipsiZ} in $\alpha$-states we have
\begin{equation}
	\big\langle\psi[J_2];\alpha_2 \big|\psi[J_1];\alpha_1\big\rangle = Z_{\alpha_1}[J_2^*,J_1] \, \delta_{\alpha_1\alpha_2}.
\end{equation}
It also follows that $\widehat{Z[J]}$ commutes with $\widehat{\psi[\tilde J]}$.

Finally, note that there is a natural map $\Upsilon$ from $\hilb_{\Sigma_1}\otimes \hilb_{\Sigma_2}$ into $\hilb_{\Sigma_1 \sqcup \Sigma_2}$ defined by concatenation of sources:
\begin{equation}
	\begin{gathered}
		\big| \psi[J_{11}]\cdots \psi[J_{1,m_{\Sigma_1}}] Z[J'_{11}]\cdots Z[J'_{1,n_1}] \big\rangle \otimes 	\big| \psi[J_{21}]\cdots \psi[J_{2,m_{\Sigma_2}}] Z[J'_{21}]\cdots Z[J'_{2,n_2}] \big\rangle  \\ \mapsto
\big| \psi[J_{11}]\cdots \psi[J_{1,m_{\Sigma_1}}] \psi[J_{21}]\cdots \psi[J_{2,m_{\Sigma_2}}]
 Z[J'_{11}]\cdots Z[J'_{1,n_1}] Z[J'_{21}]\cdots Z[J'_{2,n_2}] \Big\rangle.
	\end{gathered}	
\end{equation}
This maps acts nicely within each $\alpha$-sector, taking $\hilb_{\Sigma_1}^\alpha\otimes \hilb_{\Sigma_2}^\alpha$ into $\hilb_{\Sigma_1 \sqcup \Sigma_2}^\alpha$.  In particular, since acting on $| \HH \rangle$ with the $\widehat{Z[J]}$ yields a dense set of states in $\hbu$, one may write $| \alpha \rangle = f_\alpha(\{Z[J_i]\}) | \HH\rangle$ for some function $f_\alpha$ that takes the value $1$ on arguments $\{Z_\alpha [J_i]\}$ but which vanishes
on $\{Z_{\alpha'} [J_i]\}$ for all $\alpha' \neq \alpha$.  One then finds
\begin{equation}
\big| \alpha \big\rangle \otimes \big| \alpha' \big\rangle  = f_\alpha f_{\alpha'} \big| \HH \big\rangle = \delta_{\alpha, \alpha'} \big| \alpha \big\rangle,
\end{equation}
and more generally
\begin{equation}
\Upsilon: \hilb_{\Sigma_1}^{\alpha}\otimes \hilb_{\Sigma_2}^{\alpha'} \rightarrow \delta_{\alpha, \alpha'}  \hilb_{\Sigma_1 \sqcup \Sigma_2}^{\alpha}.
\end{equation}
Here we have used the notation $c{\cal H}$ for non-negative real $c$  to denote a Hilbert space with inner product $c$ times that of ${\cal H}$.  In particular, $c{\cal H} = \{0\}$ for $c=0$.  We will use $\Upsilon_\alpha$ to denote the restriction of $\Upsilon$ to diagonal tensor products of the form
$\hilb_{\Sigma_1}^{\alpha}\otimes \hilb_{\Sigma_2}^{\alpha}$.

It is natural to attempt to interpret $\hilb_{\Sigma}^{\alpha}$ as the Hilbert space of a dual CFT $\mathcal{C}_\alpha$ on $\Sigma$; this is the natural formulation of an isomorphism between bulk and boundary Hilbert spaces in the context of ensembles and baby universes. In this case, we would expect $\Upsilon_\alpha$ to be an isomorphism, since this property would certainly hold true in a local dual theory. But this is not always the case, as the map may not be surjective; we will discuss an explicit example in section \ref{sec:modelHwBoundaries}.
The failure of $\Upsilon_\alpha$ to be an isomorphism is a precise version of another potential `factorisation problem' \cite{Harlow:2015lma,Guica:2015zpf,Harlow:2018tqv}, which differs from the partition function factorisation problem discussed in the introduction and the start of this section.   This new issue is naturally associated with spatial wormholes while  \eqref{eq:nofac} is related to spacetime wormholes. In particular, the factorization problem of \cite{Harlow:2015lma,Guica:2015zpf,Harlow:2018tqv} occurs  when there are two-sided black hole states with a spatial wormhole (Einstein-Rosen bridge) which cannot be represented as superpositions of products of `microstates' in the corresponding one-sided Hilbert spaces. For example, in a bulk theory with a standard Maxwell field but no charged particles, there are eternal charged black holes but no one-sided counterparts. An extreme version appears in pure JT gravity, which has a two-boundary Hilbert space but no single-sided Hilbert space.   We expect that this feature is an artefact of simple toy models, and would be absent in more realistic theories.

\section{Example: a very simple topological theory}\label{sec:models}

This section further explores the structure described in section \ref{sec:wormholesandBU} in very simple theories of two-dimensional gravity.
Indeed, the model described in section \ref{sec:Zonly} is plausibly the simplest possible such theory.
Our models are inspired by recent work studying spacetimes of nontrivial topology in JT gravity \cite{Saad:2018bqo,Saad:2019lba,Stanford:2019vob}, along with the addition of `end-of-the-world brane' dynamical boundaries \cite{Penington:2019kki}.  We further simplify that class of models by removing any notion of a dynamical metric or dilaton, leaving a theory of topology alone.  The resulting models are tractable enough to be solved exactly, and for many details to be made explicit.  They thus give a surprisingly clean illustration of the ideas of section \ref{sec:wormholesandBU}, and demonstrate the type of results to which such ideas can lead.

We begin by presenting the simplest model (without end-of-the-world branes) in section \ref{sec:Zonly}.  This theory allows only one boundary condition $Z$, associated with a single operator $\widehat{Z}$ of the class described in section \ref{sec:ops}, with the path integral defined by a single bulk parameter $S_0$ determining the suppression of nontrivial topology, along with a (somewhat ad hoc) parameter $S_\partial$ associated with boundaries, whose preferred value $S_\partial=S_0$ will be determined later by a consistency analysis in section \ref{sec:fudge}.  We then evaluate its amplitudes in section \ref{sec:Zamp} and construct the Hilbert space of closed universes $\hbu$ in section \ref{sec:ZBU}.  The most interesting output of this model is that the spectrum of $\widehat{Z}$ turns out to be non-negative and discrete, and in fact takes non-negative integer values for $S_\partial=S_0$, compatible with an interpretation as the dimension of a dual Hilbert space.  The model with end-of-the-world branes is then described in section \ref{sec:EOW}, and its $\alpha$-states are described in section \ref{sec:hbuEOW}.  Here we find that, no matter how many species $k$ of end-of-the-world brane states we allow, for $S_\partial=S_0$ all $\alpha$-states define an inner product on end-of-the-world brane states with rank equal to or less than the eigenvalue $Z_{\alpha}$ of $\widehat{Z}$, compatible with states in a dual Hilbert space of dimension $Z_{\alpha}$.
This remarkable compression of the Hilbert space illustrates the importance of understanding the null states \ref{eq:nullstates} in extracting the correct physics.  It also shows in this model that results analogous to the R\'enyi entropy computations of \cite{Almheiri:2019qdq,Penington:2019kki} will hold not just for typical members of the ensemble defined by the Hartle-Hawking no-boundary state, but in fact for all allowed $\alpha$-states.

We then return to the ad hoc parameter $S_\partial$ in section \ref{sec:fudge}. First, we describe how different choices for this parameter modify the model. We find that for generic $S_\partial$ (and in particular $S_\partial=0$) the end-of-the-world brane models fail to be reflection positive, and find the set of $S_\partial$ for which reflection positivity holds true. For values of $S_\partial$ satisfying reflection positivity for any number $k$ of end-of-the-world brane states, the spectrum of $\widehat{Z}$ is a subset of the non-negative integers and the rank of the end-of-the-world brane Hilbert space is bounded as above. In particular, the reflection positive models have all the properties required to interpret $Z_\alpha$ as the dimension of a Hilbert space which contains the end-of-the-world brane states.

\subsection{A theory of topological surfaces}
\label{sec:Zonly}

We now consider a theory of purely topological two-dimensional gravity in which spacetime is a two-dimensional manifold\footnote{For definiteness, we take smooth (not just topological) manifolds, and accordingly use the language of equivalence under diffeomorphisms rather than homeomorphisms.} (surface), but the only additional structure we introduce is an orientation. We thus have neither a spacetime metric nor the conformal or complex structure that would appear in the standard model of topological gravity \cite{Dijkgraaf:1990qw}. The histories that can appear in a path integral are then the set of oriented topological surfaces with boundaries dictated by the relevant boundary conditions.  This set is discrete and (for each connected component) is famously classified by genus and number of circular boundaries \cite{mobius1863theorie,jordan1866deformation}.  Since there is no possibility to add sources in this model, we simply use $Z$ to denote the boundary condition on any circular boundary.\footnote{We take the set of boundary conditions to be a vector space, so that a general boundary condition assigns a (perhaps complex) weight to each non-negative integer $n$ enumerating the possible numbers of circular boundaries.}

In this first model, the only boundaries are those fixed by boundary conditions.  As described in section \ref{sec:moreHilbert}, such boundaries should be thought of as distinguishable even when their boundary conditions coincide.  As a result, the space of allowed configurations is the set of oriented surfaces with \emph{labelled} boundaries, and two such configurations are considered equivalent only when they are related by a diffeomorphism that preserves each boundary separately.

We therefore define our path integral as a sum over such diffeomorphism classes of surface $M$. Nevertheless, residual effects of diffeomorphism invariance can lead to a nontrivial measure $\mu(M)$ on this space. This can arise when a group $\Gamma(M)$ of residual gauge symmetries remains after gauge fixing diffeomorphisms. This naturally leads to symmetry factors in the measure, of the form $\mu(M)=\frac{1}{|\Gamma(M)|}$.  One may therefore expect to write our path integral in the tentative form
\begin{equation}\label{eq:topPI}
	\int \mathcal{D}\Phi \,e^{-S[\Phi]} := \sum_{\text{Surfaces }M} \mu(M) \; e^{-S[M]},
\end{equation}
where we sum over surfaces $M$ obeying the appropriate boundary conditions, up to diffeomorphisms acting trivially on the boundaries, weighted by an action $S[M]$.

One would ideally like to derive the measure factor $\mu(M)$ from a more complete model. Here, we will be content to define the model with a well-motivated choice of measure that leads to natural results.  Since boundaries are distinguishable, and since any two surfaces related by boundary-preserving diffeomorphisms are already considered equivalent, we will assume the trivial measure $\mu(M)=1$ for any connected manifold. It then remains to discuss only contributions to $\mu(M)$ from boundary-preserving diffeomorphisms that interchange the connected components of $M$.  These can act only on compact connected components (i.e., the ones that have no boundary).  With this understanding, the detailed form of $\mu(M)$ turns out to have little effect on the physics of interest.  It leads only to a change of the `cosmological partition functon' $\mathfrak{Z}$, the sum over compact universes, which is an overall normalisation of amplitudes (though at the end of section \ref{sec:ZBU} we will encounter a situation in which our choice of measure is physically important).  Nevertheless, we regard diffeomorphisms that permute compact connected components (necessarily with the same genus $g$) as residual gauge symmetries, and divide by the number of such permutations in the measure. This means that, if $M$ has $m_{g}$ connected components of genus $g$ with no boundary for each $g$, we have
\begin{equation}
\label{eq:ts0}
\mu(M) = \frac{1}{\prod_g  m_{g}!}.
\end{equation}

Following the principles of effective field theory, we should now write down the most general action allowed by the degrees of freedom. Fortunately, with only the topological degrees of freedom available to us,
there is a unique local such action $S(M)=-S_0\chi(M)$, proportional to the Euler characteristic $\chi$ of spacetime\footnote{Here we take locality to mean invariance under cutting and gluing surfaces. A precise version of the above statement is then that $\exp(S_0\chi)$ is the most general form for the amplitudes of a two-dimensional topological quantum field theory (TQFT) with trivial (one-dimensional) Hilbert space on the circle.}, with a unique free parameter $S_0$. This is the Einstein-Hilbert action in two dimensions, and is the topological term of the action in JT gravity.

Despite the apparent uniqueness for the action, we now introduce an additional term $-S_\partial |\partial M|$, where $|\partial M|$ denotes the number of circular boundaries of $M$.  As forewarned in the introduction to this section, for the moment the extra parameter $S_\partial$ appears completely ad hoc. In particular, while this is an intrinsic function of asymptotic boundaries, it is not a \emph{local} counterterm. Indeed, as stated above, we expect that the unique local theory of our form is given by setting $S_\partial=0$.  We discuss how this factor may arise in \ref{sec:fudge} below, perhaps most simply by introducing a new local degree of freedom residing on boundaries.  For now we simply note that the parameter effectively just rescales the definition of $Z$; i.e., it can be removed by introducing $\tilde Z = e^{S_\partial}Z$ and replacing each $Z$ in \eqref{eq:topPI} by $\tilde Z$.

Since all values of $S_\partial$ are related by this scaling, it suffices to discuss only a single value in detail, and then to use the above scaling to understand all other values.  Until section \ref{sec:fudge}, we will thus confine discussion to the particularly simple case $S_\partial = S_0$.  As an a posteriori justification, we will show in section \ref{sec:fudge} that the end-of-the world brane models fail to be reflection positive when $S_\partial=0$, and $S_\partial = S_0$ is the most natural choice to cure this failure.

Our action is thus given by
\begin{gather}
	S(M) = -S_0 \chi(M) - S_\partial \, n(M), \\
	\text{where we choose } S_\partial = S_0 \quad \text{(until section \ref{sec:fudge}).}
\end{gather}
The practical simplification of choosing $S_\partial=S_0$ is that it precisely cancels boundary contributions to $\chi$ in the action.
The amplitudes in our path integral thus take the form
\begin{equation}\label{eq:Znsum}
	\big\langle Z^n \big\rangle \quad = \sum_{\substack{M \text{ with} \\ |\partial M|=n}}\mu(M)e^{S_0 \tilde{\chi}(M)} \enspace,
\end{equation}
which we have written in terms of a modified Euler characteristic $\tilde \chi$ that does not count boundaries and which is given simply by
\begin{equation}
\label{eq:tchi}
	\tilde{\chi} = \sum_{\substack{\text{Connected}\\\text{components}}} (2-2g).
\end{equation}
Here $g$ is the usual genus of each connected component that counts handles.

It will be useful below to sometimes use an alternate presentation of the sum \eqref{eq:Znsum}.  Instead of summing over surfaces with labeled boundaries, we can write $\big\langle Z^n \big\rangle$ as a sum over ordered lists $M_L$  of \emph{connected manifolds}, and also where we choose not to label the boundaries.  The number of ways to label the boundaries is then accounted for by including a separate factor of the multinomial coefficient $\frac{n!}{\prod_i n_i!}$, where $n_i$ is the number of boundaries in the $i$th entry of the list $M_L$.  As is well known, $\frac{n!}{\prod_i n_i!}$ gives precisely the number of ways to arrange $n$ boundaries into lists of subsets that have $n_i$ boundaries in the $i$th subset.   For a list of length $m$, including a factor of $\frac{1}{m!}$ then accounts for the fact that the components are not ordered in the original sum \eqref{eq:Znsum}, and also for the factor of $\mu(M)$ that arises when some items in the list both coincide and have no boundaries (so that exchanging these items neither generates a new term in \eqref{eq:Znsum} nor generates a new partition of the $n$ boundaries).    Thus we may rewrite  \eqref{eq:Znsum} as
\begin{equation}\label{eq:Znsum2}
	\big\langle Z^n \big\rangle \quad = \sum_{\substack{\text{Ordered}\  \text{lists} \ M_L \\ \text{of \ connected \ surfaces} \\ \text{with} \ n\text{ boundaries}}} \frac{n!}{m!\prod_i n_i!} e^{S_0 \tilde \chi} \enspace,
\end{equation}
where $n$, $m$, and $n_i$ are as above.

Before computing the amplitudes \eqref{eq:Znsum}, it is useful to comment further on the interpretation of $Z$ in terms of a putative dual $0+1$-dimensional quantum mechanics (which we will sometimes call a CFT in analogy with AdS/CFT). Each $Z$ would be naturally associated with the path integral of this quantum mechanics on the circle, which would describe the partition function $\Tr e^{-\beta H}$ for a circle of length $\beta$. But since we have no metric, there is no notion of boundary length $\beta$, and invariance under diffeomorphisms of the boundary implies a vanishing Hamiltonian $H=0$. This means we have a topological quantum mechanics (a one-dimensional TQFT) where the only observable is the trace of the identity operator, which is the dimension of the Hilbert space:
\begin{equation}\label{eq:Zdim}
	Z \stackrel{?}{=} \Tr_{\hilb_\text{CFT}} 1 = \dim \hilb_\text{CFT}
\end{equation}
A unitary dual quantum mechanics is therefore characterised by $Z$ taking a value in the natural numbers $\NN$ (or perhaps by $Z$ being infinite). In the presence of spacetime wormholes connecting these boundaries, it would thus seem natural to find that $Z$ is a random variable taking nonnegative integer values. We will see below that this is precisely the case for our model.

\subsection{Evaluating the amplitudes}
\label{sec:Zamp}

We now solve for the amplitudes $\big\langle Z^n \big\rangle$ defined above. We begin by computing the no-boundary partition function $\mathfrak{Z}$ as in equation \eqref{eq:HHnorm}.  This is the case $n=0$, given by the sum over arbitrary compact spacetimes without boundary.  For this, we first compute the sum $\lambda$ over \emph{connected} compact surfaces, which are classified by genus.
The measure is trivial for a connected surface, i.e.\ $\mu(M)=1$, so we have
\begin{equation}\label{eq:lambda1}
	 \lambda :=  \sum_{\substack{\text{Connected} \\ \text{compact \ surfaces}}} e^{S_0\chi} =\sum_{g=0}^\infty e^{S_0(2-2g)} = \frac{e^{2S_0}}{1-e^{-2S_0}}.
\end{equation}
With our amplitudes defined by \eqref{eq:Znsum}, and in particular excluding boundaries from the count in the Euler character, the value of $\lambda$ is always the amplitude for any connected component of spacetime (with fixed but arbitrary boundaries) after summing over connected topologies.  This property determines all amplitudes of the model.

In the usual way, one may write $\mathfrak{Z}$  as the exponential of the sum $\lambda$ over connected surfaces. For this, it is important that we include symmetry factors in our definition of the measure $\mu(M)$. Indeed, the exponentiation is particularly explicit by using \eqref{eq:Znsum2} with $n=n_i=0$, in which lists of length $m$ contribute $\frac{1}{m!}$ times the $m$th power of the sum in \eqref{eq:lambda1}.  We thus find
\begin{equation}
\label{eq:mfZ}
	\mathfrak{Z} = \big\langle 1\big\rangle = e^\lambda \,.
\end{equation}
In particular, in our model the path integral defined by the sum over topologies converges.

We now introduce boundaries. To evaluate $\langle Z^n\rangle$, it is simplest to compute a generating function
\begin{equation}
\label{eq:genfunc}
\left \langle e^{u Z} \right\rangle= \sum_{n=0}^\infty \frac{u^n}{n!} \big\langle Z^n \big\rangle,
\end{equation}
and to extract the amplitudes from a power series in the `chemical potential' $u$. Again, we wish to write \eqref{eq:genfunc} as the exponential of a sum over connected geometries.  This is precisely the usual combinatorics familiar from Feynman diagrams, but it can also be seen explicitly from \eqref{eq:Znsum2} which gives
\begin{equation}
\label{eq:genfunc2}
\left \langle e^{u Z} \right\rangle=
 \sum_{\substack{\text{Ordered lists } M_L \\ \text{of connected surfaces}}} \frac{u^{\sum_i n_i}}{m!\prod_i n_i!} e^{S_0 \tilde{\chi}(M_L)} \enspace,
\end{equation}
where $m$ is the number of surfaces in the list $M_L$, and $n_i$ for $i=1,\ldots m$ is the number of boundaries of the $i$th surface in the list. Since $\tilde{\chi}$ for the disconnected surface $M_L$ is the sum of $\tilde{\chi}$ for the individual components, this disconnected pieces exponentiate,
\begin{equation}
\label{eq:lngenfunc}
\log \left \langle e^{u Z} \right\rangle=  \sum_{n=0}^\infty \sum_{\substack{\text{Connected } M \\  n \text{ boundaries}}} \frac{u^n}{n! } e^{S_0\tilde{\chi}(M)}.
\end{equation}
Furthermore, since the factor
$\frac{u^n}{n!}$ is determined entirely by $n$ while the factor
$e^{S_0\tilde{\chi}(M)}$ depends only on the genus $g$, the double sum in \eqref{eq:lngenfunc} may be written as the product
\begin{equation}
\label{eq:lngenfunc2}
\log \left \langle e^{u Z} \right\rangle=  \left(\sum_{g=0}^{\infty}  e^{S_0\tilde{\chi(M)}}\right) \left( \sum_{n=0}^\infty \frac{u^n}{n!} \right) =  \lambda e^u.
\end{equation}
Here the last equality has used \eqref{eq:lambda1} to identify $\lambda$ with the sum over $g$. We can extract the correlators $\Big \langle Z^n \Big \rangle$ by expanding the generating function $\exp\left( \lambda e^u\right)$ in powers of $u$.

We pause to note that there is a more direct way to compute the amplitudes $\big\langle Z^n\big\rangle$.  Here we first divide by $\mathfrak{Z}$ to remove contributions from closed manifolds and thus any mention of $\mu(M)$.  What remains is then just to simply count the relevant configurations remaining in \eqref{eq:Znsum}.  Such configurations are classified according to which of the $n$ boundaries lie in the same connected component of spacetime, and thus by a partition of the set $\{1,2,\ldots,n\}$ labelling the boundaries. For each connected component of spacetime, it then remains only to sum over genus, giving a factor of $\lambda$ from \eqref{eq:lambda1}. We may thus compute the amplitudes from a counting of partitions, graded by the number of subsets of $\{1,2,\ldots,n\}$ that the partition defines:
\begin{equation}\label{eq:Zamplitudes}
	\mathfrak{Z}^{-1}
	\big\langle Z^n\big\rangle \quad = \enspace  \sum_{\substack{\text{Partitions }p\\\text{of } \{1,2,\ldots,n\}}} \lambda^{(\text{Number of subsets in }p)} = B_n(\lambda).
\end{equation}
Here $B_n$ is known as the Bell polynomial of order $n$ (\texttt{BellB[n,$\lambda$]} in Mathematica; also called \href{https://en.wikipedia.org/wiki/Touchard_polynomials}{Touchard polynomial}).  In agreement with our previous result, these polynomials are indeed known to have the generating function $\exp(\lambda(e^u-1))$ as in \eqref{eq:lngenfunc2} after dividing by $\mathfrak{Z}=e^\lambda$.

To illustrate the counting in detail, consider the example of the third moment $\big\langle Z^n\big\rangle$; i.e., the case $n=3$. There are five distinct ways to divide the three boundaries into connected components:
\begin{equation}
\label{eq:count2}
	\begin{aligned}
	\mathfrak{Z}^{-1}  \big\langle Z^3 \big\rangle &= \vcenter{\hbox{\includegraphics[width=28pt]{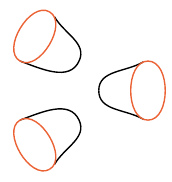}}} +
		\vcenter{\hbox{\includegraphics[width=28pt]{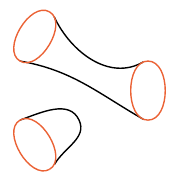}}} +
		\vcenter{\hbox{\includegraphics[width=28pt]{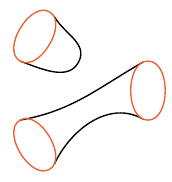}}} +
		\vcenter{\hbox{\includegraphics[width=28pt]{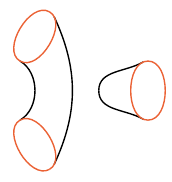}}} +
		\vcenter{\hbox{\includegraphics[width=28pt]{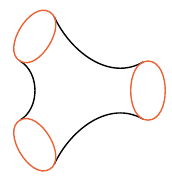}}} \\
		 &= \lambda^3 + 3\lambda^2 + \lambda
	\end{aligned}
\end{equation}
Since the boundaries are distinguishable, the three configurations with two connected components are counted separately, and there are no explicit symmetry factors in the first line above.\footnote{For indistinguishable boundaries the answer would be multiplied by $\frac{1}{3!}$, or more generally by $\frac{1}{n!}$ for $n$ boundaries).}  The alternative counting used in \eqref{eq:Znsum2} would instead list each topologically distinct term in \eqref{eq:count2} only once, but would accompany each term by the number $N_L$ of distinct ordered lists that one can construct from the connected components and the factor of $\frac{n!}{m! \prod n_i!}$ from \eqref{eq:Znsum2}.  This gives the identical result
\begin{equation}
	\begin{aligned}
	\mathfrak{Z}^{-1}  \big\langle Z^3 \big\rangle &= \frac{3!}{3!(1!)^3} \vcenter{\hbox{\includegraphics[width=28pt]{ZZZ1}}} + \frac{2 \ 3!}{2! 2! 1!}
		\vcenter{\hbox{\includegraphics[width=28pt]{ZZZ2}}}
	\frac{3!}{1!3!}	\vcenter{\hbox{\includegraphics[width=28pt]{ZZZ5}}} \\
		 &= \lambda^3 + 3\lambda^2 + \lambda,
	\end{aligned}
\end{equation}
where the first term has $(N_L,m!,\frac{n!}{\prod n_i!}) = (1,3!,\frac{3!}{1!1!1!})$ since the $3$ components are all identical but have only one boundary each, the second term has $(N_L,m!,\frac{n!}{\prod n_i!}) = (2,2!,\frac{3!}{2!1!})$ since the two  components are not homeomorphic but the cylinder has $2$ boundaries, and the third term has $(N_L,m!,\frac{n!}{\prod n_i!}) = (1,1,\frac{3!}{3!})$ since all $3$ boundaries lie in the single connected component.

We now interpret the amplitudes in terms of a probability distribution where $Z$ is regarded as a random variable. To do this, we divide the generating function $\left \langle e^{u Z} \right\rangle $ by the normalisation factor $\mathfrak{Z}$ and write the result as the Taylor series for the exponential:
\begin{equation}
\label{eq:expgfZ}
	\mathfrak{Z}^{-1} \left \langle e^{u Z} \right\rangle =  \sum_{d=0}^\infty p_d(\lambda) e^{u d},\qquad p_d(\lambda)=e^{-\lambda} \frac{\lambda^d}{d!} \,.
\end{equation}
Extracting the coefficient of $\frac{u^n}{n!}$ from \eqref{eq:expgfZ} gives
\begin{equation}
\label{eq:Zm}
	\mathfrak{Z}^{-1} \left \langle Z^n \right\rangle =  \sum_{d=0}^\infty d^n p_d(\lambda) ,\qquad p_d(\lambda)=e^{-\lambda} \frac{\lambda^d}{d!} \, ,
\end{equation}
showing that all moments can be generated from a single distribution for $Z$ with support on nonnegative integers $d$ having manifestly non-negative probabilities $\Pr(Z=d)=p_d(\lambda)$. We thus identify $Z$ as a Poisson random variable with mean $\lambda$. We may also read this off directly from \eqref{eq:lngenfunc2} using the fact that $\exp\left[\lambda (e^u-1) \right]$ is the moment generating function for a Poisson random variable. Alternatively, one can see this from the amplitudes \eqref{eq:Zamplitudes} using the fact that $B_n$ is the $n$th moment of the Poisson distribution. The appearance of the Poisson distribution can be understood from the result that all connected components of spacetime contribute the same amplitude $\lambda$ after summing over genus, independent of the number of boundaries. This corresponds to the fact that the cumulants of the Poisson distribution (that is, the completely connected correlation functions) are all equal to $\lambda$.

This is a surprising and remarkable result. As reviewed in section \ref{sec:3q} below, a perturbative description of the theory following \cite{Giddings:1988wv} (based on a Fock space labelled by number of baby universes and with wormholes treated as a small correction) would have
led to the expectation that $Z$ should have a continuous distribution supported on all real numbers. Instead, from our exact nonperturbative solution we find that the support of $Z$ is discrete, and limited to nonnegative values.

Furthermore, for our choice $S_\partial = S_0$ (or more generally for $S_\partial = S_0 + \log n$ for any positive integer $n$), since $Z$ takes nonnegative integer values $d$ we find that the result is compatible with the interpretation \eqref{eq:Zdim} in terms of an ensemble of dual Hilbert spaces.  Although at this stage this result appears to depend on fine tuning the parameter $S_\partial$, we will see in section \ref{sec:fudge} that full consistency (in particular full reflection positivity) of the model in fact favours precisely the relation $S_\partial = S_0 + \log n$.

As a final comment, it is interesting that the relation \eqref{eq:lambda1} between the `bare' parameter $e^{S_0}$ and the physically observable parameter $\lambda$ is not injective, but is instead two-to-one. This means that there for a given value of $e^{S_0}$, there is a second value $e^{\tilde{S}_0}$ that gives rise to the same $\lambda$, and hence the same theory.  In particular, we find
\begin{equation}
	e^{-\tilde{S}_0} = 1-e^{-S_0}\,.
\end{equation}
This is a strong--weak self-duality of the model in the sense that the semiclassical limit of large $S_0$ suppresses connected topologies (and thus describes weakly coupled universes), but yields the same theory as a very small value of the dual $\tilde{S}_0$. At the self-dual value $e^{S_0}=2$ we have $\lambda=4$, and smaller values of $\lambda$ correspond to complex couplings, with $e^{-S_0}\in \frac{1}{2} + i \RR $. From the point of view of the path integral in a semiclassical expansion it is surprising that such a complex coupling gives rise to reflection positive amplitudes, and hence to a unitary Hilbert space and positive probabilities.

\subsection{The baby universe Hilbert space}
\label{sec:ZBU}

We can now give a complete description of the Hilbert space of closed universes $\hbu$. Every state can be written as a linear combination of $\big|Z^m\big\rangle$ created by inserting $m$ boundaries in the past, with inner product
\begin{equation}\label{eq:ZmZnIP}
	\begin{aligned}
	\Big\langle Z^n \Big| Z^m \Big\rangle &= \Big\langle Z^{m+n} \Big\rangle \\
	&= e^\lambda B_{m+n}(\lambda) \\
	&= \sum_{d=0}^\infty \frac{\lambda^d}{d!} d^{m+n}.
	\end{aligned}
\end{equation}
A more general state $\sum_{n=0}^\infty c_n |Z^n\rangle$ can then be represented as $|f(Z)\rangle$, where $f$ is a function with Taylor coefficients $c_n$, which grow slowly enough for convergence. Demanding that the partial sums $\left\{\sum_{n=0}^N c_n |Z^n\rangle\right\}_N$ form a Cauchy sequence guarantees that $f$ defines an entire analytic function (see appendix \ref{app:convergence}). Before considering the details of the inner product, we are thus led to the idea that $\hbu$ is a space of functions $f:\RR\to\CC$ (or perhaps $f:\CC\to\CC$), with argument $Z$.

We can read off the extension of the inner product to states $|f(Z)\rangle$ from the last line in \eqref{eq:ZmZnIP}:
\begin{equation}\label{eq:fgIP}
	\Big\langle g(Z) \Big| f(Z) \Big\rangle = \sum_{d=0}^\infty \frac{\lambda^d}{d!} \overline{g(d)} f(d).
\end{equation}
This is (up to normalisation factor $e^\lambda$) the covariance of random variables $f(Z),g(Z)$ where $Z$ is Poisson distributed. But the salient feature of \eqref{eq:fgIP} is that it depends only on the vales of $f$ and $g$ evaluated at non-negative integers (also known as the set $\NN$ of natural numbers). In particular, we find that the state $\big|f(Z)\big\rangle$ has zero norm whenever the function $f$ vanishes on $\NN$:
\begin{equation}
	\Big\Vert \;\big|f(Z)\big\rangle\;\Big\Vert^2=0 \iff f(d)=0 \text{ for all }d\in\NN .
\end{equation}
To form the Hilbert space $\hbu$, we must quotient by such null states as in \eqref{eq:nullstates}. For example, since $\sin(\pi Z)$ vanishes on $\NN$ we have the otherwise surprising relation
\begin{equation}
	\left|\sin(\pi Z)\right\rangle = \sum_{n=0}^\infty \frac{(-1)^n \pi^{2n+1}}{(2n+1)!} \big|Z^{2n+1}\big\rangle =0.
\end{equation}
More generally, for any $f$ we have $\big|\sin(\pi Z)f(Z)\big\rangle=0$, so in some sense the space of null states is the same size as the total space before the quotient.
Similarly, the Hartle-Hawking state can be represented by the constant function $f(Z)=1$, or more generally by any function that has $f(d)=1$ for all $d\in \NN$ (for example, $|\HH\rangle = |e^{2\pi i j Z}\rangle$ for any integer $j$). To emphasise the impact of the quotient by null states, note that by adding vectors of the form $\left|Z^n\sin(\pi Z)\right\rangle $ we can change any finite number of coefficients $c_n$ (for $n\neq 0$) in the expansion of the state $\sum_{n=0}^\infty c_n |Z^n\rangle$ at will.  As a result, the only physical information in any finite collection of coefficients $c_n$ is the overlap with the $Z=0$ eigenstate (given by $c_0$).

These considerations reveal an enormous degeneracy in how states of $\hbu$ are represented as sums of $|Z^n\rangle$. We regard this degeneracy as a gauge equivalence. As described in section \ref{sec:BUH} this gauge symmetry is a natural modification of diffeomorphism invariance associated with allowing topology change in the functional integral.  But the enormous power of this seemingly natural modification comes as a surprise.  This indicates that the corrections to diffeomorphism invariance are not generic, but are instead highly correlated.  As a result, the corrections conspire to enhance the impact of the gauge symmetry, and thus to produce the degeneracy observed above.  Such conspiracies call out for a more fundamental explantation, and we will see in sections \ref{sec:fudge} and \ref{sec:relation} below that at least some of these conspiracies are in fact implied by reflection positivity of our path integral.

In parallel with the treatment in section \ref{sec:ops}, we can now discuss the $\alpha$-states of our model. These are the eigenstates $\Big|Z=d \Big\rangle$ of $\widehat{Z}$, labelled by $d\in\NN$, and they must form a basis for $\hbu$.  When expressed as a sum of the states $|Z^n\rangle$ states, we may choose coefficients defining the Taylor series of any analytic function taking a non-zero value at $Z=d$ but vanishing at other natural numbers, since multiplication by $Z$ acts as multiplication by the constant $d$ on such a function. One of the infinitely many ways to represent such eigenstates states is then
\begin{equation}
\label{eq:Z=d}
	\Big|Z=d \Big\rangle = \left(\frac{\lambda^d}{d!}\right)^{-1/2} \left|\frac{\sin(\pi Z)}{\pi(Z-d)}\right\rangle,
\end{equation}
where the coefficient is chosen to enforce the normalisation
\begin{equation}
	\Big\langle Z=d'\Big|Z=d \Big\rangle = \delta_{dd'}\;.
\end{equation}

Finally, we discuss the spacetime interpretation of our operator $\widehat{Z}$ and its eigenstates $\big|Z=d\big\rangle$. From \eqref{eq:fgIP}, note that projecting the states $\big| f(Z) \big\rangle$ onto the (here, one-dimensional) subspace where $\widehat{Z}$ takes the value $d$ is equivalent to restricting the sum on the right-hand side of  \eqref{eq:fgIP} to the given eigenvalue $d$, or equivalently to terms of order $\lambda^d$.  But due to \eqref{eq:lambda1} (and the fact that the analogous equations are identical for any fixed number $n>0$ of boundaries on the connected surface), these give precisely the contributions in \eqref{eq:Znsum} that arise from spacetimes with $d$ connected components. We thus find that working in the eigenspace with eigenvalue $d$ is equivalent to restricting the sum over amplitudes to terms where the universe has precisely $d$ connected components\footnote{We thank Xi Dong for discussions on this point.}.

In other words, the operator $\widehat{Z}$ counts the number of connected components of spacetime! This is quite surprising, since this is not a quantity we would naturally associate with a Cauchy slice if we were to attempt to quantise by gauge fixing diffeomorphisms (unlike the number of connected components of \emph{space}, which is a natural observable when universes cannot split and join, but is not gauge invariant when they can).

The $\alpha$-states are designed to make amplitudes factorise \eqref{eq:alphafactorise}, and it is interesting to note how our model achieves this. To work in an $\alpha$-state $\big|Z=d\big\rangle$, we can impose the nonlocal constraint that spacetime has exactly $d$ connected components. This does not exclude wormhole configurations connecting multiple boundaries, but provides additional correlations between disconnected configurations of boundaries. It thus achieves factorisation in a surprising way, which may be instructive for less simple models. Note that our choice of symmetry factors on spacetimes without boundary, which otherwise only acts to renormalise $\mathfrak{Z}$, is crucial for this simple description of $\alpha$-state correlation functions.

Since $\widehat{Z}$ takes values in $\NN$, $\hbu$ has a natural representation as a harmonic oscillator Hilbert space in which $\widehat{Z}$ acts as a number operator.\footnote{This is not to be confused with the free Fock space description of section \ref{sec:3rdQuantisation}, in which $\widehat{Z}$ is a harmonic oscillator position operator.}  We can define the annihilation operator $a$ as acting to shift functions of $Z$,
\begin{equation}
	a \big|f(Z)\big\rangle = \sqrt{\lambda}\big|f(Z+1)\big\rangle,
\end{equation}
so that we have the relations
\begin{equation}
	\begin{gathered}
		\widehat{Z} = N = a^\dag a \,, \\
		a|Z=0\rangle =0\,, \quad \text{and} \\
		|Z=d\rangle = \frac{1}{\sqrt{d!}}(a^\dag)^d |Z=0\rangle.
	\end{gathered}
\end{equation}
In this description, the Hartle-Hawking state is a coherent state, which can be represented as
\begin{equation}
	\big|\,\HH\,\big\rangle = e^{\sqrt{\lambda} a^\dag} |Z=0\rangle.
\end{equation}
The distribution of the associated ensemble then follows from the well-known fact that the number operator follows a Poisson distribution in a coherent state.

\subsection{End-of-the-world branes}\label{sec:EOW}

We now extend the model described above by introducing dynamical boundaries, which (following \cite{Penington:2019kki}) we call end-of-the-world (EOW) branes. We choose to include an arbitrary number $k$ of species of EOW brane, so each of these boundaries is labelled by an index $i\in\{1,2,\ldots,k\}$. Equivalently, we can place a topological quantum mechanics on the EOW branes, with zero Hamiltonian and a $k$-dimensional Hilbert space, so that $i$ labels an orthonormal basis of states in that Hilbert space. Apart from the species label, the only local data on an EOW brane is an orientation compatible with the spacetime it bounds.

Introducing the EOW branes has two effects. Firstly, they can appear as closed boundaries in the sum over topologies, but this is largely unimportant, only acting to change the value of $\lambda$ so that it is no longer given by \eqref{eq:lambda1}. More importantly, the EOW branes allow us to impose a new class of possible boundary conditions. Namely, we can specify that we have a boundary condition which is an oriented interval labelled at its endpoints by EOW brane species $i$ and $j$. Since the interval is oriented, we may refer to it as having a past endpoint that creates an EOW brane of type $i$ and a future endpoint that destroys an EOW brane of type $j$. We refer to both past and future labels as EOW brane sources. In a putative 0+1 dual, the condition that a boundary creates an EOW brane with label $i$ corresponds to the preparation of a certain 0+1 dual state $\psi_i$. We denote a boundary interval between EOW branes $i$ and $j$ by $(\psi_j,\psi_i)$ since the bulk path integral with this boundary condition should compute the inner product between these states.
\begin{equation}
	(\psi_j,\psi_i) = \vcenter{\hbox{\includegraphics{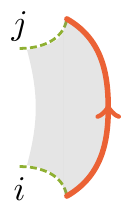}}}
\end{equation}
Since the boundaries carry an orientation, the notation distinguishes bra-vectors from ket-vectors so that $(\psi_j,\psi_i)\neq (\psi_i,\psi_j)$; in general, these are CPT conjugate boundary conditions.  This coincides with the general notation introduced in section \ref{sec:BUH}.

\begin{figure}
	\centering
	\includegraphics[width=.3\textwidth]{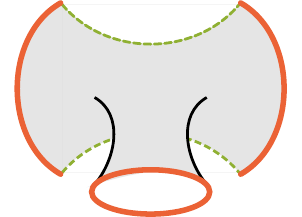}
	\caption{A spacetime contributing to an amplitude $\big\langle (\psi_j,\psi_i)(\psi_i,\psi_j)Z\big\rangle$. The solid red lines indicate asymptotically AdS boundaries, and the dashed green lines are EOW brane boundaries. The spacetime has two boundary components, each with the topology of a circle.  One (solid red circle at bottom) is a single circular asymptotically AdS boundary (a $Z$-boundary).  The other is formed by a pair of asymptotically AdS segments connected by a pair of EOW brane segments to form a topological circle.\label{fig:formbndys}}
\end{figure}

Including the $\psi_i$, the most general amplitude can now be written
\begin{equation}
	\Big\langle Z^m (\psi_{j_1},\psi_{i_1})\cdots (\psi_{j_n},\psi_{i_n})\Big\rangle.
\end{equation}
The associated boundary conditions for the path integral require $m$ circular boundaries without EOW brane sources and $n$ additional interval boundary segments labelled appropriately with EOW brane species. Since the EOW branes are dynamical, the path integral is then computed by summing over all oriented surfaces whose circular boundaries are of the following three types: 1) circular EOW brane boundaries, each labelled by an arbitrary species independent of all boundary conditions, 2) $m$ circular boundaries without EOW brane labels as dictated by the number of $Z$'s in the amplitude, and 3) additional circular boundaries formed by partitioning into subsets the oriented intervals $(\psi_{j},\psi_{i})$ dictated by the boundary conditions and, for each subset, forming a circle by connecting the $(\psi_{j},\psi_{i})$ segments using oriented EOW brane segments whose species labels match the source labels at both endpoints. See figure \ref{fig:formbndys} for an example. 

We now know the set of amplitudes to compute and the corresponding configurations over which we are to sum. It remains only to specify the measure on the configurations. As before, the Euler characteristic is the unique local action without introducing additional degrees of freedom. However, we will again include a parameter $S_\partial$ associated with each circular boundary.  We use the same $S_\partial$ for every circular boundary, no matter how it is formed from asymptotic pieces and EOW branes.  Again, we will see in section \ref{sec:fudge} that this can be obtained by introducing additional local degrees of freedom which reside on both asymptotic and EOW brane boundaries, and integrating them out. While this no longer corresponds to a simple scaling of our operators, we will nonetheless once again focus on the case $S_\partial=S_0$, resulting in an action which counts only genus and not the number of boundary components, and comment on the extension to other values in section \ref{sec:fudge}.

It remains to specify the symmetry factors that will be the analog of $\mu(M)$ in \eqref{eq:topPI}.   In doing so, it is useful to note that, since all asymptotic boundaries are treated as distinguishable, they will not contribute to symmetry factors.  The only indistinguishable boundaries are those formed by circles involving EOW branes alone.  Furthermore,  such circles are completely independent of the boundary conditions.  They thus enter all of our sums in precisely the same way as the genus $g$.  The analogue of \eqref{eq:Znsum} for our new model is then
\begin{equation}\label{eq:EOWamp}
	\Big\langle Z^m (\psi_{j_1},\psi_{i_1})\cdots (\psi_{j_n},\psi_{i_n})\Big\rangle   =
\sum_{M} \mu(M)e^{S_0 \tilde \chi} \enspace,
\end{equation}
where we sum over diffeomorphism classes of surface $M$ with the boundary conditions specified on the left hand side. The measure $\mu$ is analogous to \eqref{eq:ts0} but includes additional factors associated with counting end-of-the-world branes using Bose statistics.

We may now proceed to evaluate the above amplitudes.  As a first step, we again define $\lambda$ as the sum over connected surfaces with no asymptotic boundaries in analogy with \eqref{eq:lambda1}. However, this sum must now allow for the possibility of circular EOW brane boundaries, each with $k$ possible species labels. Since EOW brane boundaries can be specified in precisely the same way for each genus, this simply multiplies the result \eqref{eq:lambda1} by an overall factor counting the number of possible such labelled boundaries.  For a fixed number $n$ of EOW brane boundaries, including symmetry factors we count $\frac{k^{n}}{n!}$ ways to label the boundaries with $k$ species. Summing this factor over $n$ shows the new factor to be $e^k$ and we obtain
\begin{equation}\label{eq:lambda2}
	\lambda  = \frac{e^{2S_0}}{1-e^{-2S_0}} e^k.
\end{equation}

As before, we can now compute all amplitudes through a generating function, where we sum over all configurations, with any number of asymptotic boundaries, and fugacities $u$ and $t_{ij}$ (with $i=1,\cdots,k$) for the $Z$ and $(\psi_j,\psi_i)$ boundaries respectively.  As we explain below, this yields
\begin{equation}\label{eq:gf2}
	\Bigg \langle \exp \bigg( u Z + \sum_{i,j=1}^k t_{ij} (\psi_j,\psi_i)\bigg) \Bigg \rangle = \exp\left[\lambda \frac{e^u}{\det(I-t)}\right],
\end{equation}
where $t$ is the $k\times k$ matrix with entries $t_{ij}$, and $I$ the $k\times k$ identity matrix.

Once again, we compute this result by writing it as the exponential of a sum over connected spacetimes, each weighted by a factor of $\lambda$ from summing over genus and closed EOW branes.  The connected contribution is a sum over all possible boundaries we could insert on a given connected spacetime (excepting circular EOW brane boundaries, which have already been absorbed into $\lambda$). This sum is itself given as the exponential of a sum over distinct types of boundaries:
\begin{equation}
	\frac{e^u}{\det(I-t)} = \exp\left[u + \sum_{n=1}^\infty \frac{1}{n} \Tr t^n \right]
\end{equation}
The $u$ accounts for insertions of circle boundaries $Z$ as before. The $n$th term in the sum comes from boundary components consisting of $n$ intervals corresponding to some $(\psi_j,\psi_i)$, alternating with $n$ EOW branes. Summing over species of EOW branes results in the matrix product and trace, and the factor of $\frac{1}{n}$ avoids overcounting equivalent configurations where the $n$ component intervals are cyclically permuted.

For an alternative route to this result where various factors are more explicit, we can present \eqref{eq:EOWamp} as a sum over ordered lists of connected manifolds.  This is readily obtained from \eqref{eq:Znsum2} by recognizing that the circular EOW brane boundaries enter every sum on the same footing with the genus $g$. We have
\begin{align}\label{eq:EOWamp2}
	\Big\langle Z^m (\psi_{j_1},\psi_{i_1})\cdots (\psi_{j_n},\psi_{i_n})\Big\rangle & \\  =
\sum_{\substack{L, \{D_i,I_i\} \\ 1\le i\le L \ \text{with} \ \substack{\sum_i I_i=n_I \\ \sum_i D_i=D}}} & \sum_{\substack{\text{Ordered}\  \text{lists} \ M_L \\ \text{of} \ L \ \text{connected \ surfaces} \\ \text{where \ entry \  } i \ \text{has} \ D_i,I_i \\ \text{distinguishable/indistinguishable} \\ \text{boundaries}}} C(D) \frac{u^m}{L!} \ \frac{k^{\sum_i I_i}}{\prod_i I_i!} \  \frac{D!}{\prod_i D_i!} \ e^{S_0 \tilde \chi} \;,\nonumber
	\end{align}
where the factor $\frac{k^{I_i}}{I_i!}$ for each connected manifold counts the number of ways (including symmetry factors) to assign EOW brane labels to $I_i$ indistinguishable circular boundaries and the factor $\frac{D!}{\prod_i D_i!}$ again counts partitions of the $D$ distinguishable boundaries into (labelled) subsets of size $D_i$.  Finally, the factor $C(D)$ represents the number of ways to form $D$ distinguishable boundaries from the specified boundary conditions (together with interpolating EOW brane segments).

In comparing with \eqref{eq:EOWamp2}, the relation to the exponential of \eqref{eq:gf2} is clear from the factor of $1/L!$ in \eqref{eq:EOWamp2}, the inclusion of factors of $\frac{k^I_i}{I_i!}$ in \eqref{eq:gf2}, and the defining property of generating functions.  By this last feature, we mean the fact that the definition of the generating functions \eqref{eq:gf2} converts the factors $C(D)\frac{D!}{\prod_i D_i!} $ counting the number of ways to match distinguishable boundaries to boundary conditions into the above-described weighted sum over all possible boundary conditions for each connected component.

We now interpret the amplitudes as describing an ensemble, for which \eqref{eq:gf2} is the (unnormalised) generating function for moments of random variables $Z$ and $(\psi_j,\psi_i)$. Let us first set $t=0$ in order to consider the marginal distribution of $Z$. We then recover the old result \eqref{eq:lngenfunc2} without EOW branes, so $Z$ is again Poisson distributed, though with a new value of $\lambda$ given by \eqref{eq:lambda2}.

We can now characterise the distribution of $(\psi_j,\psi_i)$ by conditioning on $Z=d$ for each fixed $d\in\NN$. To find the corresponding conditional generating functions, we Taylor expand the exponential in \eqref{eq:gf2} and write each term as an average over the Poisson probabilities $p_d(\lambda)=e^{-\lambda} \frac{\lambda^d}{d!}$:
\begin{gather}
	\Bigg \langle \exp \bigg( u Z + \sum_{i,j=1}^k t_{ij} (\psi_j,\psi_i)\bigg) \Bigg \rangle = e^\lambda \sum_{d=0}^\infty e^{ud} p_d(\lambda) \Bigg \langle \exp \sum_{i,j=1}^k t_{ij} (\psi_j,\psi_i) \Bigg \rangle_{\!\!Z=d} \nonumber \\
	 \implies \left\langle \exp \Bigg(\sum_{i,j=1}^k t_{ij} (\psi_j,\psi_i) \Bigg)\right\rangle_{\!\!Z=d}   = \det(I-t)^{-d} \label{eq:Wishart}.
\end{gather}
The result is the generating function for a standard complex Wishart distribution \cite{Wish} with $d$ degrees of freedom.

To make this more transparent, and to simultaneously explain this distribution to the uninitiated reader, we can rewrite the generating function by introducing $kd$ `auxiliary' complex variables $\psi_i^a$, arranged in a $d\times k$ matrix. The index $i=1,\ldots,k$ labels the EOW brane states, and we will interpret $a=1,\ldots,d$ as labels for an orthonormal basis of the boundary Hilbert space $\hilb_\text{CFT}$ (which is $d$-dimensional based on our interpretation \eqref{eq:Zdim} of $Z$). The $\psi_i^a$ variables will be interpreted as the components of the EOW brane states $\psi_i$ in this orthonormal basis.

In terms of the $\psi_i^a$ variables, our Wishart generating function \eqref{eq:Wishart} can now be written as a Gaussian integral:
\begin{equation}
\label{eq:WishartGaussian}
	\det(1-t)^{-d} =  \int \prod_{i=1}^k\prod_{a=1}^d \left(\frac{1}{\pi}d\psi_i^a d\bar{\psi}_i^a \,  e^{-\bar{\psi}_i^a \psi_i^a}\right) \exp\left(\sum_{i,j=1}^k t_{ij}\sum_{a=1}^d   \bar{\psi}_j^{a}\psi_i^a \right)
\end{equation}
Comparing with the expectation value \eqref{eq:Wishart} we are computing, we can read off the distribution by identifying the matrix of inner products $(\psi_j,\psi_i)$ as
\begin{equation}
\label{eq:split}
	(\psi_j,\psi_i) = \sum_{a=1}^d \bar{\psi}_j^a\psi_i^a
\end{equation}
from the final factor in the integral. The remainder of the integral gives the measure for the $\psi_i^a$, as independent random variables, each chosen from a complex normal (Gaussian) distribution with unit variance:
\begin{equation}
	\psi_i^a \sim\text{ independent standard complex normal random variables}.
\end{equation}

In the 0+1 dual interpretation, this means that the wavefunction of each EOW brane states is selected independently and uniformly at random from the unit sphere of a $d$-dimensional Hilbert space $\mathcal{H}_\text{CFT}$, and then multiplied by a random normalization so that its squared norm is drawn from an appropriate $\chi^2$-distribution. In particular, the number of linearly independent states, given by the rank of the matrix of inner products, is bounded by $Z$: with probability one we have
\begin{equation}
\label{eq:rankpsiipsij}
	\operatorname{rank} (\psi_j,\psi_i) = \min \{k,Z\}.
\end{equation}

This is another surprising and remarkable result from such a simple model, since in the semiclassical limit (without the exponentially small effects of spacetime wormholes) the $k$ EOW brane states appear to be orthogonal, and we can choose $k$ to be as large as we like. As discussed below in section \ref{sec:3rdQuantisation}, even if we include Euclidean wormholes there is an expansion in $e^{-S_0}$ which for a finite number of amplitudes at any finite order gives no obvious sign that apparently distinct EOW brane states must in fact be linearly dependent. Nonetheless, in the complete solution after summing all nonperturbative effects, we find that the number of linearly independent states is truncated. As in \cite{Penington:2019kki}, as and discussed further in section \ref{sec:relation}, this is a version of the semiclassical Page curve \cite{Page:1993df}.

At first sight, this appears to require an enormous conspiracy in the nonperturbative contributions, which might lead one to suspect that it is an artefact of studying particularly simple models. We will show below that this is not the case, since it follows from a more primitive principal, namely reflection positivity of the path integral. For this, we must study the Hilbert space interpretation of the model with EOW branes.

\subsection{Baby universe Hilbert space with EOW branes}
\label{sec:hbuEOW}

We now incorporate the EOW branes into the baby universe Hilbert space. This enlarges the space relative to that of section \ref{sec:ZBU} because, along with circular closed universes, we also have $k^2$ new types of universe whose spatial slice is an interval bounded by EOW branes, say with labels $i$ and $j$ (where the orientation defines a preferred order).  On the other hand, the above-mentioned conspiracies will also imply the existence of new null states.

It is most straightforward to construct $\hbu$ from the $\alpha$-states. These are eigenstates of the $\widehat{Z}$ operator as before, but now are simultaneously eigenstates of the $k^2$ operators $\widehat{(\psi_j,\psi_i)}$ as well; note that Hermitian conjugation acts on these operators by swapping $i,j$. We label the corresponding eigenvalues by $Z_\alpha$ and $(\psi_j,\psi_i)_\alpha$, so we have
\begin{equation}
	\begin{aligned}
		\widehat{Z}\big|\alpha\big\rangle &= Z_\alpha\, \big|\alpha\big\rangle \\
		\widehat{(\psi_j,\psi_i)}\, \big|\alpha\big\rangle &= (\psi_j,\psi_i)_\alpha\, \big|\alpha\big\rangle.
	\end{aligned}
\end{equation}
The set of $\alpha$-states is determined by the allowed sets of eigenvalues, which is constrained by \eqref{eq:rankpsiipsij}.

As in section \ref{sec:ZBU}, the eigenvalues $Z_\alpha$ of $\widehat{Z}$ are given by the nonnegative integers $d$. Indeed, we can still define states $|Z=d\rangle$ by any of the means discussed in that section, for example by \eqref{eq:Z=d}. However, they are now not full $\alpha$-states, since they are eigenstates only of $\widehat{Z}$ and not of $\widehat{(\psi_j,\psi_i)}$. Instead they are the projections of the Hartle-Hawking state onto the corresponding eigenspace of $\widehat{Z}$. We can generate the rest of this eigenspace by acting with the operators $\widehat{(\psi_j,\psi_i)}$ on $|Z=d\rangle$.

In each such eigenspace, we can now diagonalise the operators $\widehat{(\psi_j,\psi_i)}$. Their simultaneous eigenvalues correspond to Hermitian $k\times k$ positive definite matrices of rank at most $d$ (though any rank other than $\min(d,k)$ has probability zero in any normalizable state). The baby universe Hilbert space therefore decomposes as a direct sum:
\begin{equation}
\label{eq:Z=dsum}
	\begin{gathered}
		\hbu = \bigoplus_{d=0}^\infty \mathcal{H}_{Z=d} \\
		\mathcal{H}_{Z=d} = L^2(M_k^d)\\
		M_k^d =\{\text{Hermitian p.d. }k\times k \text{ matrices}, \rank\leq d\}.
	\end{gathered}
\end{equation}
The summands $\mathcal{H}_{Z=d}$ are the usual $L^2$ spaces of square integrable functions on the relevant space of restricted rank matrices $M_k^d$ (defined with any convenient smooth measure). For $d\leq k$, $M_k^d$ forms a $(2kd-d^2)$-dimensional manifold; we can write $(\psi_j,\psi_i) = \sum_{a=1}^d \bar{\psi}_j^a\psi_i^a$ as in \eqref{eq:split} so that the $2kd$ counts the number of independent real parameters in $\psi_i^a$ while the $d^2$ subtracts for the invariance under unitary rotations of the $a$ directions. For $d\geq k$, the restriction on rank is vacuous.

With this description, the $\alpha$-states are delta function wavefunctions living in the subspaces $\mathcal{H}_{Z=d}$, supported on some particular matrix $(\psi_j,\psi_i)_\alpha\in M_k^d$. In particular, we write their inner product as
\begin{equation}
	\langle \alpha'|\alpha\rangle = \delta_{\alpha\alpha'},
\end{equation}
where $\delta_{\alpha\alpha'}$ is the product of a Kronecker delta $\delta_{Z_\alpha Z_{\alpha'}}$ for the eigenvalue of $\widehat{Z}$ with an appropriate Dirac delta function on $M^d_k$ associated with the choice of $L^2$ measure in \eqref{eq:Z=dsum}.

Finally, the wavefunction of the Hartle-Hawking state in this description is given by
\begin{equation}
	\big\langle \alpha\big|\HH\big\rangle = \sqrt{\frac{\lambda^{Z_\alpha}}{Z_\alpha!} f_{Z_\alpha}\big((\psi_j,\psi_i)_\alpha\big)} \enspace ,
\end{equation}
where $f_{Z_\alpha}$ is the probability density function of the complex Wishart distribution with $Z_\alpha$ degrees of freedom with respect to the measure on our $L^2$ space; this is the overlap $\big\langle \alpha\big|Z=Z_\alpha\big\rangle=f_{Z_\alpha}$. For $Z_\alpha\geq k$, this density is given explicitly in \eqref{eq:Wishartdensity}.

\subsection{Hilbert spaces with boundaries}
\label{sec:modelHwBoundaries}
Our discussion of Hilbert spaces is not yet complete. In particular, other Hilbert spaces of interest arise when we insert complete sets of states on Cauchy slices that intersect `asymptotically AdS' boundaries. Here there are two types of boundary, distinguished by their orientation; we call them `left' and `right' boundaries of space. In a 0+1 dual, the two types of boundaries would correspond to CPT conjugate theories.

In our model, the most general slice $\Sigma$ of the asymptotically AdS boundaries will consist of
$n_L$ left boundaries and $n_R$ right boundaries.
We thus denote the associated Hilbert space $\hilb_\Sigma$ from section \ref{sec:moreHilbert} as $\hilb_{n_L,n_R}$.  Reversing the orientation of all boundaries gives the dual (Hermitian conjugate) Hilbert space, so $\hilb_{n_L,n_R}^* = \hilb_{n_R,n_L}$. The simplest of these is $\hbu=\hilb_{0,0}$, which we have already discussed. We will be primarily interested in the one-sided Hilbert space $\hilb_{0,1}$ (related to $\hilb_{1,0}$ by duality) and the two-sided space $\hilb_{1,1}$.

We begin by considering the single boundary Hilbert space $\hilb_{0,1}$, which is spanned by states of the form $|\psi_i; Z^m \,(\psi_{j_1},\psi_{i_1})\cdots (\psi_{j_n},\psi_{i_n})\rangle$. Recall that the operator $\widehat{\psi_i}$  maps $\hbu$ to $\hilb_{0,1}$ (or more generally $\hilb_{n_L,n_R}\to \hilb_{n_L,n_R+1}$). All of the above states can be produced by acting with the operator $\widehat{\psi_i}$ on a state of closed baby universes. In particular, we can span $\hilb_{0,1}$ by acting with one of the $k$ operators $\widehat{\psi_i}$ (for $i=1,\ldots k$) on $\alpha$-states of $\hbu$. The inner product on such states is
\begin{equation}
	\big\langle \psi_j;\alpha'\big| \psi_i;\alpha\rangle = \big\langle \alpha'\big|\widehat{(\psi_j,\psi_i)}\big|\alpha\rangle = \delta_{\alpha\alpha'} (\psi_j,\psi_i)_\alpha \,,
\end{equation}
so in particular, the different $\alpha$-sectors are orthogonal, and $\hilb_{0,1}$ admits a direct sum decomposition
\begin{equation}
	\hilb_{0,1} = \bigoplus_\alpha \hilb_{0,1}^\alpha ,
\end{equation}
(where this is to be understood in the appropriate sense given that some of the parameters defining $\alpha$ are continuous). The inner product on each sector $\hilb_{0,1}^\alpha$ is simply given by the matrix of eigenvalues $(\psi_j,\psi_i)_\alpha$. On sectors with $Z_\alpha<k$, this is degenerate, and $\hilb_{0,1}^\alpha$ is $Z_\alpha$-dimensional:
\begin{equation}
	\dim\hilb_{0,1}^\alpha = \min\{k,Z_\alpha\}.
\end{equation}

Next, we look at the two-boundary sector $\hilb_{1,1}$. In the same way, this Hilbert space can be populated by acting with boundary creating operators on states of $\hbu$, for example on $\alpha$-states. We have the same direct sum structure as before, $\hilb_{1,1} = \bigoplus_\alpha \hilb_{1,1}^\alpha$. States within each $\hilb_{1,1}^\alpha$ can be created by acting with separate EOW brane states on left and right boundaries using $\widehat{\psi_j^*}\widehat{\psi_i}$. But we now have an additional possibility where we introduce a single asymptotic boundary that connects left and right.  In a general theory, one might call this the cylinder boundary (with topology $\Sigma$ times an interval), and one might think of it as obtained by cutting in half a partition function on $\Sigma\times S^1$.  By acting on $\big|\HH \big \rangle$, it thus creates a state that one expects to interpret as a `thermofield double' in some CFT dual. In our case the cylinder degenerates to a line segment (since $\Sigma$ is a point), which we can think of as half of a $Z$ circle. We denote the boundary condition by $\cyl$, the associated operator by $\widehat{\cyl}$, and the resulting state by $\Big| \cyl \Big \rangle = \widehat{\cyl}\Big|\HH \Big \rangle$. Thus,
\begin{equation}
	\hilb_{1,1}^\alpha \text{ is spanned by }\big|\psi_j^*,\psi_i;\alpha\big\rangle, \;  \big|\cyl ;\alpha\big\rangle,
\end{equation}
and the inner products of these states are given by
\begin{equation}
	\begin{aligned}
		\big\langle \psi_{j_2}^*, \psi_{i_2};\alpha'\big|\psi_{j_1}^*,\psi_{i_1};\alpha\big\rangle &= \delta_{\alpha\alpha'} (\psi_{i_2},\psi_{i_1})_\alpha \, (\psi_{j_1},\psi_{j_2})_\alpha ,\\
		\big\langle\cyl;\alpha'\big|\psi_{j_1}^*,\psi_{i_1};\alpha\big\rangle &= \delta_{\alpha\alpha'} (\psi_{j_1},\psi_{i_1})_\alpha ,\\
		\big\langle\cyl;\alpha'\big|\cyl;\alpha\big\rangle &= \delta_{\alpha\alpha'} Z_\alpha .
		\end{aligned}
\end{equation}
From the first of these, we see that for fixed $\alpha$ the states $|\psi_{j}^*,\psi_{i};\alpha\big\rangle$ span a subspace isomorphic to the tensor product of two single boundary subspaces, so this tensor product embeds naturally in $\hilb_{1,1}^\alpha$; i.e.,  $\hilb_{0,1}^\alpha\otimes\hilb_{1,0}^\alpha\subseteq\hilb_{1,1}^\alpha$. This inclusion could be an exact equality, but only if the new state $|\cyl;\alpha\big\rangle$ can be built from a linear combination of factorised states $|\psi_{j}^*,\psi_{i};\alpha\big\rangle$. This suggests that we look for a linear combination
\begin{equation}
	\big|\Delta\big\rangle =\big|\cyl;\alpha\big\rangle - \sum_{i,j=1}^k c_{ij}\big|\psi_{j}^*,\psi_{i};\alpha\big\rangle
\end{equation}
with zero norm. Such a vector would be projected out of the Hilbert space $\hilb_{1,1}$, giving  $|\Delta\rangle=0$ and providing an identity relating the cylinder state $\big| \cyl \big \rangle $ to a superposition of one-sided states.

Before computing the norm of our ansatz $|\Delta\rangle$, we first change to a more convenient basis diagonalising the EOW brane inner product in the $\alpha$-state in question (with eigenvalues $(\psi_j,\psi_i)_\alpha$). Specifically, we pick linear combinations $\phi_a$ of the $\psi_i$ boundary conditions for which $(\phi_b,\phi_a)_\alpha=\delta_{ab}$, with the index $a=1,\ldots,r$ running up to the rank of the matrix of inner products. In this basis, we rewrite our candidate null state and compute its norm:
\begin{gather}
	\big|\Delta\big\rangle =|\cyl;\alpha\big\rangle - \sum_{a,b=1}^r c_{ab}|\phi_{b}^*,\phi_{a};\alpha\big\rangle \\
	\begin{aligned}
		\big\langle \Delta\big|\Delta\big\rangle &= \big\langle \cyl;\alpha\big|\cyl;\alpha\big\rangle - \sum_{a,b=1}^r c_{ab}\big\langle \cyl;\alpha\big|\phi_b^*,\phi_a;\alpha\big\rangle \\
		&\qquad -\sum_{a,b=1}^r c_{ab}^*\big\langle\phi_b^*,\phi_a;\alpha \big|\cyl;\alpha\big\rangle + \sum_{a,b,a',b'=1}^r c_{ab}c_{a'b'}^*\big\langle\phi_{b'}^*,\phi_{a'};\alpha \big|\phi_b^*,\phi_a;\alpha\big\rangle \\
		&= Z_\alpha -2\sum_{a=1}^r \Re c_{aa} + \sum_{a,b=1}^r |c_{ab}|^2 \\
		&= Z_\alpha-r \qquad\qquad (c_{ab}=\delta_{ab}).
	\end{aligned}\nonumber
\end{gather}
In the last line we have chosen the coefficients $c_{ab}=\delta_{ab}$ to be $\delta_{ab}$, as this minimizes $\big\langle \Delta\big|\Delta\big\rangle$.

The above calculation teaches us two things. Firstly, for the norm to be nonnegative we have an inequality which applies in all $\alpha$ states:
\begin{equation}\label{eq:Page1}
	\text{Reflection positivity}\implies Z_\alpha \geq \rank  (\psi_j,\psi_i)_\alpha .
\end{equation}
This explains our empirical result \eqref{eq:rankpsiipsij} that the rank of the EOW brane inner product is bounded by $Z_\alpha$, in terms of reflection positivity of the path integral. The same argument can be used in much more general models, and we repeat it with the inclusion of a conserved energy in section \ref{sec:relation}, where we also connect it with the Page curve \cite{Page:1993df}.

Secondly, we find that if the inequality \eqref{eq:Page1} is saturated, we have $|\Delta\rangle=0$, and hence an identity
\begin{equation}
\label{eq:TFDrelation}
	|\cyl;\alpha\big\rangle = \sum_{a=1}^r |\phi_{a}^*,\phi_{a};\alpha\big\rangle .
\end{equation}
Since the `factorized states' $|\psi_{j}^*,\psi_{i};\alpha\big\rangle$ then span the two-sided Hilbert space $\hilb_{1,1}^\alpha$, we also find an equivalence between Hilbert spaces
\begin{equation}
	\hilb_{0,1}^\alpha\otimes\hilb_{1,0}^\alpha \equiv \hilb_{1,1}^\alpha\, .
\end{equation}
This factorization holds in our model for sectors with $Z_\alpha\leq k$; i.e., when there are enough EOW branes to populate a one-sided Hilbert space of dimension $Z_\alpha$.

To emphasise the importance of $\alpha$-states in this argument, we examine how it fails in a more general (normalised) state $|\Psi\rangle\in \hbu$, such as the Hartle-Hawking state. Specifically, let us choose linear combinations $\phi_a$ of EOW brane states $\psi_i$ to diagonalise the expectation value of the inner product in the state $|\Psi\rangle$; i.e., we take
\begin{equation}
	\big\langle \Psi \big| \widehat{(\phi_b,\phi_a)} \big|\Psi\big\rangle = \delta_{ab}\,,
\end{equation}
where $a,b=1,\cdots,r$, with $r=\rank \big\langle \Psi \big| \widehat{(\psi_j,\psi_i)} \big|\Psi\big\rangle$. If we now compute the norm of the state
\begin{equation}
\label{eq:Deltastate}
	\big|\Delta\big\rangle = \big|\cyl;\Psi\big\rangle - \sum_{a=1}^r \big|\phi_a^*,\phi_a;\Psi\big\rangle,
\end{equation}
we find an extra term, coming from the overlaps $\big\langle \phi_b^*,\phi_b;\Psi\big|\phi_a^*,\phi_a;\Psi\big\rangle$:
\begin{equation}
\label{eq:wVar}
	\big\langle \Delta\big|\Delta\big\rangle=\big\langle\Psi\big|\widehat{Z}\big| \Psi\big\rangle - r+\sum_{a,b=1}^r \operatorname{Var}_\Psi\left[(\phi_b,\phi_a)\right].
\end{equation}
Here we have defined the variance of boundary condition $X$ as the connected amplitude for $XX^\dag$,
\begin{equation}
	\operatorname{Var}_\Psi[X] = \big\langle\Psi\big|\widehat{X}\widehat{X}^\dag\big| \Psi\big\rangle-\big\langle\Psi\big|\widehat{X}\big|\Psi\big\rangle\big\langle\Psi\big|\widehat{X}^\dag\big| \Psi\big\rangle.
\end{equation}
This vanishes in $\alpha$-states, though is generically non-zero.

For example, in the Hartle-Hawking state, the expectation value of the overlaps of EOW brane states is already diagonal,
\begin{equation}
	\frac{\Big\langle\HH\big|\widehat{(\psi_j,\psi_i)}\big|\HH\big\rangle}{\Big\langle\HH\big|\HH\big\rangle} = \lambda \delta_{ij}
\end{equation}
so we can define $\phi_a =\lambda^{-1/2} \psi_a$, and we have $r=k$. The variance of the individual terms $(\phi_b,\phi_a)$ is small,
\begin{equation}
	\operatorname{Var}_\HH[(\phi_b,\phi_a)] = \lambda^{-1} (1+\delta_{ab}),
\end{equation}
but there are $k^2$ such terms, so they are collectively important when $k$ is of order $\lambda$ or larger. As a result, \eqref{eq:wVar} gives no meaningful bound relating the rank of the inner product $(\psi_j,\psi_i)$ to the partition function $Z$.  Note that this is not really an issue of fluctuations in the particular parameter $Z_\alpha$, as the same discussion applies to the states $|Z=d\rangle$, which fix the eigenvalue of $\widehat{Z}$ but not those of $\widehat{(\psi_j,\psi_i)}$.

Returning to the issue of reflection positivity, we should also discuss the Hilbert spaces $\hilb_{n_L,n_R}$ associated with arbitrary numbers of left and right boundaries.  But in our model all possible boundary conditions creating such states can be formed by combining $\cyl$ with the above $\psi_i$.  In superselection sectors with $Z_\alpha \le k$, the above result then implies
$\hilb_{n_L,n_R} = \hilb_{1,0}^{\otimes n_L} \otimes \hilb_{0,1}^{\otimes n_R}$ and the inner product on $\hilb_{n_L,n_R}$ is positive definite.    In superselection sectors with $Z_\alpha > k$ the higher Hilbert spaces are not tensor products of the lower Hilbert spaces. But much as above, considering states similar to \eqref{eq:Deltastate} again shows the inner product to be positive for $Z_\alpha > k=r$.  We thus see by direct calculation that our path integral satisfies reflection positivity.

\subsection{The boundary parameter $S_\partial$}\label{sec:fudge}

We now discuss the parameter $S_\partial$, contributing an action proportional to the number of boundaries. First we describe how changing $S_\partial$ from its preferred value $S_\partial=S_0$ alters the physics, and thus in particular explain why this value is preferred. We then discuss how we might naturally incorporate such a parameter in the model.

Let us first consider  the model without EOW branes, discussed in sections \ref{sec:Zonly}, \ref{sec:Zamp} and \ref{sec:ZBU}.  There the only effect of $S_\partial$ is to rescale the quantities and operators associated with the $Z$ boundaries. We thus find an ensemble interpretation in which $Z$ is $e^{S_\partial-S_0}$ times a Poisson random variable, so that the $\alpha$-states are characterised by $\widehat{Z}$ eigenvalues $Z_\alpha\in e^{S_\partial-S_0}\NN$. From the gravitational perspective, there is nothing wrong with this model for any positive value of $S_\partial$. In particular, reflection positivity is preserved for all Hilbert spaces. Complex values are excluded by reflection positivity on $\hilb_{1,1}$, which is spanned by orthogonal states $|\cyl;\alpha\rangle$ with norm $\langle\cyl;\alpha|\cyl;\alpha\rangle=Z_\alpha$. From the boundary perspective, there is a good dual interpretation only when $e^{S_\partial-S_0}$ is a nonnegative integer, so that $Z_\alpha$ takes nonnegative integer values which can be interpreted as the dimension of a dual Hilbert space. Nothing from the bulk perspective appears to prefer such values, so our choice $S_\partial=S_0$ appears to be rather artificial.

This changes once we introduce the EOW brane states. The bulk then provides a principled reason to prefer particular values of $S_\partial$, as the inner product on EOW brane states will otherwise fail to be positive semidefinite. To see this, we focus on a sector of $\hbu$ with fixed $\hat{Z}$ eigenvalue $Z=d$, in which our EOW brane amplitudes are given by the generating function \eqref{eq:Wishart}, reproduced below with fugacities $t_{ij}$ rescaled by a factor of $i$ for later convenience and with the matrix of EOW brane inner products encoded in a $k\times k$ Hermitian matrix $M$, $M_{ij}=(\psi_j,\psi_i)$:
\begin{equation}\label{eq:chidk}
	\chi_{d,k}(t) = \left\langle e^{i\Tr(tM)}\right\rangle_{Z=d} = \det(1-it)^{-d}
\end{equation}
For $d\in\NN$, by introducing $dk$ auxiliary Gaussian variables we showed in \eqref{eq:WishartGaussian} that this gives a probability distribution for $M$, and hence a reflection positive inner product on $\hbu$. This argument does not apply for $d\notin \NN$, so we must find a different way to determine whether we have a positive semidefinite inner product.

If $M$ is to be interpreted as a random variable selected from some probability distribution, \eqref{eq:chidk} defines $\chi_{d,k}$ as the characteristic function of the distribution. This is the Fourier transform of the probability density function $p_{d,k}$, which is in general a distribution on the space of $k\times k$ Hermitian matrices. It thus determines our inner product, which acts on a space of functions $f,g$ of $k\times k$ Hermitian matrices $M$:
\begin{gather}
	\langle g| f\rangle = \int dM p_{d,k}(M) g(M)^* f(M), \\
	\text{where } \chi_{d,k}(t) = \int dM e^{i\Tr(tM)} p_{d,k}(M).
\end{gather}
The distribution $p_{d,k}$ is determined uniquely from the inverse Fourier transform of $\chi_{d,k}$.\footnote{Our integration measure on Hermitian matrices is defined as the flat measure on independent real components, $dM=\prod_i dM_{ii} \prod_{i<j} d\Re M_{ij}d\Im M_{ij}$, and here we take $t$ to be a Hermitian matrix so that $\Tr(tM)$ is real.} For this to define a positive semidefinite inner product, we need $p_{d,k}$ to be a nonnegative distribution (that is, it gives positive values when integrated against positive test functions such as $|f(M)|^2$). The question of whether the $Z=d$ subspace of $\hbu$ has a positive semidefinite inner product is equivalent to the existence of a probability distribution with characteristic function $\chi_{d,k}$.

A succinct summary answering this question is contained in \cite{graczyk2003complex}, to which we refer the reader for the results we now use. For $d>k$, the inverse Fourier transform of $\chi_{d,k}$ is a continuous function of $M$, taking non-zero values only on positive-definite matrices:
\begin{equation}\label{eq:Wishartdensity}
\begin{gathered}	
	p_{d,k}(M) = \mathcal{N}_{d,k}\det(M)^{d-k}e^{-\Tr M}, \quad M \text{ positive definite}, \\
	\mathcal{N}_{d,k}^{-1}= \pi^{\frac{k(k-1)}{2}}\Gamma(d)\Gamma(d-1)\cdots \Gamma(d-(k-1)).
\end{gathered}
\end{equation}
This is manifestly nonnegative and so defines a probability distribution. This result extends to $d>k-1$, where the probability density diverges at the edge where $M$ becomes degenerate, but is still integrable. This is easiest to see from the density in terms of the eigenvalues of $M$; fixing $k-1$ positive eigenvalues and taking the last $\lambda\to 0$, the density goes as $\lambda^{d-k}$. The important result for us is that this range $d>k-1$, along with the smaller nonnegative integer values of $d$ already covered by \eqref{eq:WishartGaussian}, turns out to exhaust the values of $d$ for which the inner product on $\hbu$ is positive semidefinite:
\begin{equation}
\begin{aligned}
	\chi_{d,k}(t) = \det(1-it)^{-d}& \text{ defines a probability distribution} \\
	 &\iff d\in \{0,1,2,\ldots,k-2\}\cup [k-1,\infty).
\end{aligned}
\end{equation}
We can intuit this from \eqref{eq:Wishartdensity} by analytic continuation of the density in $d$. As $d$ approaches $k-1$, the density goes to zero for any fixed positive definite matrix from the zero in normalisation factor $\mathcal{N}_{d=k-1,k}=0$, but the probability density piles up near $\det M=0$ and we end up with a probability density supported on the submanifold of singular matrices with rank $k-1$. However, if we try to go further to $k-2<d<k-1$, the probability density becomes negative. Even for values of $d<k-1$ at which the probability density appears to be positive, the density is not integrable near $\det M=0$.  On the other hand, since $\chi_{d,k}$ is analytic (so its Fourier transform decays exponentially) and $\chi_{d,k}(t=0)=1$, the integral of the distribution $p_{d,k}$ over all $M$ is well-defined and equal to unity.  The resolution is that  $p_{d,k}$ becomes a singular distribution which must be defined by a principal value prescription, and which is not positive definite on the singular submanifold $\det M=0$.

As a result, the inner product on $\hbu$ can be positive definite only when all sectors with $d \notin \NN$ have $d\ge k-1$.  For a given $S_\partial$, this requirement is most stringent for the smallest non-zero eigenvalue of $Z_\alpha$, namely $d = e^{S_\partial-S_0}$.  We thus find that reflection positivity can hold only when either $S_\partial-S_0$ is the logarithm of a positive integer, or $S_\partial>S_0+\log(k-1)$.

We can use the arguments of the last section to slightly strengthen our restrictions on $S_\partial$ by considering positivity in Hilbert spaces with boundaries, and in particular in $\hilb_{1,1}$. The discussion leading to \eqref{eq:Page1} shows that  positivity in $\hilb_{1,1}$ requires $\rank M \leq d$ for the matrix of inner products $M$ in each sector $Z=d$. This is violated by the distribution \eqref{eq:Wishartdensity} in the range $k-1<d<k$, since $M$ has probability density supported on matrices with full rank, $\rank M=k$. This gives us our final result:
\begin{equation}
	\text{Reflection positivity} \implies e^{S_\partial-S_0}\in\NN \text{ or } S_\partial >S_0+\log k.
\end{equation}
For any non-zero number of EOW brane species, we find that a non-zero value of $S_\partial$ is required; the absence of a boundary action $S_\partial=0$ does not lead to a reflection positive theory. The most natural choice is the minimal value $S_\partial=S_0$, which is the definition of the theory we used throughout the rest of this section.

The failure of models with $S_\partial=0$ motivates us to explain the physics that might lead to an action counting the number of boundary components $|\partial M|$. This is nontrivial, because $|\partial M|$ is not a local action. For example, if we take a cylinder (with two boundaries), we can slice it in two along its length, and glue together the two edges of each piece so that we form two separate cylinders. The resulting manifold has four boundaries, so $|\partial M|$ is not preserved by this cut and paste.

 However, we can achieve the same effect with a local action by introducing a new degree of freedom on each boundary. This should propagate along both asymptotic and EOW brane boundaries. Note that we regard this as part of the bulk dynamics that happens to be localised at the boundary, and not part of the dual `CFT' dynamics. Most simply, this can be a topological quantum mechanics with Hilbert space $\hilb_\partial$. In that case, each boundary provides a factor of $\dim\hilb_\partial$, and we can regard $-S_\partial |\partial M| = -\log \dim\hilb_\partial$ as a nonlocal effective action from integrating out this dynamics. This gives a local definition of our theory, but only if $e^{S_0}$ is an integer. This is not entirely satisfactory: besides the somewhat artificial restriction on $S_0$, it seems that this degree of freedom should allow for additional boundary conditions that project onto a particular state of this boundary quantum mechanics, in which case we are again left with the theory $S_\partial=0$.

 A slightly different possibility is that some local bulk dynamics gives rise to a path integral localised at the boundary, but one which cannot be described by any quantum mechanics. This seems like a strange situation at first sight, but we note that precisely this phenomenon occurs for JT gravity. In that theory, a local bulk theory gives rise to a degree of freedom associated with asymptotic boundaries, described by the Schwarzian path integral \cite{Maldacena:2016upp,Engelsoy:2016xyb}. The Schwarzian alone is not a consistent quantum mechanics, since the path integral on the circle cannot be interpreted as $\Tr e^{-\beta H}$ for any Hamiltonian $H$ \cite{Stanford:2017thb,Harlow:2018tqv}. This possibility arises from a quotient by residual gauge symmetries acting nontrivially on the boundary (in that case, an $SL(2,\RR)$).  Nonetheless, the gravitational theory (for example, the Lorentzian theory on a spacetime lying between two boundaries, has a good Hilbert space interpretation. While we do not have a concrete proposal to make at this time, we speculate that some analogous dynamics (or an appropriate accounting of residual gauge freedom) could naturally give rise to a theory of topology which includes a boundary effective action $S_\partial$. In particular, we hope that our model might be obtained as a limit of a theory with more dynamics, and that this construction might offer insight into this possibility.

\subsection{Spacetime `D-branes'}\label{sec:Dbranes}

We conclude the discussion of the model with some interpretative remarks for some of the results in terms of `spacetime D-branes,' which we call SD-branes below. An SD-brane means an object on which spacetime can end, and as such is seen from spacetime as D-branes are seen from the worldsheet in string theory. In particular, they are not localised in spacetime in any way. This will be similar in spirit to the discussion of D-branes and `eigenbranes' in \cite{Saad:2019lba,Blommaert:2019wfy}, though the framework of the Hilbert space of baby universes provides a new interpretation. We will focus on the model without EOW branes.

To study the theory in the presence of an SD-brane, we should introduce a new type of boundary of spacetime, interpreted as spacetime ending on the SD-brane. We will assign a free (possibly complex) parameter $g$ to these boundaries, interpreted as a coupling to the SD-brane. To compute an amplitude in the presence of an SD-brane, we should allow for any number (including zero) of these additional boundaries; i.e., the spacetime is allowed to end many times on the same SD-brane. But for the purposes of computing amplitudes, each SD-brane boundary acts much the same as a $Z$ boundary, so we can account for them by inserting factors of $g Z$. To avoid overcounting different spacetimes connecting to the SD-brane, we must divide by factorials of the number of boundaries, treating the new boundaries as indistinguishable and introducing further symmetry factors where appropriate.  We thus have the following recipe for computing the amplitude in the presence of an SD-brane with coupling $g$:
\begin{align}
	\Big\langle f(Z) \boxed{\text{SD-brane}_g} \Big\rangle &= \Big\langle f(Z) \Big\rangle + \Big\langle f(Z) gZ \Big\rangle + \Big\langle f(Z) \tfrac{1}{2}(g Z)^2  \Big\rangle + \Big\langle f(Z) \tfrac{1}{3!}(g Z)^3  \Big\rangle+ \cdots \nonumber \\
	&= \Big\langle f(Z) e^{gZ}\Big\rangle.
\end{align}
As before, the notation on the left-hand side indicates the boundary conditions for the path integral. But from the right-hand side we learn that the insertion of an SD-brane is equivalent to inserting the operator $e^{g\hat{Z}}$. In other words, the SD-brane is not a new object at all! Instead, a state $\Big|\boxed{\text{SD-brane}_g}\Big\rangle$ containing an SD-brane was already present in $\hbu$ as a coherent state  $\big|e^{gZ}\big\rangle$ of baby universes.  We may thus identify the corresponding boundary conditions:
\begin{equation}
\label{eq:SDexpZ}
	 \boxed{\text{SD-brane}_g} = e^{gZ} .
\end{equation}
This exponential of $Z$ is somewhat analogous to the determinant $\det(E-H)$ introduced in \cite{Saad:2019lba}, where it was interpreted as a brane in JT gravity. The determinant is analogous because it can be written as the exponential $\exp\left(\Tr\log(E-H)\right)$ of the single boundary object $\Tr\log(E-H)$ (single-trace in the dual matrix integral).

Now, what do the amplitudes actually look like in the presence of an SD-brane? To answer this, we compute the generating function \eqref{eq:genfunc} in an SD-brane state:
\begin{equation}\label{eq:Dbrane1}
	\begin{aligned}
	\Big\langle\boxed{\text{SD-brane}_g}\Big| e^{uZ}\Big|\boxed{\text{SD-brane}_g}\Big\rangle &= \Big\langle e^{g^* Z} e^{uZ} e^{gZ}\Big\rangle\\
	&=  \Big\langle e^{(u+2\Re g)Z} \Big\rangle \\
	&= \exp\left(\lambda e^{u+2\Re g}\right) \\
	&= \exp\left(\tilde{\lambda} e^{u}\right),\qquad \tilde{\lambda}=e^{2\Re g}\lambda\,.
\end{aligned}
\end{equation}
We here used the result $\Big\langle e^{uZ}\Big\rangle=\exp(\lambda e^u)$ of \eqref{eq:lngenfunc2} in the Hartle-Hawking state, with a shifted value of $u$ due to the presence of the SD-brane. The result \eqref{eq:Dbrane1} tells us is that amplitudes in the presence of an SD-brane are the same as amplitudes in the Hartle-Hawking state, but with a different value of the coupling $\lambda$. In fact, we can move between any positive real values of $\lambda$ by adding an appropriate SD-brane. This is a familiar situation from worldsheet string theory, where different values of an apparently free parameter (e.g.\ the coupling of the string to the Euler characteristic) turn out to describe different states of the same theory (e.g. \ coherent states of the dilaton).

We can also make use of these SD-branes in yet one more way by considering the effect of the imaginary part of the coupling $\theta= \Im g$. This has no effect in the amplitude \eqref{eq:Dbrane1}, and to see its relevance we must allow for a different kind of SD-brane state in which $g$ is not fixed but instead has a superposition of different values for $\theta$. First, we note that the representation of the SD-brane as $e^{gZ}$ and the integer spectrum for  $Z$ imply that $\theta$ should be understood to be periodic with period $2\pi$. A natural basis of states superposing different values of $\theta$ is thus defined by the Fourier transformed states,
\begin{equation}
	\Bigg| \boxed{\widetilde{\text{SD-brane}}_d} \Bigg\rangle := \int_{-\pi}^{\pi}\frac{d\theta}{2
	\pi} e^{-id\theta} \Big| \boxed{\text{SD-brane}_{i\theta}} \Big\rangle,\quad d\in \NN,
\end{equation}
where for simplicity we will now focus on the case $g=i\theta$, or $\Re g=0$.
In particular, the above basis diagonalizes the inner product:
\begin{equation}\label{eq:Dbrane2}
	\Bigg\langle\boxed{\widetilde{\text{SD-brane}}_{d'}}\Bigg|\boxed{\widetilde{\text{SD-brane}}_d}\Bigg\rangle = \delta_{dd'} \frac{\tilde{\lambda}^d}{d!} \qquad (d,d'\in\NN ).
\end{equation}
For $d<0$, this inner product vanishes, indicating that the resulting state is null.

To understand these states better, we may use the representation \eqref{eq:SDexpZ} of the SD-brane states as an exponential to write them as
\begin{equation}
	\Bigg| \boxed{\widetilde{\text{SD-brane}}_d} \Bigg\rangle := \int_{-\pi}^{\pi}\frac{d\theta}{2
	\pi} e^{-id\theta} \Big| e^{i\theta Z}\Big\rangle = (-1)^d\Bigg| \frac{\sin(\pi Z)}{\pi(Z-d)}\Bigg\rangle.
\end{equation}
But this is precisely the expression we gave in \eqref{eq:Z=d} for the $\alpha$-state $\big|Z=d\big\rangle$! Furthermore, it is now clear that taking $\Re g \neq 0$ simply rescales the resulting state $\big|Z=d\big\rangle$.

This means that we can give a somewhat geometric description of a given $\alpha$-sector by including a particular (Fourier transformed) $\widetilde{\text{SD}}$-brane. This $\widetilde{\text{SD}}$-brane is not a new fundamental object, but is built from a coherent state of interacting baby universes. The $\widetilde{\text{SD}}$-brane description of $\alpha$-states is at first sight rather different from the alternative geometric interpretation given in section \ref{sec:ZBU} where the $Z=d$ sector arose after constraining the path integral to spacetimes with $d$ connected components.  However, we see that the two are equivalent in the end.  We expect a similar equivalence to arise in the model with EOW branes, and correspondingly in the JT gravity contexts of \cite{Saad:2019lba,Blommaert:2019wfy}.

\section{Entropy bounds and the Page curve}
\label{sec:relation}

A remarkable property of our models above was the strong role played by null states, and in particular the bound \eqref{eq:rankpsiipsij} on the rank of the inner product in any $\alpha$-sector with $Z_\alpha =d$.  In section \ref{sec:modelHwBoundaries} we showed this bound to follow from an abstract argument involving the cylinder state $\Big| \cyl \Big \rangle$ in the Hilbert space $\hilb_{1,1}$ associated with a pair of disconnected boundaries.  As the reader may already realize, it is straightforward to generalize this argument so as to apply to very general reflection positive gravitational path integrals. More realistic models will likely have an infinite number of states in any $\hilb_\Sigma$, so to obtain a meaningful bound on the number of states we must impose a constraint. We will achieve this here by bounding the entropy of mixed states in $\hilb_\Sigma$ with a given expected energy $E$.

\subsection{Entropy bounds}

We now state this form of the argument using the more general notation from section \ref{sec:wormholesandBU}.  The ideas are closely related to those in \cite{Jafferis:2017tiu}.
As before, we work in some definite (but arbitrary) $\alpha$-sector of the given theory and also choose a spatial boundary manifold $\Sigma$; i.e., we consider a particular Hilbert space $\hilb_\Sigma^\alpha$ from section \ref{sec:moreHilbert}.

One property we require of our theory is that there is a notion of time evolution, here in Euclidean time. This means that the allowed boundary conditions include  Euclidean `cylindrical' boundary manifolds $C_\beta  = \Sigma \times I_\beta$ for intervals $I_\beta$ of arbitrary length $\beta>0$.  According to the general principles of section \ref{sec:wormholesandBU}, this boundary condition describes an operator on $\hilb_\Sigma^\alpha$ that we may call $e^{-\beta H}$ and for which $e^{-\beta_1 H}e^{-\beta_1 H}=e^{-(\beta_1 +\beta_2) H}$.   For a given state $\Big | \psi[J] \Big \rangle$ defined by sources $J$ on a boundary manifold $\dmanifold$ (with $\partial \dmanifold = \Sigma$), the action of $e^{-\beta H}$ on $\Big | \psi[J] \Big \rangle$ simply defines a new source $J_\beta$ on a larger boundary manifold $\dmanifold_\beta=I_\beta \dmanifold$ constructed by gluing $I_\beta$ to $\dmanifold$,
\begin{equation}
e^{-\beta H} \Big | \psi[J] \Big \rangle = \Big | \psi[J_\beta] \Big \rangle .
\end{equation}

The final property we require of our theory is that
the CPT conjugation acting on boundary conditions acts trivially on $I_\beta$. When
$e^{-\beta H}$ is trace-class, this condition
ensures that states $\phi_a \in \hilb_{\Sigma}^\alpha$ define a Hermitian matrix $(\phi_b,e^{-\beta H}\phi_a)_\alpha$ which can be diagonalized to yield discrete eigenvalues with finite degeneracy.
We will take this to be the case for now and return later to the possibility that $e^{-\beta H}$ might fail to be trace-class.

The above semi-group property of $e^{-\beta H}$ then implies that the eigenvectors can be chosen to be independent of $\beta$. Together with Hermiticity, it also implies the relation $e^{-\beta H} = \left( e^{-\beta H/2}\right)^\dagger e^{-\beta H/2}$ so that the eigenvalues must be non-negative.  Henceforth, we thus take $\phi_a$ to denote such an orthonormal eigenbasis of $\hilb_\Sigma^\alpha$ with eigenvalues $e^{-\beta E_a}$.

The key fact is then that the boundary conditions $e^{-\beta H}$ must also define an operator on the baby universe Hilbert space $\hbu$, which we can use to define cylinder states by acting on the $\alpha$-states $\Big | \alpha \Big \rangle \in \hbu$ in direct analogy with section \ref{sec:modelHwBoundaries}:
\begin{equation}
\widehat{e^{-\beta H}} \Big |\alpha \Big \rangle = \Big | \cyl_\beta ; \alpha \Big \rangle \in \hilb_{\Sigma^* \sqcup\Sigma}^\alpha\ \ .
\end{equation}

We will be interested in forming mixed states on $\hilb_{\Sigma}^\alpha$, which can be thought of as elements of the Hilbert space $\hilb_{\Sigma^*}^\alpha\otimes\hilb_{\Sigma}^\alpha$, spanned by products $\phi_b^*\otimes \phi_a$ of our  eigenstates $\phi_a\in \hilb_{\Sigma}^\alpha$ and their CPT conjugates. This space of density matrices is isometrically embedded via states $|\phi_b^*, \phi_a; \alpha \rangle$ into the `two-sided Hilbert space' $ \hilb_{\Sigma^* \sqcup\Sigma}^\alpha$ associated with two copies of our spatial boundary $\Sigma$. Since these latter states were built from orthonormal eigenstates of $e^{-\beta H}$ on  $\hilb_\Sigma^\alpha$, the overlaps are given by
\begin{align}
	\big\langle\phi_{b}^*,\phi_{a};\alpha\big|\cyl_{\beta/2};\alpha\big\rangle &=  \delta_{ab}e^{-\beta E_a/2} \,,\\
	\big\langle\phi_{b'}^*,\phi_{a'};\alpha\big|\phi_{b}^*,\phi_{a};\alpha\big\rangle &= \delta_{ab'}\delta_{a'b} \,.
\end{align}
The last overlap we require is the norm of the state $\big|\cyl_{\beta/2};\alpha\big\rangle$. This involves gluing two cylinders of length $\beta/2$ to create boundary conditions with a circle of length $\beta$: we have $\widehat{\cyl_{\beta/2}}^\dag \widehat{\cyl_{\beta/2}} = \widehat{Z(\beta)}$, where the operator $\widehat{Z(\beta)}$ acting on $\hbu$ is defined by boundary conditions $\Sigma\times S^1_\beta$, with a thermal circle $S^1_\beta$ of length $\beta$. The norm of our cylinder state is then given by
\begin{equation}
	\big\langle\cyl_{\beta/2};\alpha \big|\cyl_{\beta/2};\alpha\big\rangle = Z_\alpha(\beta),
\end{equation}
where $Z_\alpha(\beta)$ is the eigenvalue of $\widehat{Z(\beta)}$ in the $\alpha$ state, $\widehat{Z(\beta)}\big|\alpha\big\rangle= Z_\alpha(\beta)\big|\alpha\big\rangle$.

We now introduce a state
\begin{equation}
\label{eq:DeltaDef}
	\big|\Delta\big\rangle =\big|\cyl_{\beta/2};\alpha\big\rangle - \sum_a e^{-\beta E_a/2}\big|\phi_{a}^*,\phi_{a};\alpha\big\rangle	,
\end{equation}
and impose that its norm is nonnegative,
\begin{equation}
\label{eq:genDelta}
	\langle\Delta|\Delta\rangle = Z_\alpha(\beta) - \sum_a e^{-\beta E_a} \geq 0.
\end{equation}
As in section \ref{sec:modelHwBoundaries}, it is important that this computation was performed in a fixed $\alpha$-sector.  While we arrived at \eqref{eq:DeltaDef} under the assumption that $e^{-\beta H}$ is trace class, a similar argument using approximate eigenvectors would in any case bound the trace of   $e^{-\beta H}$ by $Z_\alpha(\beta)$.  Thus the case where $e^{-\beta H}$ fails to be trace class cannot occur and we can use \eqref{eq:DeltaDef} and \eqref{eq:genDelta} as written.

We can use the inequality \eqref{eq:genDelta} to make some more direct statements about the spectrum of states in $\hilb^\alpha_\Sigma$. Firstly, we can use it to bound the number of orthogonal states $N(E)$ with bounded energy $E_a\leq E$. In a thermodynamic limit, we would usually expect this to be dominated by states with energy close to the maximum, so $N(E)$ is controlled by the density of states at energy $E$. To bound this quantity, note that $\sum_a e^{-\beta E_a} \geq N(E) e^{-\beta E}$, by dropping all states with $E_a>E$ in the sum. From the result \eqref{eq:genDelta} we can then say that $ N(E) \leq e^{\beta E} Z_\alpha(\beta)$ for any $\beta$. The sharpest bound is obtained by minimising over all $\beta$, finding
\begin{equation} \label{eq:Nbound}
	\log N(E) \leq  S_\alpha(E),
\end{equation}
where
\begin{equation}
	S_\alpha(E) := \inf_\beta\{\beta E +\log Z_\alpha(\beta)\}. \label{eq:Salphadef}
\end{equation}
This quantity is nothing but the Legendre transform of $\log Z_\alpha(\beta)$, which is the usual way of obtaining the canonical entropy from a partition function. In a semiclassical theory, and in the overwhelming majority of $\alpha$-states, we expect $S_\alpha(E)$ to be approximately the Bekenstein-Hawking entropy of an appropriate black hole. This is because $Z_\alpha(\beta)$ is defined by the Gibbons-Hawking path integral with periodic Euclidean boundary conditions \cite{Gibbons:1976ue}, computed semiclassically by the on-shell action of a classical Euclidean black hole. The associated entropy $S_\alpha(E)$, defined as the Legendre transform of $\log Z_\alpha(\beta)$, is then given by the Bekenstein-Hawking formula. This remains accurate in typical $\alpha$ states (in the measure of the Hartle-Hawking ensemble) as long as the variance of the $\widehat{Z(\beta)}$ operator is small. This is the case if connected wormhole configurations between two asymptotic $Z(\beta)$ boundaries are suppressed.

The same quantity $S_\alpha(E)$ appears in a stronger bound, constraining  the von Neumann entropy $S(\rho)$ of any mixed state $\rho$ on $\hilb^\alpha_\Sigma$. This constraint depends on the energy expectation value $E = \Tr(\rho H)$, where from our earlier considerations we can define $H$ on $\hilb^\alpha_\Sigma$ by matrix elements $(\phi_b,H\phi_a)_\alpha=E_a\delta_{ab}$. Specifically, we prove that
\begin{gather}
	S(\rho) \leq S_\alpha(E) \quad \text{for $\rho$ any density matrix on }\hilb^\alpha_\Sigma \text{ with } \Tr(\rho H)=E. \label{eq:Sbound}
\end{gather}
It suffices to show this for the density matrix that maximises $S(\rho)$ subject to the energy constraint. This is simply a Gibbs state,
\begin{equation}
	\rho_\text{Gibbs}(\beta) = \frac{e^{-\beta H}}{Z_\text{Gibbs}(\beta)},\quad Z_\text{Gibbs}(\beta)= \Tr(e^{-\beta H}) = \sum_a e^{-\beta E_a},
\end{equation}
where we choose $\beta$ to fix the desired energy,
\begin{equation}
	E = -\frac{\partial}{\partial \beta} \log Z_\text{Gibbs}(\beta).
\end{equation}
Note that $Z_\text{Gibbs}$ is precisely the quantity we bounded in \eqref{eq:genDelta}, with the inequality $Z_\text{Gibbs}(\beta) \leq Z_\alpha(\beta)$. Now, we can compute the von Neumann entropy of $\rho_\text{Gibbs}$ as the Legendre transform of $Z_\text{Gibbs}$:
\begin{align}
	S(\rho_\text{Gibbs}(E)) &= \inf_\beta\{\beta E + \log Z_\text{Gibbs}(\beta)\} \\
	&\leq S_\alpha(E)
\end{align}
The inequality follows because $S_\alpha(E)$ is defined in \eqref{eq:Salphadef} by the same minimisation as used here to obtain $S(\rho_\text{Gibbs}(E))$, after replacing $Z_\text{Gibbs}(\beta)$ by the larger function $Z_\alpha(\beta)$. This demonstrates the claimed entropy bound \eqref{eq:Sbound}.

\subsection{Consequences and interpretations}

Our results \eqref{eq:Nbound} and \eqref{eq:Sbound} show that, for theories defined by reflection positive path integrals, the density of states in any $\hilb_\Sigma^\alpha$ is bounded by $S_\alpha(E)$ from \eqref{eq:Salphadef}, which generically we expect to be given by the Bekenstein-Hawking entropy of an appropriate black hole.

We interpret this result as a semiclassical Page curve. The class of mixed states $\rho$ on $\hilb_\Sigma^\alpha$ that we can prepare by asymptotic sources includes old black holes. For example, we can create pure state black holes by collapse, couple to an auxiliary `bath' system into which the Hawking radiation escapes, and trace out the bath. In the usual semiclassical description, it seems that this process can produce states of a given energy with arbitrarily large entropy. This entropy comes from the large interior which grows with time (in particular linearly with time along a `nice slice' \cite{Polchinski:1995ta}), which can be populated with a growing number number of naively distinct possible low energy states. Our result shows that in an alpha sector of a reflection positive path integral, nonperturbative effects giving exponentially small overlaps between these states must conspire to produce surprising linear relations between them. Such relations must occur after the Page time so that the entropy of the black hole is bounded by the Bekenstein-Hawking entropy, to satisfy \eqref{eq:Sbound}. If this inequality is (approximately) saturated, the entropy of the black hole (i.e.\ the density matrix on $\hilb_\Sigma^\alpha$) and of the radiation will follow the Page curve.

We expect that in contexts where the naive number of states in $\hilb_\Sigma$ can be made arbitrarily large, one will find that the bound $S(\rho)\leq S_\alpha(E)$ of \eqref{eq:Sbound} can be saturated, as in our model with large $k$. In particular, we expect this to hold for the old black holes in the discussion above. This requires saturation of the inequality in $\eqref{eq:genDelta}$ for all $\beta$, and so $|\Delta\rangle$ becomes a null state.  Note that $|\Delta\rangle=0$ is equivalent to the statement that $Z_\alpha(\beta)$ is equal to the actual thermal partition function $\Tr e^{-\beta H}$ on $\hilb_{\Sigma,\alpha}$. The result that the function $Z_\alpha(\beta)$ can be written as a thermal trace is a strong constraint on the eigenvalues of $\widehat{Z(\beta)}$, which should  be viewed as generalizing the result $Z_\alpha\in \NN$ from our models in section \ref{sec:models}.

In the case of saturation, the statement that $|\Delta\rangle$ is null leads to a gauge equivalence
\begin{equation}\label{eq:TFD}
	\big| \cyl_{\beta/2};\alpha\big\rangle = \sum_a e^{-\beta E_a/2} \big|\phi_a^*,\phi_a;\alpha\big\rangle.
\end{equation}
Following \cite{Maldacena:2001kr}, the cylinder state is naturally associated with a two-sided black hole with an Einstein-Rosen bridge joining the two boundaries. We see the familiar equivalence between this and a superposition of product states emerging as an example of our gauge equivalence.

To connect further with our desire to understand  black hole evaporation, we recall from section \ref{sec:wormholesandBU} that for any state $\rho$ prepared with asymptotic sources, the R\'enyi (and von Neumann) entropies $S_n(\rho)$ of $\rho$ again define operators on $\hbu$ and take definite values in $\alpha$-sectors.  These entropies are then subject to versions of the above bound in each $\alpha$-sector, and as a result so are their expectation values $\Big \langle S_n(\rho) \Big \rangle$ in the Hartle-Hawking state. In the context of black holes,  any such entropies will then reproduce an appropriate Page curve defined by the Bekenstein-Hawking entropy.  In particular, the final result will then be much as in the recent discussions of replica wormholes \cite{Almheiri:2019qdq,Penington:2019kki} which in our language are indeed the most natural saddle points contributing to the average entropy $\Big \langle S_n(\rho) \Big \rangle$.\footnote{More properly, the replica wormholes are saddle points for $\Big \langle\Tr(\rho^n)\Big \rangle$, but the distinction is unimportant as long as the variance of these quantities is small.}  The argument above shows that similar results will then hold when one computes the full result of any reflection positive gravitational path integral. Further, it tells us that these bounds hold not just on average, but in every $\alpha$-state. This puts additional constraints on higher moments of the entropy.

It is, however, important to note the precise sense in which the entropies $\Big \langle S_n(\rho) \Big \rangle$ have just been defined.  From our perspective, the basic quantities are the eigenvalues $S_{n,\alpha}(\rho)$ of $\widehat{S_n(\rho)}$ in the various $\alpha$-states. These are entropies defined separately on each $\hilb_{\Sigma,\alpha}$.  Working in the Hartle-Hawking state then computes the average $\Big \langle S_n(\rho) \Big \rangle$  of such entropies over the $\alpha$-states in the Hartle-Hawking ensemble.  In particular, while this  $\Big \langle S_n(\rho) \Big \rangle$ is computed by replica wormholes (to a first approximation), it manifestly
does \emph{not} include entanglement with the baby universe sector.

This is a physically useful notion of entropy as the $\alpha$-sectors are superselected from the standpoint of asymptotic observers, and entanglement with superselection sectors is in principle unobservable.  Nevertheless, if one wishes to consider the entropy of some density matrix on the full space $\hilb_\Sigma$ (and not just on a single $\alpha$-sector) defined by some fixed set of sources, entanglement with baby universes will generally lead to much larger entropies that exceed the Bekenstein-Hawking entropy and thus do not reproduce the expected Page curve.  In this more mathematical sense, Hawking was correct \cite{Hawking:1976ra} that information is lost in black hole evaporation.  This is all in direct parallel with the conclusions of \cite{Coleman:1988cy,Giddings:1988cx,Giddings:1988wv,Polchinski:1994zs} from long ago.  We will also discuss such connections in more detail in a forthcoming companion paper.

\section{On third-quantized perturbation theory}
\label{sec:3q}
\label{sec:3rdQuantisation}

\subsection{Formulating a wormhole perturbation theory}

We have been interested above in contexts where spacetime wormholes provide the dominant effects.  But in most circumstances spacetime wormholes are not the minimum action configurations.  In such cases, it is natural to expect other configurations to dominate, and for the contributions of spacetime wormholes to be  nonperturbatively suppressed by a factor of the form $e^{-S}$, where $S$, of order $G_N^{-1}$, is the action of a wormhole. This holds for computing simple amplitudes in our models of section \ref{sec:models}, for which higher topologies are suppressed by factors of the large parameter $\lambda$. In such cases it is natural to use an approximation where different universes evolve independently at leading order, and where spacetime wormholes are included as perturbative interactions between universes. The resulting perturbation theory is the `third quantised' formalism of \cite{Giddings:1988wv}. This approximation was also emphasized in other contemporaneous literature on wormholes \cite{Coleman:1988cy,Giddings:1988cx,Fischler:1989ka,Polchinski:1994zs}.

We now describe an analogous approximation in our framework. This will serve both to complete the connection to the above literature and to provide a better understanding of the interesting circumstances described above in which this approximation fails.  Nevertheless, this section represents a distraction from the main line of inquiry presented here, and some readers may wish to skip directly to section \ref{disc}.

The early works \cite{Coleman:1988cy,Giddings:1988cx,Giddings:1988wv} focused on studying microscopic wormholes, with the intent of describing physics on distances scales much larger than the wormhole's characteristic size (say, Planck scale). The relevant scale is the `width' of the wormhole mouth, thought of as some length scale associated with the cross-sectional area.  In contrast, the separation between the spacetime regions associated with the wormhole mouths can be much larger. In that context, it is most natural to describe the physics using the operators of the low energy effective field theory, studying the effect of integrating out the microscopic wormholes.   In contrast, we have wormhole mouths which, as with replica wormholes, are determined by a classical or quantum extremal surface.  As a result, our wormholes will typically have a size similar to some black hole horizon, which may be both macroscopic and large.  For us it thus will be more natural to discuss CFT boundary operators $\widehat{Z[J]}$ in place of the low energy bulk fields.  This captures much of the same physics, and is analogous to using an S-matrix description in place of an effective Lagrangian.\footnote{In the language of \cite{Preskill:1988na}, the effects of higher topology we study are more closely analogous to `wormhole interactions', as opposed to the `instanton interactions' arising from nearby wormhole mouths of primary interest in that work.}  The effects on the bulk effective field theory that arise from integrating out macroscopic wormholes will be explored in section \ref{disc}.

Suppose then that, for some theory and amplitude of interest, the contribution from topologies connecting many boundaries is suppressed relative to disconnected topologies. This holds for familiar simple amplitudes in theories of interest, including the model discussed in section \ref{sec:models}, as well as for JT gravity --- though it does not hold for all amplitudes, as we will discuss below. In a case where it does, at zeroth order of approximation we may neglect the connected contributions, obtaining an amplitude that approximately factorizes:
\begin{equation}
	 \mathfrak{Z}^{-1}\Big\langle Z[J_1]\cdots Z[J_n]\Big\rangle \approx \mathfrak{Z}^{-n}\Big\langle Z[J_1]\Big\rangle\cdots \Big\langle Z[J_n]\Big\rangle
\end{equation}
Identifying an asymptotically AdS boundary $Z[J]$ with an operator $\widehat{Z[J]}$ acting on the baby universe Hilbert space $\hbu$ as in
\eqref{eq:opdef}, at this leading order of approximation we can simply replace $\widehat{Z[J]}$ with a multiple of the identity operator $\mathfrak{Z}^{-1}\langle Z[J]\rangle$.  In particular, at this level of approximation, acting with any $\widehat{Z[J]}$ on $\big| \HH \big\rangle$ yields another state proportional to $\big| \HH \big\rangle$, so the baby universe Hilbert space defined in section \ref{sec:BUH} collapses to a single dimension.

To incorporate nontrivial wormhole physics, we must go to next order in the approximation, allowing contributions to the path integral from spacetimes that connect either one or two asymptotic boundaries, but not more.   The contributions from spacetimes with one asymptotic boundary are then analogous to quantum field theory tadpoles, while the two boundary contributions are analogous to quantum field theory propagators.  In particular, the Hilbert space $\hbu$ becomes nontrivial, and takes the form of a Fock space.  To see this, we define `single universe states' by subtracting the `tadpole contributions' from one boundary states; i.e., one need only introduce the modified (tilded) states
\begin{equation}
	\widetilde{|Z[J]\rangle}= |Z[J]\rangle - \mathfrak{Z}^{-1} \langle Z[J]\rangle|\HH\rangle,
\end{equation}
and similarly for states involving larger numbers of universes. Loosely speaking, the spacetime created by the operator $\widehat{Z[J]}$ is most likely to immediately cap off, failing to create a closed universe. It is natural to subtract this possibility, in which case we are most likely to create a single closed universe which can propagate to another asymptotic boundary, justifying the name of `single universe state'. Going to higher orders in the approximation would require additional subtractions for this description to remain valid.

The resulting Fock space structure can be used to define baby universe creation and annihilation operators $a^\dagger_J, a_{J^*}$, where in particular we have
\begin{align}
a_J|\HH\rangle&=0; \\
	a^\dag_J|\HH\rangle&= |Z[J]\rangle - \mathfrak{Z}^{-1} \langle Z[J]\rangle|\HH\rangle,
\end{align}
and the algebra $\left[a_{J_1},a_{J_2}\right]=0$,
\begin{equation}\label{eq:BUaadag}
	\left[a_{J_1},a^\dag_{J_2}\right] = \Big\langle Z[J_1^*]Z[J_2]\Big\rangle - \mathfrak{Z}^{-1}\Big\langle Z[J_1^*]\Big\rangle \Big\langle Z[J_2]\Big\rangle.
\end{equation}
One can then write corrections to the boundary operators $\widehat{Z[J]}$ in terms of baby universe creation and annihilation operators:
\begin{equation}\label{eq:freeZ}
	\widehat{Z[J]} \sim \mathfrak{Z}^{-1}\big\langle Z[J]\big\rangle\; +\; a^\dag_J+a_{J^*} + \cdots,
\end{equation}
where $\cdots$ indicates higher order terms.

One is then tempted to think of the states $\widetilde{|Z[J]\rangle}$ as (approximations to) states of a single closed baby universe, with  a wavefunction for the metric and other fields determined by the source $J$ (and by varying $J$ we would expect to obtain an overcomplete set of coherent states). We can diagonalise the inner product on the single-universe Hilbert space, taking linear combinations of $\widehat{Z[J]}$ for different $J$ to give operators $\widehat{Z_i}$ which are chosen to be Hermitian and give amplitudes satisfying
\begin{equation}
	\mathfrak{Z}^{-1}\langle Z_i Z_j\rangle - \mathfrak{Z}^{-2} \langle Z_i\rangle\langle Z_j\rangle	=\delta_{ij}\,.
\end{equation}
We can then write $\widehat{Z_i} = \langle Z_i\rangle + a^\dag_i+a_i+ \cdots$, with a more conventional oscillator algebra $[a_i,a^\dag_j]=\delta_{ij}$ labelled by an orthonormal basis of single-universe states. Repeated applications of $a^\dag_i$ are then said to create more universes, which can interact through topologies connecting three or more boundaries and into which we could incorporate as higher order terms in \eqref{eq:freeZ}. As long as these higher topologies are suppressed, we can thus construct a useful perturbation theory, where the inner product in \eqref{eq:BUaadag} gives the `free propagator' for single universe states, with higher topologies contributing vertices.

In particular, as noted above, based on the validity of the free approximation $\hbu$ appears to be well described by a Bosonic Fock space built on the single-universe Hilbert space. The Hartle-Hawking state provides the oscillator ground state, and multi-universe states are built by acting with $a^\dag_i$ operators. Alternatively, in the free approximation we can think of $\hbu$ in terms of the wavefunction $\Psi(Z_i)$, a function of the real variables $Z_i$. The operator $\widehat{Z}_i$ then acts as a position operator (or a free field operator in QFT, where the label $i$ could be momentum, for example), multiplying by $Z_i$. As the oscillator vacuum, the Hartle-Hawking state has a Gaussian wavefunction for each $Z_i$, shifted to be centred on $\langle Z_i\rangle$.

It is now tempting to use this free Fock space description to describe the spectrum of $\widehat{Z[J]}$, and hence the dual ensemble and the $\alpha$-states. We are led to expect that the spectrum of $\{Z_i\}$ has continuous support on the whole of $\RR$, independently for every $i$. In the resulting ensemble the $Z_i$, and hence the $Z[J]$, are normally distributed at the first nontrivial order described above, with covariance matrix given by the single-universe inner product\footnote{This is equivalent to the statement that the vacuum state of a free field theory is Gaussian with corresponding covariance matrix.} in \eqref{eq:BUaadag}. At each higher order, corrections from interactions would then appear to contribute only small non-Gaussian corrections to the measure, the conclusion reached in \cite{Preskill:1988na}, for example. However, in this respect, we have been misled by the free `approximation' \ref{eq:freeZ}. It turns out to be invalid because, while perturbation theory is accurate in many circumstances, it is not applicable in $\alpha$-states, as we will argue in a moment. The true, nonperturbative spectrum is smaller because the Fock space description of the Hilbert space is invalid once we take into account the null states \eqref{eq:nullstates} by which we must quotient by to obtain $\hbu$. Due to the null states, the `universe number' which grades the Fock space is not a diffeomorphism invariant observable.

Before we describe the breakdown of third-quantised perturbation theory, we clarify that it is not necessarily signalled by the dominance of spacetime wormhole effects. It may happen that the most important contribution to an amplitude comes from a nontrivial topology, but higher topologies remain negligible. This occurs prominently in two recent examples. The first is the spectral form factor $\langle Z(\beta+it)Z(\beta-it)\rangle$ of JT gravity \cite{Cotler:2016fpe,Saad:2019lba,Saad:2018bqo}, for which the contribution from the disconnected topology decays in time, while the connected topology gives a contributions that is exponentially suppressed but growing.   Eventually, the connected topology dominates, giving the `ramp'.  A second example is the $n$th R\'enyi entropy of an evaporating black hole after the Page time, which can be described as a sum of $n$-boundary amplitudes; the dominant configuration is a `replica wormhole', a spacetime which connects the $n$ boundaries \cite{Almheiri:2019qdq,Penington:2019kki}. However, higher topologies continue to be suppressed in such cases, and a similar perturbation theory remains valid; it simply happens to be dominated by $n$-universe vertices, so requires their inclusion.\footnote{This perturbation theory is also useful for discussing the average entanglement spectrum close to the Page time \cite{Penington:2019kki}, though it requires summation of a class of `tree-level' diagrams involving vertices of all valences.}

Instead, we are interested in cases when the third quantised perturbation theory fails entirely, and many topologies must be considered at once. For example, this occurs when we compute amplitudes with a parametrically large number of boundary components, giving very large moments of $\widehat{Z[J]}$. Equivalently, we can describe these amplitudes as the overlaps of states with very large universe occupation number\footnote{This notion is well-defined only in the third quantised perturbation theory, but can nonetheless be used to diagnose whether that perturbation theory is self-consistent.}. While any particular process of splitting and joining universes is suppressed, the total amplitude of such interactions is enhanced by combinatorial factors counting the number of processes with many possible universes (or joining many possible boundaries).  This allows higher topologies to become important.

Crucially, this breakdown of perturbation theory applies to $\alpha$-states and so is vitally important for understanding the spectrum of $\widehat{Z[J]}$.   The approximation of weakly interacting baby universes is thus not a reliable guide to the details of the spectrum. In the free theory, the $\alpha$-states are like position eigenstates in the harmonic oscillator.  They thus have infinite expectation value for the number operator. As we reduce the uncertainty in the $\alpha$ parameters and create a baby universe wavefunction with a more narrow spread, the mean universe occupation number increases, and eventually becomes exponentially large. At that point, the above approximation is not self-consistent for studying such states.

In retrospect, it should not be surprising that perturbation theory is of limited use for determining the spectrum of observables. As a simple example of similar behavior, if we perturb around the minimum of a potential in quantum mechanics, we cannot at any finite order tell whether the configuration space is compact, and hence if the momentum should be quantised.\footnote{We mentioned above the natural third quantization interpretation of $\widehat{Z[J]}$ as a position-like operator, but we could equally well have interpreted it as an analogue of free particle momentum}

The truncation of the spectrum of $\widehat{Z[J]}$ is invisible at any finite order in the third-quantised perturbation theory.  Thus in that description it could be seen only via some nonperturbative effect, or in an exact solution if one turns out to be available. Our models of \ref{sec:models} provide a simple example of the latter.  Recall that, in terms of the usual bulk perturbation theory in $G_N$, the spacetime wormholes describing third-quantised interactions are already nonperturbative, so the relevant expansion parameter is of the form $e^{-S}$ for an action $S$ of order $G_N^{-1}$. From this point of view,  the compression of the Hilbert space is then a \emph{doubly} nonperturbative effect, contributing to simple amplitudes as $e^{-c\, e^{-S}}$ for some (possibly imaginary) constant $c$.

\subsection{Perturbation theory in the topological model}

To give some insight into the validity of third quantised perturbation theory, we discuss its applicability in the context of the model of section \ref{sec:models}. We will restrict our considerations to the model without EOW branes.

The small parameter that suppresses topology is $e^{-S_0}$, with $S_0$ multiplying the Euler characteristic. It is natural to organise the third quantised perturbation theory as an expansion in that parameter, with higher genus topologies appearing as loops. However, the details of such an expansion (particularly accounting for diffeomorphisms of connected surfaces) are not necessary for the point we wish to illustrate.  To simplify the discussion, we thus instead assume that the full connected correlators (and thus any sums over connected surfaces with given boundaries) have already been computed exactly. These are all given by the same number $\lambda$, so our perturbation theory will be an expansion in inverse powers of $\lambda$.  As noted in section \ref{sec:models}, this expansion is organised by counting the number of connected components of spacetime.

Let us begin by noting a precise sense in which the free Gaussian approximation is appropriate at large $\lambda$. This follows from first observing that a sum of $N$ independent Poisson distributions with parameter $\lambda/N$ is again a Poisson distribution, with parameter $\lambda$. Taking $\lambda$ and $N$ large with fixed ratio then implies that we can apply the central limit theorem to the Poisson distribution as $\lambda\to\infty$. Specifically, we may define
\begin{equation}\label{eq:XZ}
	X = \frac{Z-\lambda}{\sqrt{2\lambda}},
\end{equation}
which has mean zero and variance unity. This $X$ is just new encoding of the boundary condition $Z$, with the shift by $\lambda$ acting to subtract the `tadpole' and set $\langle X\rangle=0$, and with an additional rescaling to fix the variance $\mathfrak{Z}^{-1}\langle X^2 \rangle=\frac{1}{2}$. The central limit theorem then implies that as $\lambda\to \infty$ the distribution of $X$ converges to a normal (and thus Gaussian) distribution.   In particular, at large $\lambda$ any amplitudes $\langle f(X)\rangle$ for bounded continuous functions $f$ (fixed independently of $\lambda$) approach those computed by integrating against a Gaussian. These are the vacuum amplitudes of a harmonic oscillator, with  wavefunction $\propto e^{-\frac{x^2}{2}}$, so this defines the `free' Gaussian approximation mentioned above.

We will return to the discussion of this wavefunction later.  Before doing so, we the large $\lambda$ expansion to study the moments $\mathfrak{Z}^{-1}\langle Z^n \rangle= B_n(\lambda)$ and note both when and how that expansion fails as we also take $n$ to be large. For fixed $n$, the leading order contribution at large $\lambda$ comes from completely disconnected spacetimes, giving $B_n(\lambda)\sim \lambda^n$. At the next order, we have spacetimes with $n-1$ disconnected components, which requires one `cylinder', a component joining two boundaries.\footnote{For simplicity of language, we will call this a cylinder even though it packages a sum over surfaces of all genus with two boundaries. A more precise language might refer to it as a renormalized cylinder.} There are $\binom{n}{2}=\frac{n(n-1)}{2}$ choices of which boundaries to join, so we have
\begin{equation}
	B_n(\lambda) = \lambda^n + \frac{n(n-1)}{2} \lambda^{n-1}+\cdots \quad\lambda\to\infty, \text{ fixed }n.
\end{equation}
At the next order, we have spacetimes with $n-2$ components, which means either two cylinders, or a `pair of pants' connecting a trio of boundaries to the same component of spacetime. We can continue in this way to any desired order $\lambda^{n-k}$ in the expansion by accounting for possible topologies with $n-k$ connected components.

Now, let us consider what happens when $n$ also becomes large. The first sign of trouble occurs when $n$ if of order $\sqrt{\lambda}$, when the second term in the above expansion is no longer smaller than the first. There are roughly $n^2/2$  ways to choose pairs of boundaries to join by a cylinder (neglecting the correction from choosing the same boundary twice), which is sufficiently large to overcome the suppression by $\lambda$. But this does not apply only for a single cylinder; terms with any number of cylinder components again contribute at the same (leading) order. In some sense our free approximation has failed.

However, it turns out that the large $\lambda$ expansion remains useful because we can explicitly account for the sum over configurations with $k$ cylinder components. For $2k\ll n$, there are approximately $\frac{1}{k!}\left(\frac{n^2}{2}\right)^k$ ways to select $k$ pairs of boundaries to join with a cylinder, where we have neglected the correction from `interactions', where the same boundary is chosen more than once. Summing over this `free gas of cylinders' gives us a multiplicative correction to the $n$th moment of $Z$,
\begin{equation}
	B_n(\lambda) \sim \lambda^n e^{\frac{n^2}{2\lambda}} \quad\lambda,n\to\infty, \text{ fixed } \frac{n^2}{\lambda}. \label{eq:Bncyl}
\end{equation}
In this regime, we can now systematically correct \eqref{eq:Bncyl} in powers of $\lambda^{-1}$ as before.  Such corrections can account for including higher topologies with more boundaries as well as compensating for the overcounting of cylinder configurations.

From \eqref{eq:Bncyl}, we see that $\langle Z^n\rangle$ is dominated by contributions with roughly $\frac{n^2}{\lambda}$ cylinder components. This can be much greater than one and the analysis will remain applicable, though it should certainly remain much less than $n$, so we must have $n\ll \lambda$. If this is the case, the correction from the cylinders is small in the sense that it is subleading to the $\lambda^n$ term when expressed as an expansion of $\log B_n(\lambda)$.

Taking $n$ larger still, \eqref{eq:Bncyl} remains accurate until $n$ is of order $\lambda^{2/3}$. At that point we find significant corrections from including any number of connected components having three boundaries  each (`pairs of pants'), and also from certain aspects of the overcounting of configurations of multiple cylinders.   In the latter context, the relevant configurations are those in which two cylinders end on the same boundary.  We previously included these configurations for simplicity (and to obtain a definite power of $\lambda$), but since they are not allowed we must now compensate by subtracting off their contributions.  Together, these two effects multiply \eqref{eq:Bncyl} an extra factor of $e^{-\frac{n^3}{3\lambda^2}}$. This pattern continues, with similar $e^{\# \frac{n^k}{\lambda^{k-1}}}$ corrections appearing whenever $n$ becomes of order $\lambda^{1-\frac{1}{k}}$ for $k=2,3,4,\ldots$. As discussed in appendix \ref{app:largelambda}, this structure is also apparent from a direct asymptotic expansion of $B_n(\lambda)$.

In summary, in the regime $\lambda \ll n$ the large $\lambda$ expansion remains a tractable way to compute the moments $\langle Z^n\rangle$ and is organized by types of contributing geometries. However, once $n$ is of order $\lambda$, this perturbation theory breaks down catastrophically, since there is no longer any suppression of connected topologies with many boundaries. This is the regime in which the novel effects of null states and gauge invariance become relevant, truncating the spectrum of $Z$ and making its discreteness apparent.

To explain this last statement in more detail, we first describe the state $|Z^n\rangle$ in the free approximation. We begin by translating to the harmonic oscillator position variable variable $X$ introduced in \eqref{eq:XZ}, writing $Z^n = \lambda^n\left(1+\sqrt{\frac{2}{\lambda}}X\right)^n$. Expanding $\log Z^n$ at large $\lambda$ (but any fixed $n$), this gives $\log Z^n = n \log \lambda + \sqrt{\frac{2}{\lambda}}n X + O(n\lambda^{-1})$.  We may thus approximate $Z^n \sim \lambda^n
\exp \left(\sqrt{\frac{2}{\lambda}}n X\right)$. For sufficiently small $n$ that the free approximation is applicable, we therefore have an approximate equivalence between the following states:
\begin{gather}
	\big|Z^n\big\rangle \simeq {\mathfrak{Z}}^{1/2} \lambda^n  e^{\sqrt{\frac{2}{\lambda}}n \hat{X}}\big|0\big\rangle \simeq (e\lambda)^n\Big|e^{\frac{n}{\lambda} Z}\Big\rangle
\end{gather}
Here the final equality uses \eqref{eq:XZ}, and the middle state lives in the harmonic oscillator Hilbert space of the free approximation.  In particular,  $|0\rangle$ is the (normalized) oscillator vacuum with wavefunction $\psi(X)\propto e^{-\frac{X^2}{2}}$.   After applying the exponential operator, the resulting wavefunction is a shifted Gaussian, which is a coherent state of the harmonic oscillator with average occupation number (here, `universe number') $\frac{n^2}{\lambda}$. From the above analysis, it follows that the free approximation is valid for universe numbers $N \ll \lambda$.

Now, a wavefunction of width $\Delta X$ in the $X$ variable has an occupation number that scales as $N \simeq(\Delta X)^{-2}$ as the width goes to zero, where the leading contribution comes from writing occupation number in terms of the Harmonic oscillator Hamiltonian and focusing on the kinetic term.  In terms of the width $\Delta Z$ in $Z$, this is $N\simeq \lambda (\Delta Z)^{-2}$. But resolving the natural integer discreteness in the spectum of $Z$ requires $\Delta Z\sim 1$, and hence $N$ of order $\lambda$.  As a result, and as one might expect, the discreteness of the $Z$ spectrum is thus associated with the complete breakdown of third quantised perturbation theory.

We can also see directly that this regime is connected with the appearance of null states, and thus the appearance of new gauge equivalences. Perhaps the simplest equivalence is that between the Hartle-Hawking state and the exponential $\big|e^{2\pi i Z}\big\rangle$. Note that any state $\big|e^{\alpha Z}\big\rangle$ is described in the free approximation by a coherent state with average occupation number $N\sim |\alpha|^2\lambda$. But for $\alpha$ of order one (for example, for $\alpha = 2 \pi i$) this is of order $\lambda$ and the free approximation fails.

All these phenomena occur when the state of baby universes has unsuppressed interactions with a given boundary. Roughly speaking, if we have a state of $\hbu$ containing $N$ closed universes and introduce a new boundary, the new boundary will connect to any given universe with amplitude $\lambda^{-1}$.  Hence it will connect to \emph{some} universe with amplitude $N/\lambda$. This effect becomes of leading order at $N$ of order $\lambda$, when the free description breaks down. We emphasise that this heuristic is appropriate for $N\ll \lambda$ when the free approximation can be used, but that $N$ itself becomes ill-defined once it becomes of order $\lambda$.  At that point,  null states appear and, furthermore, the null states are not preserved by any notion of universe number operator $\hat{N}$.

\section{Discussion}
\label{disc}

As with many works motivated by the black hole information problem, various readers may wish to focus on either the technical aspects of the above results or, alternatively, on their further significance for quantum gravity.  For this reason, we separate our discussion below into more technical remarks in section \ref{sec:summary} and a broader consideration of implications in section \ref{sec:implications}

\subsection{Summary and future directions}
\label{sec:summary}

We have seen that combining features of AdS asymptotics with the basic perspective of Coleman \cite{Coleman:1988cy} and of Giddings and Strominger \cite{Giddings:1988cx,Giddings:1988wv} from the late 1980's leads to a sharp structure in which states in a `baby universe Hilbert space' $\hbu$ control an ensemble of results for quantities $Z[J]$ computed at asymptotically AdS boundaries.  This version of the argument uses only manifest properties of the path integral and makes no further assumptions about locality.

Nevertheless, the final result is much the same as in
\cite{Coleman:1988cy,Giddings:1988cx}.  In particular,
the full bulk theory naturally includes both $\hbu$ and what one may call asymptotically AdS states, and there is a sense in which the two sectors interact.  However, the theory has superselection sectors for the algebra of operators on the asymptotically AdS states, so that an observer with no access to $\hbu$ naturally experiences an ensemble. The superselection sectors are associated with a complete orthonormal basis $\{ \big| \alpha \big \rangle\}$ of $\hbu$ in which the $Z[J]$ take definite values and exhibit factorization.  Thus for a given state $\big | \Psi \big \rangle \in \hbu$, the probability of outcome $Z_\alpha[J]$ is
$p_\alpha = \big| \big\langle \Psi \big| \alpha \big\rangle \big|^2$.  Furthermore, all properties of the full spectrum of superselection sectors can at least in principle be computed from correlators in the Hartle-Hawking no-boundary state $\big|\HH \big\rangle \in \hbu$.

We then explored this construction in detail in simple topological models inspired by Jackiw-Teitelboim gravity with and without end-of-the-world branes (EOW branes,  see e.g. \cite{Kourkoulou:2017zaj,Penington:2019kki}), and perhaps also with an extra boundary degree of freedom.  Without EOW branes, there is a single asymptotically AdS boundary condition $Z$, for which the associated operator $\widehat{Z}$ is naturally interpreted as the dimension of the CFT Hilbert space.  This operator is also present in the model with EOW branes.  Interestingly, the models predict this operator to have a quantized spectrum with eigenvalues $Z_{\alpha}\in e^{S_\partial-S_0}\NN$, where $S_\partial$ is a parameter associated with the extra boundary degree of freedom. The potential eigenstates associated with other potential eigenvalues turn out to be null states.  Perhaps even more intriguingly, unless $S_\partial$ is taken to be larger than $S_0+\log k$, the models with EOW branes are reflection positive only when all $Z_{\alpha}$  are nonnegative integers, and thus only when $e^{S_\partial-S_0} \in \NN$.  The particular ensemble defined by the Hartle-Hawking no-boundary state gives a Poisson distribution for the $Z_{\alpha}$.

Models with EOW branes have additional boundary conditions $(\psi_j,\psi_i)$ for $i,j=1,\dots k$.  The $(\psi_j,\psi_i)$ are naturally interpreted as the matrix of inner products between EOW brane states in a dual boundary quantum mechanics.  For given (integer) $Z_{\alpha}$, the eigenvalues of $\widehat{(\psi_j,\psi_i)}$ take the form $\sum_a \bar{\psi}^a_j\psi^a_i$ for some rectangular matrix $\psi^a_i$ of size $k \times Z_{\alpha_k}$.  As a result, the rank of any $(\psi_j,\psi_i)_\alpha$ cannot exceed either $k$ or $Z_{\alpha}$. The ensemble defined by the Hartle-Hawking no-boundary state arises from choosing independent complex Gaussian random entries for each of the $\psi^a_{i}$.

For $k \gg Z_{\alpha}$, this structure $(\psi_j,\psi_i)_\alpha=\sum_a \bar{\psi}^a_j\psi^a_i$ requires a sizeable compression of the naive the CFT Hilbert space (which would have had dimension $k$).   In particular, any list of more than $Z_{\alpha}$ states in the CFT Hilbert space turns out to be linearly dependent due to the presence of null states.  We also argued that a similar constraint on the number of linearly dependent states must arise in any theory where the gravitational path integral defines a positive semi-definite physical inner product.  Our general argument is closely related to ideas in \cite{Jafferis:2017tiu}, and various related suggestions can be found in e.g. \cite{Lowe:1995ac,Goheer:2002vf,Maloney:2015ina,Almheiri:2018xdw,Fu:2019oyc}.  But the result is deeply related to recent successes \cite{Penington:2019npb,Almheiri:2019psf,Almheiri:2019qdq,Penington:2019kki} in reproducing various forms of the Page curve associated with the black hole information problem.  With hindsight one can say that it was implicit in all of these works, and in fact moderately explicit in \cite{Penington:2019kki}.  But here we see that it is an exact statement at finite $Z$ in every possible baby universe state.

Indeed, in order to explain the R\'enyi computations of \cite{Penington:2019kki} for typical members of the Hartle-Hawking ensemble some version of this compression must occur whenever the number of a priori independent states inside a quantum extremal surface exceeds the generalized entropy defined by the region outside.  And due to a maximin argument \cite{Penington:2019npb,Almheiri:2019psf}, one expects this to occur whenever the number of a priori independent quantum states that can exist inside a given bulk domain of dependence with fixed exterior geometry exceeds the area of the codimension-2 surface where the past and future boundaries of this domain of dependence intersect;  see also \cite{Akers:2019lzs} for more on quantum maximin surfaces.

In the context of black hole evaporation, for general baby universe states $\big| \Psi \big\rangle$ this picture gives a sense in which interactions with baby universes formally lead to loss of information during the evaporation of black holes.  But as described previously in \cite{Coleman:1988cy,Giddings:1988cx,Giddings:1988wv,Polchinski:1994zs}, since the $\alpha$-states define superselection sectors for asymptotic observers, any given asymptotic observer can find no operational signs of this information loss.  In particular, while the observer may not be able to predict the exact outcome of an experiment involving black holes, they may simply consider the experiment to be a partial measurement of the previously unknown value of (in this interpretation unique) value of $\alpha$ describing the universe in which they live.  To the extent that $\alpha$ has been measured, no further information is then lost.

At the technical level there remain many interesting generalizations to explore in the future.  For example, even in the models discussed here, it would be useful to understand if one can formulate the Hilbert spaces $\hbu$ using slices at `finite time', or in other words without reference to asymptotic boundaries.    Moving beyond the current model, one would like to add topological matter, and also to explore a similarly topological version of the de Sitter models of \cite{Cotler:2019dcj} and \cite{Penington:2019kki}.  Work along these lines is in progress and we hope to report soon.  In the longer term, it is also clearly of interest to study more realistic models.

\subsection{Transcending the ensemble: implications and interpretations for each $\alpha$-sector}
\label{sec:implications}

We now turn to more speculative comments concerning the implications of our results above.

A key lesson from this work appears to be that, at least in sufficiently simple models, gravitational path integrals by themselves succeed in describing a great deal of microscopic information.  In particular, in our models the bulk path integral leads to a definite construction of the possible boundary theories --- defined by simultaneous eigenvalues $Z_\alpha[J]$ --- and also of the ensemble defined by the Hartle-Hawking state. However, this was possible only due to the exact solubility of the model, and in particular the convergence of the sum over topologies. In more realistic models, we will surely not be so fortunate.

Even in the simple case of JT gravity and its cousins \cite{Saad:2019lba,Stanford:2019vob,Penington:2019kki}, the gravitational path integral fails to converge. Though the model is sufficiently simple that the path integral for any given topology is exactly computable, the sum over topologies is an asymptotic series with zero radius of convergence in the expansion parameter $e^{-S_0}$. While there is an extremely natural completion of the model defined by a dual double-scaled matrix integral, it remains unclear whether the gravitational path integral uniquely selects this completion, or how it is realised in the bulk. This completion is associated with nonperturbative effects in the sum over topologies, which are \emph{doubly} nonperturbative in $G_N$. The same doubly nonperturbative scale was associated with truncation of the baby universe Hilbert space in our model, suggesting a tantalising connection to explore in more generality.

If we apply the ideas of this paper to more conventional `top-down' examples of AdS/CFT duality, such as type IIB supergravity (or string theory) with AdS$_5 \times S^5$ boundary conditions, there are several possible outcomes. The first possibility, suggested by our simple model and JT gravity, is that a nonperturbatively complete bulk theory defines a large Hilbert space $\hbu$ of baby universes. The eigenstates $\big| \alpha \big\rangle$ would then be associated with a menagerie of dual CFTs, and the Hartle-Hawking state again defines an ensemble of them. However, this is in tension with the established statement of the duality, which uniquely selects $\mathcal{N}=4$ Yang-Mills theory as a CFT dual.\footnote{Recall that a given $\alpha$-state determines partition functions for all possible boundary conditions on the bulk fields. These boundary conditions include specifications the flux on $S^5$ and the asymptotic dilaton, associated with the rank $N$ of the dual $U(N)$ gauge group and the 't Hooft coupling $\lambda$ respectively. An $\alpha$-state would specify a family of theories labelled by these parameters.} % (as well as theories for other asymptotics, such as AdS$_5 \times \RR\mathbb{P}^5$ giving other gauge groups).}
A nontrivial ensemble would require surprising new families of maximally supersymmetric CFTs; in particular, since $\mathcal{N}=4$ Yang-Mills is the unique such theory at weak coupling, these new CFTs must be strongly coupled throughout their moduli space.

Perhaps the more likely scenario is that $\mathcal{N}=4$ Yang-Mills is the unique dual and there is no ensemble. The baby universe Hilbert space interpretation is that $\hbu$ is one-dimensional, so the Hartle-Hawking state is the unique state of closed universes. The nonperturbative diffeomorphism invariance that produced null states is then required to act in the most emphatic possible fashion, rendering every possible state gauge equivalent.  This unique state must then also be an $\alpha$-state, and must exhibit factorization despite the existence of spacetime wormholes.  Nevertheless, in analogy with typical $\alpha$-states in our model, it remains possible that simple spacetime wormhole configurations still give excellent approximations to certain amplitudes.  Of course, in analogy with highly atypical $\alpha$-states in our model, it is also possible that
that simple spacetime wormhole configurations always receive large corrections.

An intermediate position is that the bulk theory leads to an ensemble interpretation in an asymptotic (say, large $N$) expansion, but there is a unique theory at any finite $N$. This is consistent with the observation \cite{Heemskerk:2009pn} that essentially any effective field theory in AdS solves the bootstrap order by order in large $N$ perturbation theory. We can thus emulate a consistent CFT in a large $N$ expansion, which nevertheless need not exist at any given finite $N$.

In any case, the suggestion is that the gravitational path integral should contain the full physics in each consistent $\alpha$-sector. And since the baby universe state in such sectors does not change, there is no room in a given sector for information loss.  As a result, the gravitational path integral should teach us how each consistent $\alpha$-sector transfers information to the outgoing Hawking radiation.

With this in mind, we recall that a key feature of the discussion in \cite{Coleman:1988cy,Giddings:1988cx,Giddings:1988wv} was the idea that one could integrate out the spacetime wormholes and describe their effects in terms of a modified effective action in which the detailed couplings were controlled by the $\alpha$-states.  In other words, the original theory with specified couplings and spacetime wormholes was equivalent (from the asymptotic point of view) to a theory with an ensemble of bulk couplings but where spacetime wormholes were forbidden. The same construction will apply in our context, but with one important distinction.  Namely, \cite{Coleman:1988cy,Giddings:1988cx,Giddings:1988wv}  focussed on wormholes with Planck-sized cross-sections under the assumption that microscopic wormholes would dominate in any physical process.  But the mouths of the replica wormholes in \cite{Almheiri:2019qdq,Penington:2019kki} are determined by the location of a quantum extremal surface.  As a result, they approximately coincide with the relevant black hole horizons and thus are macroscopic in size.  Integrating out such wormholes thus induces an ensemble of highly non-local couplings in the effective action.  Indeed, the couplings naturally mediate transitions in which any given interior configuration specifying the geometry and matter fields arbitrarily far inside the black hole can be replaced by any other, no matter deep the black holes throat may have become.  At least for replica numbers $n$ near $1$, the action for a replica wormhole whose mouth has area $A$ is of order $\frac{A}{4G}$ \cite{Lewkowycz:2013nqa}, so the amplitude for such processes should be exponentially small in this quantity.  However, in an old black hole the large number of internal states can lead to a large effect as seen directly above and in \cite{Penington:2019kki} (and as foreshadowed in \cite{Mathur:2015nra,Giddings:2017mym,Mathur:2017fnw}).

The exact location and nature of the above non-local interactions is clearly of some interest.  In particular, while quantum extremal surfaces may appear outside the black hole's event horizon \cite{Almheiri:2019yqk}, for black holes evaporating into a vacuum they should always lie inside \cite{Penington:2019npb,Almheiri:2019psf}.  Were all of the physics determined by replica wormholes confined far enough inside the horizon, there would be no possibility of affecting the exterior, and in particular no way it could purify the emitted Hawking radiation.  However, any separation of the QES from the horizon arises from time dependence, which is typically associated with quantum effects.  The backreaction of such effects on the spacetime is then suppressed by a power of $G$.  As a result, the  QES tends to be adiabatically close to any horizon, and thus separated by an amount only of order $G$. In addition, since the QES is determined by balancing the quantum effect of evaporating against a classical effect, the saddle-point is somewhat broad. A rough estimate of the width of the saddle-point suggests that the typical fluctuations of the area are also of order $G$.\footnote{\label{QESfootnote}For example, we can perform the path integral over replicated geometries and matter, while leaving unfixed the location of the QES where branching between replicas occurs. This leaves a final integral over the QES location to compute, which is roughly $\int e^{-S_\text{gen}}$ for $n$ close to $1$, where $S_\text{gen}$ is the generalised entropy of the QES and we integrate over its location. The integral over the area of the QES (fixing ingoing time, for example) is then $\int dA \,e^{-S_\text{gen}(A)}$, with $S_\text{gen}(A)\sim \frac{A}{4G} + \# \log(A_0-A)$ \cite{Penington:2019npb,Almheiri:2019psf}, where $A_0$ is the area of the (stretched) horizon. At the saddle point, where $A_0-A$ is of order $G$, we have $S_\text{gen}''(A)$ of order $G^{-2}$ leading to a width $\Delta A$ of order $G$.} This places the QES outside the horizon with order one amplitude. The associated non-local interactions will then naturally transfer information from the deep black hole interior into the outgoing Hawking radiation in much the form suggested in \cite{Giddings:2012gc,Giddings:2013kcj}.

However, for a full understanding of the physics associated with such interactions it appears one must take into account the corrections they imply for the theory's physical inner product.  As described in section \ref{sec:relation}, such corrections are associated with extending the familiar diffeomorphism invariance of gravitational systems to a more general slicing invariance of the path integral with topology change.   Extending this to arbitrary Euclidean time evolution --- even involving processes that change the topology of the slice used to define the quantum state --- implies spacetimes of different topologies to be gauge related.  In other words, this is a restatement of the old maxim that for gravitational systems time evolution is a gauge symmetry unless it involves evolution along an asymptotic boundary.  This then directly implies that the path integral computes the gauge invariant physical product as one would expect from general arguments \cite{Halliwell:1990qr,Marolf:1996gb,Reisenberger:1996pu,Hartle:1997dc}  (though admittedly those arguments are most direct in contexts where it is not obvious that topology change should be included).

As a result, one may think of the induced nonlocal interactions as modifying the gravitational constraints; i.e., with new terms in the Wheeler-DeWitt equation.  The interesting feature, however, is that these modifications are highly non-generic.  In the regime that in our models corresponds to $k \gg Z_\alpha$ , there are a large number of strongly correlated small corrections, where the correlations conspire to give a large number of null states; i.e., they make the physical inner product highly degenerate so that a priori independent states are in fact linearly dependent in the physical Hilbert space, and so that the dimension of the physical Hilbert space is bounded by $Z_\alpha$.  Furthermore, following ideas related to \cite{Jafferis:2017tiu}, we argued in section \ref{sec:relation} that null states must enforce a similar bound in a general reflection positive gravitational path integral.

It is this bound that leads to the Page curve, and which thus determines the rate at which the above interactions transfer information out of the black hole. As a result, while the above non-local interactions are intimately tied to this change in the inner product, it is natural to think of the former as secondary and the latter as primary.  In particular, it is in terms of the inner product that (for reflection positive path integrals) we find a clean statement of the correlations and conspiracies inherent in the details of the induced interactions; see again section \ref{sec:relation}.

We believe the explicit demonstration of such a large number of null states to be a lesson of fundamental importance.  It implies that --- due to the above mentioned conspiracies --- the gauge symmetry of gravitational systems is much larger and more powerful than had been previously established.  The idea that bounds on entropy might be related to such a gauge symmetry date back at least to the early 1990's, when such suggestions arose in discussions of black hole complementarity proposals (see e.g.\ comments in \cite{Lowe:1995ac}) and cosmological analogues in de Sitter space.  It is also much like the truncation of the bulk Hilbert space implicit in random tensor network models \cite{Hayden:2016cfa,Qi:2018shh} in which the disorder is implemented by inserting randomly chosen projections into the bulk.  However, we now see this to be a direct result of the gravitational path integral.

The physics of this enlarged gravitational gauge invariance remains to be understood in detail, especially in the context of more realistic models.  Nevertheless, the argument of section \ref{sec:relation} indicates that the long discussed relation \cite{Maldacena:2001kr,VanRaamsdonk:2009ar,VanRaamsdonk:2010pw,Jafferis:2017tiu} between two-sided bulk black holes and bulk thermofield double states \eqref{eq:TFD} should be understood as an example of this gauge equivalence. In particular, we now see that the so-called ``superselection sectors" of \cite{Marolf:2012xe} --- which were argued there to be physically distinct --- are in fact gauge equivalent.\footnote{This gauge equivalence resolves a problem noted in that work concerning how such superselection sectors in transform under permutations.}

We now speculate further on the implications of this enhanced gauge invariance for issues involving black hole information and the connection to other works.  It seems clear that in sufficiently old black holes (where the number of a priori independent internal states is sufficiently large), this gauge invariance implies that vast numbers of a priori independent states must in fact to be regarded as physically equivalent.    Furthermore, at least in our model, this happens in an essentially random way that does not respect any additional structure\footnote{In particular, the spectrum of possibilities allows \emph{any} Hermitian inner product of the appropriate rank.}.  Extrapolating this result to more complicated models suggests that one will find many states which a priori seem to have very different physics --- and in particular in which infalling observers have vastly different experiences --- but which are nevertheless gauge equivalent.  For example, just as there can be gauge equivalence between Alice meeting Bob and Alice finding only empty space, there is no reason for the physical inner product to respect Alice's notion of particle number (as distinguished, say, from total charges coupled to a gauge field), or even her notion of particle number in a given mode. As a result, even for pure state black holes, the experience of observers inside the black hole may fundamentally fail to be well-defined as a gauge invariant concept. One may view this as a variant of the firewall-like possibility described in \cite{Marolf:2013dba} that black holes may have `no interior', or at least no interior from which familiar physics can be extracted.

Nevertheless, as with any gauge symmetry, one is free to fix a gauge in order to define a language (i.e., a set of observables) with which to describe the physics.  In particular, as noted above, at the level of Hilbert spaces any gauge invariance is naturally associated with what one may roughly call a projection $P$ from some kinematic Hilbert space $\hilb_\text{kin}$ to a physical Hilbert space\footnote{A structure of this general sort is inherent in Dirac's constraint quantization of gauge systems \cite{Dirac}, though the interested reader can consult \cite{Landsman:1993xe,Marolf:1994wh,Ashtekar:1995zh,Marolf:2000iq,Shvedov:2001ai} for a variety of more technical treatments.} $\hilb_\text{phys} \subset \hilb_\text{kin}$.   In this sense, one may think of a general gauge fixing procedure as a choice of linear subspace $\hilb_\text{GF} \subset \hilb_\text{kin}$ such that $P$ defines a bijection between $\hilb_\text{GF}$ and $\hilb_\text{phys}$.  Within a given such gauge fixing scheme, it may then be that the experiences of infalling observers become well-defined.  For example, in describing the interior of a black hole of radius $R_0$ that recently formed from collapse, it would be natural to choose a gauge in which the interior is of size comparable to $R_0$ (even if such small interiors are gauge equivalent to certain much larger interiors that might form when an initially much larger black hole decays to size $R_0$), and in particular in which standard effective field theory is a good approximation.

With this in mind, we recall that the discussions of  \cite{Penington:2019npb,Almheiri:2019hni,Almheiri:2019yqk,Almheiri:2019psy,Chen:2019uhq,Almheiri:2019qdq,Penington:2019kki,Chen:2019iro} described a close parallel between old black holes that have been radiating into an external system (`the bath') and the ER=EPR paradigm of \cite{Maldacena:2013xja}. In particular, these works suggested that infalling observers experience only standard physics even at the horizon of black holes that have been evaporating for longer than the Page time. At first sight such statements may seem to be in great tension with our bound on the number of linearly independent states inside the black hole.  But this tension can be resolved by interpreting the comments of
\cite{Penington:2019npb,Almheiri:2019hni,Almheiri:2019yqk,Almheiri:2019psy,Chen:2019uhq,Almheiri:2019qdq,Penington:2019kki,Chen:2019iro}
as providing a gauge fixed description, where in this case the choice of gauge depends on the state of the bath.  In other words, if the black hole system with physical Hilbert space $\hilb_\text{phys}$ is considered in the presence of another system with Hilbert space $\hilb_\text{bath}$ then, even if the bath system by itself has no gauge invariance, one is free to gauge fix by choosing a general linear subspace
$\hilb_\text{GF, joint} \subset \hilb_\text{kin} \otimes \hilb_\text{bath}$ for which $P$ defines a bijection to
$ \hilb_\text{phys} \otimes \hilb_\text{bath}$.   Note that there is no requirement for $\hilb_\text{GF, joint}$ be a tensor product $\hilb_{\text{GF}_0} \otimes \hilb_\text{bath}$ for any fixed subspace $\hilb_{\text{GF}_0} \subset \hilb_\text{kin}$.  Instead, one is free to effectively let the choice of subspace $\hilb_{\text{GF}_0} \subset \hilb_\text{kin}$ vary with the choice of state in $\hilb_\text{bath}$.

The connection with the above works is particularly clear in the discussion of Petz reconstruction in \cite{Penington:2019kki}.  There one wishes to reconstruct an operator $\mathcal{O}$ on $\hilb_\text{kin}$ using an operator $\mathcal{O}_R$ on $\hilb_\text{bath}$.   Now, since
$\mathcal{O}_R$  is an operator on  $\hilb_\text{bath}$, it is automatically gauge invariant.  However, since the operators $\mathcal{O}$ discussed in that work were constructed without regard to the (random) physical inner product, they are not gauge invariant.  This is consistent, as  $\mathcal{O}_R$  reconstructs $\mathcal{O}$ only on a subspace $\hilb_\text{code} \subset \hilb_\text{kin} \otimes \hilb_\text{bath}$ that similarly fails to be gauge invariant.  However, at least to good approximation we can think of $\hilb_\text{code}$ as defining a partial gauge fixing (meaning that we could choose some $\hilb_\text{GF,joint} \supset \hilb_\text{code}$.  In particular,  we may use any bath bra-state $\langle \psi_\text{bath}|$ to define a linear map from $\hilb_\text{code}$ to $\hilb_\text{kin}$ via its natural action on $\hilb_\text{bath}$.  And for any choice of $\langle \psi_\text{bath}|$, the image defines a subspace $\hilb_{\psi} \subset \hilb_\text{kin}$ with at most dimension $d_\text{code} \ll e^{S_\text{BH}}$, i.e., where this dimension is much less than the dimension of $\hilb_\text{phys}$.  As a result, with high probability distinct states in $\hilb_{\psi}$ will project to distinct states of $\hilb_\text{phys}$.  In this sense $\hilb_\text{code}$ approximately satisfies the requirements for a partial gauge fixing; a complete gauge fixing would result from extending $\hilb_\text{code}$ to make the projection of each $\hilb_{\psi}$ isomorphic to $\hilb_\text{phys}$.

We note that such a gauge fixed interpretation allows all of the hallmarks of what is often called state dependence \cite{Nomura:2012sw,Papadodimas:2012aq,Verlinde:2012cy,Verlinde:2013uja} and which is naturally associated with the ER=EPR paradigm.  In particular, in contexts where one expects to find only a small number of black hole states (states in $\hilb_\text{phys}$) for each bath state, it will be possible to choose a partial gauge fixing of the form described above that selects only states in $\hilb_\text{kin}$ with no drama at the horizon.  In particular, one will be able to choose a code subspace within which the evolution can be well-described by standard local effective field theory.   In addition, we note that standard objections \cite{Almheiri:2013hfa,Marolf:2013dba,Bousso:2013wia,Bousso:2013ifa,Marolf:2015dia} to state dependence focus on non-uniqueness of the predicted physics, and that such objections are clearly moot in a context where the state dependence is simply a choice of gauge (so that non uniqueness of $\hilb_\text{GF}$ is to be expected, and so that the gauge invariant predictions are in fact identical).

Nevertheless, the non-uniqueness arguments of \cite{Almheiri:2013hfa,Marolf:2013dba,Bousso:2013wia,Bousso:2013ifa,Marolf:2015dia} then show the sort of states that, while they appear at first sight to be physically distinct, must in fact be related by the enlarged gauge symmetry described above.  In particular, tracing through such leads to other gauges in which infalling observers experience varying amounts and types of drama at the horizon, as well as to gauges where the observer simply fails to exist in the interior of the black hole.\footnote{If one imposes the constraint that the observer survives (in a recognizable form) for a given proper time behind the black hole horizon, then one would expect a generic gauge consistent with this constraint to predict the maximum amount of such drama consistent with the observer's survival to that point.}  Furthermore, just as there is a particular gauge (or class of gauges) realizing ER=EPR-like scenarios, it seems likely that one can also find gauges realizing fuzzball scenarios (see e.g. \cite{Mathur:2005zp,Bena:2007kg,Balasubramanian:2008da,Skenderis:2008qn,Mathur:2008nj,Chowdhury:2010ct,Bena:2013dka,Mathur:2012jk,Mathur:2013gua}, the non-violent non-locality proposal\footnote{The non-locality scale $L_d$ in spacetime dimension $d$ is set by the condition $\Delta A \sim G$ described in footnote \ref{QESfootnote}.  On a Killing slice of a static black hole of area-radius $R$, the corresponding proper distance from the event horizon would be $L_d \sim \left(\frac{\ell_p}{R}\right)^{\frac{d-4}{2}} \ell_p$. With respect to the definitions of \cite{Giddings:2012gc}, $L_d$ then gives ``non-violent" physics for $d<4$.} \cite{Giddings:2011ks,Giddings:2012gc,Giddings:2013kcj}), proposals emphasizing the bulk Wheeler-DeWitt equation \cite{JACOBSON:2013ewa,Jacobson:2019gnm}, the black hole final state proposal \cite{Horowitz:2003he}, and perhaps other proposals as well.

On the other hand, the above discussion immediately raises the question of how different experiences of a given observer could possibly be gauge related, and thus how the above scenario could possibly be realized in models that are sufficiently realistic to describe our own universe.    While there is surely more to be said about this issue, we note that any gauge fixing scheme can be used to \emph{define} an associated gauge invariant observable.  I.e., just as one can use Coulomb gauge in electromagnetism to define gauge invariant operators (``the potential in Coulomb gauge"), in the above scenario one can use any gauge to define a notion of observer inside the black hole.  The variety of possible gauges would then mean that there are a variety of possible gauge invariant definitions of the observer which happen to coincide (or nearly coincide) under familiar conditions outside old black holes but which differ greatly inside old black holes.  One may then rephrase the above statement in a less surprising manner:  While we may well-enough understand how to define an observer at the leading semi-classical level, there may be many possible extensions of this definition at the level of non-perturbative physics, and predictions for the observer inside old black holes may depend sensitively on the choice of this extension\footnote{Note that if there is a priori no mechanism for selecting one such definition as preferred, then it is natural to adopt a Bayesian approach and declare that all such extensions are realized with equal probability (or more generally that they are realized according to some probability measure describing the priors of the given theorist studying the system).    The question of `what does an observer experience when falling into a black hole' would then be an inherently probabilistic one, somewhat akin to asking `what does an observer experience when they are decohered into many Everett branches of the wavefunction of the universe?'  We have already conjectured above that with high probability the observer simply fails to exist inside the black hole in a generic gauge, and that post-selecting only on existence of the observer would lead to high drama.}. The scenario described above (in which apparently distinct observer experiences are gauge related) may thus be considered to be just another version of this idea. It will likely be of great interest to further explore such conjectures and related physics in future work.

\paragraph{Acknowledgments}
This work was motivated and facilitated by three specific conversations, first with Geoffrey Pennington, second with Xi Dong, and third Steve Giddings, as well as by a long history of discussing black hole information with the entire UCSB High Energy and Gravity group.  We also acknowledge interesting conversations with Daniel Harlow, Gary Horowitz, Ted Jacobson, Javier Mag\'an, Juan Maldacena, Xiaoliang Qi, Steve Shenker, Mark Srednicki, Douglas Stanford, Herman Verlinde and Edward Witten. We are grateful for support from NSF grant PHY1801805 and funds from the University of California.  H.M.~was also supported in part by a DeBenedictis Postdoctoral Fellowship, and D.M. thanks UCSB's KITP for their hospitality during the final portions of this work.   As a result, this research was also supported in part by the National Science Foundation under Grant No. NSF PHY-1748958 to the KITP.

\appendix

\section{Limits of moments of the Poisson distribution}

In this appendix, we study the moments $\langle Z^n \rangle$ of a Poisson random variable $Z$ with mean $\lambda$ in various limits. This is useful to ascertain the convergence properties of sums $\sum_n c_n|Z^n\rangle$ constructing states of $\hbu$ in section \ref{sec:models}, and to illustrate the failure of the third quantised perturbation theory of section \ref{sec:3q} in our model.

The moments are given by the Bell polynomials,
\begin{equation}
	\mathfrak{Z}^{-1} \langle Z^n \rangle = B_n(\lambda),
\end{equation}
defined by
\begin{equation}
	B_n(\lambda) = e^{-\lambda}\sum_{d=0}^\infty \frac{\lambda^d}{d!} d^n \,.
\end{equation}
From this, one can check the recurrence relation
\begin{equation}
	B_{n+1}(\lambda) = \lambda (B_n'(\lambda)+B_n(\lambda))
\end{equation}
and $B_0(\lambda)=1$, from which we can see that $B_n(\lambda)$ is a monic polynomial of order $n$. In particular this gives us the scaling at large $\lambda$ and fixed $n$,
\begin{equation}
	B_n(\lambda) \sim \lambda^n, \qquad \lambda\to\infty,\enspace n \text{ fixed.}
\end{equation}

\subsection{Large $n$ and convergence}\label{app:convergence}

For studying convergence of $\sum_n c_n|Z^n\rangle$, we require the moments at large $n$ and fixed $\lambda$. For this, observe that the ratio of consecutive terms in the sum defining $B_n(\lambda)$ is
\begin{equation}
	\frac{\lambda}{d}\left(\frac{d}{d-1}\right)^n \sim \frac{\lambda}{d} e^{n/d},
\end{equation}
where the asymptotic form applies for $1\ll n \ll d^2$. For large $n$, the ratio is unity and hence the $d$th term in the sum is maximal when $d\sim \frac{n}{\log n}$. Substituting this value back into the sum, we can find an estimate of $B_n(\lambda)$ at large $n$, which we can write as
\begin{equation}
	\frac{B_n(\lambda)}{n!} \sim e^{-n \log\log n+ o(n)}, \quad n\to \infty, \quad \lambda \text{ fixed}.
\end{equation}
\begin{equation}
	\log B_n(\lambda) \sim n \log n -n \log\log n - n+o(n), \quad n\to \infty, \quad \lambda \text{ fixed}.
\end{equation}
For a more carful derivation and many more terms in the expansion, it is convenient to write $d = \frac{n}{\log n}\left(1+\frac{x}{\sqrt{n}}\right)$ and take the limit of the terms in the sum as $n\to\infty$ at fixed $x$. In this limit, the series becomes a Gaussian integral in $x$. From this, we can estimate the norm of the basis state  $\Vert |Z^n\rangle \Vert =\sqrt{\langle Z^n |Z^n \rangle} = e^{-\lambda/2}\sqrt{B_{2n}(\lambda)}$:
\begin{equation}\label{eq:normAsymp}
	\log \Vert |Z^n\rangle \Vert =n\log n -n\log\log n - n(1-\log 2) +o(n) \qquad \text{as }n\to\infty.
\end{equation}

Now we can begin to characterise convergence of sums $\sum c_n |Z^n\rangle$ in the baby universe Hilbert space of section \ref{sec:ZBU}. By definition, the series converges if the partial sums form a Cauchy sequence. That is,
\begin{equation}
	\sum_{n=0}^\infty c_n |Z^n\rangle \quad \text{converges} \iff \left\Vert \sum_{n=n_1}^{n_2} c_n |Z^n\rangle \right\Vert \to 0 \text{ as }n_1,n_2\to\infty,
\end{equation}
where in this limit we can take $n_1,n_2$ to infinity separately at different rates.\footnote{It may not be that every element of the completion can be represented by such a Cauchy sequence of partial sums. It is false for the `free' version where we allow only discs and cylinders,  replacing the Poisson distribution by its Gaussian approximation: in that case, this class of Cauchy sequences yields only analytic wavefunctions.}  We will not characterise such series completely, but find a sufficient condition to give us a class of convergent series, and a necessary condition to constrain them.

First, a necessary condition for convergence (coming from $n_1=n_2$) is that the norm of individual terms go to zero
\begin{equation}
	\text{Convergence} \implies |c_n| \big\Vert |Z^n\rangle \big\Vert \to 0 \text{ as }n\to \infty.
\end{equation}
Now, from \eqref{eq:normAsymp}, we see that $\big\Vert |Z^n\rangle \big\Vert$ is eventually larger than $R^n$ for any $R>0$, so $|c_n|R^n$ is bounded, which implies that $f(z):= c_n z^n$ converges in the disc $|z|<R$. Since this holds for all $R$, we find that our series defines an entire analytic function,
\begin{equation}
	\sum_{n=0}^\infty c_n |Z^n\rangle \quad \text{converges} \implies f(z)=\sum c_n z^n \text{ is entire analytic.}
\end{equation}
We can thus characterise convergent series in terms of the class of allowed analytic functions. Improving on the analyticity result, we can bound the growth of allowed functions $f$. To do this, we introduce the order of an analytic function, which is the infimum over all $\rho$ such that $|f(z)|<\exp(|z|^\rho)$ for sufficiently large $z$. We can strengthen our necessary
condition to
\begin{equation}
	\sum_{n=0}^\infty c_n |Z^n\rangle \quad \text{converges} \implies f(z)=\sum c_n z^n \text{ has order $\leq 1$,}
\end{equation}
which means that for every $\epsilon>0$, we have $|f(z)|< \exp(|z|^{1+\epsilon})$ for sufficiently large $|z|$. To show this, we use a result expressing the order in terms of the Taylor coefficients, namely $\operatorname{order}(f)= \limsup_{n\to\infty} \frac{n \log n}{\log(1/|c_n|)}$. For the norm of the terms in the series to go to zero, we must have $\log(1/|c_n|) - \log \big\Vert |Z^n\rangle \big\Vert$ go to infinity, so for sufficiently large $n$ we have $\log(1/|c_n|)> \log \big\Vert |Z^n\rangle \big\Vert$.  From \eqref{eq:normAsymp}, for any $\epsilon>0$ and sufficiently large $n$ we have $\log \big\Vert |Z^n\rangle \big\Vert > (1-\epsilon) n\log n$. In turn, this means that $\log(1/|c_n|)>(1-\epsilon) n\log n$ for large enough $n$, and hence $\limsup_{n\to\infty} \frac{n \log n}{\log(1/|c_n|)} \leq 1$.

Our sufficient condition is absolute convergence, which means that the sum of norms converges, and follows from the triangle inequality for the norm.
\begin{equation}
	\sum_n |c_n| \big\Vert |Z^n\rangle \big\Vert \text{ convergent} \implies \sum_n c_n |Z^n\rangle \text{ convergent}.
\end{equation}
Now, from \eqref{eq:normAsymp}, we have the result that $\big\Vert |Z^n\rangle \big\Vert$ decays faster than $n! a^n$ for any $a$. From this, we can find a simple sufficient bound on the coefficients for convergence,
\begin{equation}
	|c_n| < A \frac{x^n}{n!} \text{ for some $A,x$ }\implies \sum_n c_n |Z^n\rangle \text{ convergent}.
\end{equation}
In particular, this means that any exponential function $|e^{x Z}\rangle$, or more generally a function of exponential type, defines a convergent series by its Taylor expansion.

The gap between our sufficient and necessary conditions (order one functions that are not of exponential type) is small but nonempty, for example containing $\frac{1}{\Gamma(-z)}$.

\subsection{Large $\lambda$ and $n$}\label{app:largelambda}

Here, we study a limit of $\lambda\to\infty$ and $n\to\infty$ at fixed ratio $\nu=\frac{n}{\lambda}$, which will interpolate between the large $\lambda$ fixed $n$ and large $n$ fixed $\lambda$ results. We could proceed from the same series expression, but we use an alternative method, starting from an integral representation of $B_n(\lambda)$. This expression extracts the moments from the generating function \eqref{eq:genfunc} by a contour integral
\begin{equation}
	\frac{B_n(\lambda)}{n!} = \frac{1}{2\pi i} \oint \frac{du}{u^{n+1}} e^{\lambda (e^u-1)},
\end{equation}
where the contour encircles the origin. We can evaluate this by steepest descent, looking for stationary points of
\begin{equation}
	S(u)=e^u -1 - \nu \log u\,.
\end{equation}
The stationary points $S'(u)=0$ solve $\nu = u e^u$, and the relevant saddle point is the unique positive solution, which defines the Lambert $W$ function or product logarithm,
\begin{equation}
	 u_* = W(\nu).
\end{equation}
Applying the steepest descent method at this saddle point gives us
\begin{equation}
	\frac{B_n(\lambda)}{n!} \sim \frac{e^{\lambda S(u_*)}}{u_*\sqrt{2\pi S''(u_*) \lambda}}\,.
\end{equation}
This result in fact interpolates between our two previous results for large $\lambda$ fixed $n$ (by taking $\nu\ll 1$) and large $n$ fixed $\lambda$ (by taking $\nu \gg 1$).

It is interesting in particular to see how the large $\lambda$ result breaks down when $n$ becomes large. Taking $\nu\ll 1$ we have $u_* = \nu -\nu^2 +O(\nu^3)$, so $S(u_*)\sim -\nu\log\nu + \nu +\frac{1}{2}\nu^2 + \cdots$, with higher terms all integer powers of $\nu$. Substituting this into the steepest descent result, we have
\begin{equation}
	\frac{e^{\lambda S(u_*)}}{u_*\sqrt{2\pi S''(u_*) \lambda}} \sim \frac{e^{n\log\lambda-n\log n +n + \frac{n^2}{2\lambda}+\cdots}}{\sqrt{2\pi n}} \sim \frac{\lambda^n}{n!} e^{\frac{n^2}{2\lambda}+\cdots},
\end{equation}
where we applied Stirling's approximation to the factorial. The terms in the exponential are of the form $\frac{n^{k}}{\lambda^{k-1}}$ for $k=2,3,\ldots$, and become relevant when $n$ is of order $\lambda^{1-1/k}$. The first correction occurs from the $k=2$ term shown explicitly, first relevant when $n$ is of order $\sqrt{\lambda}$, when it contributes an order one rescaling of $B_n(\lambda)$:
\begin{equation}
	B_n(\lambda) \sim \lambda^n e^{\frac{n^2}{2\lambda}},\qquad \lambda\to\infty, \frac{n^2}{\lambda} \text{ fixed.}
\end{equation}
Higher order terms in the exponential are given by higher orders in the expansion of $S(u_*)$ at small $\nu$.

\bibliographystyle{JHEP}
\bibliography{biblio}

\end{document}